\documentclass[12pt,a4paper]{article}
\usepackage{amsmath,amsfonts,amsthm,amssymb}
\usepackage{bm,booktabs}
\usepackage{color}
\usepackage{dashrule}
\usepackage{enumitem}
\usepackage{float,fancyhdr}
\usepackage{graphicx,graphics}
\usepackage{indentfirst,listings}
\usepackage{mathrsfs,multirow,multicol}
\usepackage{natbib,nicefrac}
\usepackage{rotating,relsize}
\usepackage{setspace,subfigure,scalefnt, verbatim}
\usepackage{tabularx,tabu}
\usepackage[paper=a4paper,lmargin=2cm,rmargin=2cm,tmargin=2cm,bmargin=2cm]{geometry}
\usepackage[utf8x]{inputenc}
\usepackage[colorlinks=true,linkcolor=red,citecolor=blue]{hyperref}
\usepackage[english]{babel}
\usepackage[left]{lineno}
\usepackage{arydshln}
\usepackage{supertabular}
\usepackage{adjustbox}
\usepackage{array}
\usepackage{float}

\allowdisplaybreaks

\graphicspath{}

\usepackage{caption}

\begin{document}

\onehalfspacing

\title{\textbf{The Bivariate Defective Gompertz Distribution Based on Clayton Copula with Applications to Medical Data} \\
[.5cm]}
\author{\small \textbf{Marcos Vinicius de Oliveira Peres$^{(1)}$} \small\textbf{Ricardo Puziol de Oliveira$^{(2)}$} \\
\small\textbf{Jorge Alberto Achcar$^{(1)}$} 
\small\textbf{Edson Zangiacomi Martinez$^{(1)}$} \\
\\
$^{(1)}$ \small Ribeirão Preto Medical School, University of Sao Paulo (USP), Ribeirao Preto, SP, Brazil\\
$^{(2)}$ \small Department of the Environment, State University of Maringá (UEM), Umuarama, PR, Brazil.\\
}
\date{}
\maketitle

\begin{abstract}
	In medical studies, it is common the presence of a fraction of patients who do not experience the event of interest. These patients are people who are not at risk of the event or are patients who were cured during the research. The proportion of immune or cured patients is known in the literature as cure rate. In general, the  traditional existing lifetime statistical models are not appropriate to model data sets with cure rate, including bivariate lifetimes. In this paper, it is proposed a bivariate model based on a defective Gompertz distribution and also using a Clayton copula function to capture the possible dependence structure between the lifetimes. An extensive simulation study was carried out in order to evaluate the biases and the mean squared errors for the maximum likelihood estimators of the parameters associated to the proposed distribution. Some applications using medical data are presented to show the usefulness of the proposed model.

\bigskip

\end{abstract}


\thispagestyle{empty}

\section{Introduction}
The use of survival statistical models for time-to-event data is common in several areas of study, especially in medical research. Traditional parametric and non-parametric tools, such as, Kaplan-Meier estimator for the survival function, log-rank and Wilcoxon tests and the semi-parametric Cox proportional hazard model, are widely used in medical data analysis (see, e.g., \citealp{kleinbaum2010survival}).  These methods assume that all individuals are susceptible to the event of interest. However, for example in clinical studies,  there may be patients who will not experience the event under investigation, that is, these patients are immune to the event or they were cured during the research. This situation is suggested when a Kaplan-Meier estimator plot for the survival function describes a behavior with stable plateau and large censored data at the right of the curve \citep{corbiere2009penalized,wienke2010frailty}. In this way, the use of models that incorporate this plateau, named cure rate models, could be a better alternative to predict or to identify prognostics factors that affects the survival probability.

According to \cite{Vahidpour2016}, there are at least two kinds of models for data with cure fraction: the mixture cure rate models, also known as standard cure rate models (see, for example, \citealp{de1999mixture,Tsodikov2003,lambert2006estimating}), and the non-mixture cure rate models, which are not so popular (see \citealp{achcar2012cure,Vahidpour2016}). Let us denote by $ T $, the time for the occurrence of the event of interest. Following \cite{Maller1996}, the standard cure rate model  assuming that the probability of the time-to-event to be greater than a specified time $t$ is given by the survival function,
\begin{equation}\label{curemixture}
	S(t) = \rho + (1 - \rho)S_0(t)
\end{equation}
where $\rho \in (0,1)$ is the mixing parameter which represents the proportion of ``long-term survivors'', ``non-susceptible'' or ``cured patients'', and $S_0(t)$ denotes a proper survival function for the non-cured or susceptible group in the population. Observe that if $t\rightarrow \infty$, then $S(t)\rightarrow \rho$, that is, the survival function has an asymptote at the cure rate $\rho$. 

On other hand, the non-mixture model defines an asymptote for the survival function, that is associated to the cure rate (see, \citealp{Tsodikov2003}). In this case, the survival function for the non-mixture cure rate model is given by,
\begin{equation}\label{curenonmixture}
	S(t) = \rho^{F_0(t)} = \exp\{\ln(\rho)F_0(t)\}
\end{equation} 
where $\rho \in (0,1)$ is the probability of cured patients and $F_0(t) = 1 - S_0(t)$ denotes a proper distribution function for the non-cured or susceptible group in the population. 

Different approaches have been presented in the literature to model cure rate, especially for univariate lifetime data: \cite{boag1949maximum}, \cite{ghitany1992asymptotic}, \cite{de1999mixture}, \cite{chen2002bayesian}, \cite{lambert2006estimating}, \cite{castro2009bayesian}, \cite{chen1999new}, \cite{achcar2012cure} and \cite{martinez2013mixture}. However, the cure rate models are not the only ones to deal with long-term survivors, we also could use, as an alternative, the defective models.  The main property of a proper probability distribution is that $  \lim\limits_{t\rightarrow\infty} F(t) =1$ and, consequently, $  \lim\limits_{t\rightarrow\infty} S(t) =0$. For defective models, the survival function, $ S(t) $, converges to a value $ \rho $, where $ \rho $ denotes the cure rate.  Some approaches for defective models can been found in: \cite{cancho2001modeling}, \cite{balka2011bayesian}, \cite{da2014bayesian}, \cite{dos2017bayesian}, \cite{rocha2017new}, \cite{martinez2018new}, among others.

In some studies, the main objective may be related to analyze the lifetime data assuming two time-to-event variables. As a special situation, we could be interested in the times of occurrence of a specified event, that could be reinfection, in the treatment of both lungs where we could use univariate lifetime models assuming independence between both time-to-event variables. However, in this situation, the times could not be independent since the patient needs both lungs working to survive. In this case, there may be the presence of a dependence structure that is not present when using univariate analyzes associated with each response, which is a motivation for the use of bivariate models. Different bivariate parametric models are introduced in the literature for the  analysis of bivariate lifetime 
data:  \cite{marshall1967generalized}, \cite{block1974continuous}, \cite{vaupel1979impact} and \cite{block1974continuous}, \cite{wienke2003bivariate}, \cite{yu2008mixture}, \cite{achcar2013block}, \cite{fachini2014bivariate} and \cite{de2019discrete}. As an alternative, the dependence structure can be specified by using a Copula function due to its simplicity. According to \cite{hofert2019elements} a copula is a multivariate distribution function with standard uniform univariate marginals. Many copula functions are considered to model data with cure rate: \cite{wienke2006modelling}, \cite{li2007mixture}, \cite{fachini2014bivariate}, \cite{martinez2014bayesian}, \cite{coelho2016bivariate} and \cite{achcar2016bivariate}.
More recently, \cite{PERES2020e03961} conducted a comprehensive review of fifteen different copula functions that can be used to model survival data.

The main goal of this paper is to explore the use of the Clayton copula in the analysis of bivariate lifetime data assuming a bivariate defective Gompertz distribution to estimate the cure rate. Different correlation values between the time-to-event variables are considered in a simulation study that was done in order to describe the behavior of the dependence structure of the proposed model. The maximum likelihood method using existing numerical optimization algorithms was considered to get the inferences of interest under a frequentist approach and MCMC (Markov Chain Monte Carlo) simulation methods, as the popular Gibbs sampling and Metropolis-Hastings algorithms, were used to get the posterior summaries of interest under a Bayesian approach \citep{Gelfand1990,chib1995}.  The paper is organized as follows: in Section 2, it is presented the proposed methodology using the Clayton copula as well the inference methods. The simulation procedures and the obtained results are showed in Section 3. In Section 4, four applications related to real medica data are presented, using the proposed methodology. Finally, Section 5 closes the paper with some concluding remarks.

\section{Statistical Methods}

\subsection{Univariate Defective Gompertz Distribution}

The main property of defective models is a survival function $ S(t) $ that converges to a value $\rho$ as $ t $ tends to infinity, where $\rho$ denotes the cure rate parameter. \cite{cantor1992parametric} introduced the two-parameter defective Gompertz (DG) distribution also studied by \cite{gieser1998modelling} and \cite{dos2017bayesian}. The survival function for the DG distribution is given by,
\begin{equation}\label{SDG}
S(t)=\exp\left\lbrace - \dfrac{\alpha}{\beta}\left[1-\exp(-\beta t)\right]  \right\rbrace,
\end{equation}
where $t > 0$ and $\alpha > 0$ is the shape parameter and $\beta > 0$ is the scale parameter. Taking the limit of the survival function from the DG distribution, the cure rate parameter $\rho$ is given by
\begin{equation}\label{rhoDG}
\rho=\lim\limits_{t\rightarrow\infty}S(t)=\exp\left\lbrace - \dfrac{\alpha}{\beta}\right\rbrace.
\end{equation}
The correspondent probability density and hazard function are respectively given by
\begin{equation}
f(t)=\alpha\exp(-\beta t)\exp\left\lbrace - \dfrac{\alpha}{\beta}\left[1-\exp(-\beta t)\right]  \right\rbrace \quad\text{and}\quad h(t)=\alpha\exp(-\beta t).
\end{equation}
Note that the hazard function has only decreased shape, and this is an important limitation of a model based on the DG distribution.

\subsection{Copula Functions}\label{copulas}
Copula functions are used to create a joint distribution function of two or more marginal univariate distributions following standard uniform distribution $ \text{U}(0,1) $ to form a multivariate distribution \citep{nelsen2007introduction}. Considering a $ m $-variate function $ F $, the respective copula is a function $ C:[0,1]^m\rightarrow[0,1]$ that satisfies
\begin{equation}\label{mcop}
F(y_1,\ldots,y_m)=C(F_1(y_1),...,F_m(y_m);\phi)=C_\phi(F_1(y_1),...,F_m(y_m)),
\end{equation}
where $\phi$ is a parameter that measures the dependence between the marginals. The join probability density is given by
\begin{equation}
f(y_1,\ldots,y_m)=c_\phi(F_1(y_1),...,F_m(y_m))\prod_{i=1}^{m}f_i(y_i),
\end{equation}
where $ f_i(y_i),\; i=1,...,m$,  are the marginal density functions and $ c_\phi(F_1(y_1),...,F_m(y_m)) $  is the derivative of order $ m $ of \eqref{mcop} in relation to  $ y_1,...,y_m $. If the random variables are independent, then $c_\phi(F_1(y_1),...,F_m(y_m))=1 $.

For the bivariate case $ (m=2) $ and under the context of survival analysis, considering $S_1(t_1)$ and $S_2(t_2)$ as the univariate survival functions, the bivariate joint survival function $S(t_1,t_2)$ is defined by a copula function given by
\begin{equation}
S(t_1,t_2)=C_\phi(S_1(t_1),S_2(t_2)),
\end{equation}
for $  t_1> 0 $ and $ t_2> 0 $, with the respective joint probability density function given by
\begin{equation}
f(t_1,t_2)=\frac{\partial^2S(t_{1},t_{2})}{\partial t_{1}\partial t_{2}}=f_1(t_1)f_2(t_2)c_\phi(S_1(t_1),S_2(t_2)),
\end{equation}
where $ c_\phi(u,v) $ is the copula density function defined by
\begin{equation}
c_\phi(u,v)=\frac{\partial^2}{\partial u \partial v}C_\phi(u,v),
\end{equation}
where $u=S_1(t_1)$ and $v=S_2(t_2)$.

The estimation of the correlation between two random variables using copula functions usually is made using the Kendall's tau ($ \tau_k $) and Spearman's rho ($\tau_s$). According to \cite{joe2014dependence}, those coefficients can be expressed by the equations
\begin{equation}
\tau_k=1-4\int_{0}^{1}\int_{0}^{1}\dfrac{\partial C_\phi(u,v)}{\partial u}\dfrac{\partial C_\phi(u,v)}{\partial v}\,\text{d}u\text{d}v
\end{equation}
and
\begin{equation}\label{corr}
\tau_s=12\int_{0}^{1}\int_{0}^{1}C_\phi(u,v)\,\text{d}u\text{d}v-3
\end{equation}

The literature introduces many copula functions which could be considered to build different bivariate lifetime distributions. However, it is important to choose copula functions suitable for each type of dependence structure in the applications. In each application, it is possible to obtain some information on the dependence structure by an exploratory graphical analysis, but unfortunately this can be difficult in some cases. Another framework  that can help in choosing the copula function is to determine the empirical correlation between the random variables. This correlation can be obtained through iterative multiple imputation \citep{schemper2013estimating}.  In the present study, we explore the Clayton copula function as a special case when appropriate in the data analysis. The Clayton copula is a popular choice to be fitted by bivariate time-to-event data, due its ability to describe positive dependence.

The Clayton copula was first introduced by \cite{clayton1978model} and later studied by \cite{cook1981family} and \cite{oakes1982model}. Assuming this copula function, the joint survival function $S(t_1,t_2)$ is given by
\begin{equation}\label{clayton}
S(t_1,t_2)=\left\lbrace \left[ S_1(t_1) \right]^{-\phi} +\left[S_2(t_2) \right]^{-\phi}-1 \right\rbrace^{-1/\phi} ,
\end{equation}
where $ S_1(t_1) $ and $ S_2(t_2)$  are, respectively, the marginal survival functions for the random variables $T_1$ and $T_2$  and $\phi \in (0,\infty)$. When $ \phi \rightarrow 0 $, there is an indication that $T_1$ and $T_2$ are independent. The relationship between the copula parameter $ \rho $ and the dependence structure can be interpreted  by the Spearman's correlation $\tau_s(\phi) $. However, obtaining this measure using the equation \eqref{corr} can be a difficult task. The Kendall's correlation coefficient is given by
\begin{equation}\label{corrClayton}
\tau_k(\phi)=\frac{\phi}{\phi+2}.
\end{equation}
Note that $ 0<\tau(\phi)\leq 1$, where  if $ \phi \rightarrow \infty $, we have total dependence between $T_1$ and $T_2$. The Clayton copula, is thus adequate to model positive dependences and it has the advantage of measuring a wide range of positive correlations. The respective joint probability density function for $T_1$ and $T_2$ is given by
\begin{equation}\label{fClayton}
f(t_1,t_2)=f_1(t_1)f_2(t_2)(1+\phi)\left[ S_1(t_1)S_2(t_2)\right] ^{-1-\phi} \left\lbrace \left[ S_1(t_1) \right]^{-\phi} +\left[S_2(t_2) \right]^{-\phi}-1\right\rbrace^{-2-1/\phi},
\end{equation}
where $ f_1(t_1) $ and $ f_2(t_2)$  are, respectively, the marginal probability density functions for the random variables $T_1$ and $T_2$. 


	
\subsection{Bivariate Defective Gompertz Distribution}
The marginal probability density  and survival functions for the lifetimes $T_j\, (j=1,2)$ considering the DG distribution are given, respectively, by
\begin{equation}\label{fj}
	f_j(t_j)=\alpha_j\exp(-\beta_j t_j)\exp\left\lbrace - \dfrac{\alpha_j}{\beta_j}\left[1-\exp(-\beta_j t_j)\right]  \right\rbrace
\end{equation}
and
\begin{equation}\label{Sj}
	S_j(t_j)=\exp\left\lbrace - \dfrac{\alpha_j}{\beta_j}\left[1-\exp(-\beta_j t_j)\right]  \right\rbrace.
\end{equation}
Thus, the correspondent cure rates are given by
\begin{equation}\label{rhos}
	\rho_j=\exp\left\lbrace - \dfrac{\alpha_j}{\beta_j}  \right\rbrace,
\end{equation}
where $ j $ is equal to 1 or 2, corresponding to the time-to-event variables $ T_1 $ and $ T_2 $, respectively.

The joint survival and density functions for the bivariate defective Gompertz distribution using a Clayton copula function \eqref{clayton} (BDGD) are given, respectively, by
\begin{equation}\label{SBDG}
S(t_1,t_2)=\left\lbrace \left[ \exp\left\lbrace - \dfrac{\alpha_1}{\beta_1}\left[1-\exp(-\beta_1 t_1)\right]  \right\rbrace \right]^{-\phi} +\left[\exp\left\lbrace - \dfrac{\alpha_2}{\beta_2}\left[1-\exp(-\beta_2 t_2)\right]  \right\rbrace \right]^{-\phi}-1 \right\rbrace^{-1/\phi} ,
\end{equation}
and,
\begin{eqnarray}\label{FBDG}
f(t_1,t_2)&=& 
\alpha_1\alpha_2\exp(-\beta_1 t_1-\beta_2 t_2)\exp\left\lbrace - \dfrac{\alpha_1}{\beta_1}\left[1-\exp(-\beta_1 t_1)\right]  - \dfrac{\alpha_2}{\beta_2}\left[1-\exp(-\beta_2 t_2)\right]  \right\rbrace \nonumber\\
&\times & (1+\phi)\left[\exp\left\lbrace - \dfrac{\alpha_1}{\beta_1}\left[1-\exp(-\beta_1 t_1)\right]   - \dfrac{\alpha_2}{\beta_2}\left[1-\exp(-\beta_2 t_2)\right]  \right\rbrace\right]^{-1-\phi}\nonumber\\
& &\left\lbrace \left[ \exp\left\lbrace - \dfrac{\alpha_1}{\beta_1}\left[1-\exp(-\beta_1 t_1)\right]  \right\rbrace \right]^{-\phi} +\left[\exp\left\lbrace - \dfrac{\alpha_2}{\beta_2}\left[1-\exp(-\beta_2 t_2)\right]  \right\rbrace \right]^{-\phi}-1 \right\rbrace^{-2-1/\phi}
\end{eqnarray}

%
%

\subsection{Inference Methods}
\subsubsection{Maximum Likelihood Estimation}
To obtain the bivariate likelihood function, let us assume a random sample of size $ n $, where each sample has two lifetimes $ T_1 $ and $ T_2 $. Let us consider that both $ T_1 $ and $ T_2 $ can be right-censored and that this censoring is independent of each time-to-event. For each $ i^{th} $ observation $ (i = 1, \ldots, n) $  it is possible to classify the data  into one of four classes given by,

\begin{description}
	\item[(1)] $ C1:$ both $ t_{1i} $ and $ t_{2i} $ are uncensored lifetimes;
	
	\item[(2)] $ C2: t_{1i} $ is a complete lifetime and $ t_{2i} $ is a censored lifetime;
	
	\item[(3)] $ C3: t_{2i} $ is a complete lifetime and $ t_{1i} $ is a censored lifetime;
	
	\item[(4)] $ C4: t_{1i} $ and $ t_{2i} $ are censored lifetimes.
\end{description}

Thus, the likelihood function is given by
\begin{equation}
L=\prod_{i \in C_1}\left[ f(t_{1i},t_{2i}) \right] \prod_{i \in C_2}\left[-\frac{\partial S(t_{1i},t_{2i})}{\partial t_{1i}} \right] \prod_{i \in C_3}\left[-\frac{\partial S(t_{1i},t_{2i})}{\partial t_{2i}} \right] \prod_{i \in C_4}\left[S(t_{1i},t_{2i}) \right],
\end{equation}
where $ f(t_{1},t_{2}) $ is the joint probability function of $ T_1 $ and $ T_2 $, given in equation (\ref{fClayton}) and $ S(t_{1},t_{2}) $ is the joint survival function given by equation (\ref{clayton}) considering the Clayton copula.

Let us consider two indicator variables, denoted by $\delta_{1i} $ and  $\delta_{2i} $, where $\delta_{ki} =1$ when $t_{ki}$ is an observed lifetime and $\delta_{ki} =0$ when $t_{ki}$ a censored observation, $ k = 1, 2 $ and $ i = 1, . . ., n $. In this way, it is possible to rewrite the likelihood function as
\begin{eqnarray}\label{Lik}
L=\prod_{i = 1}^{n}\left[f(t_{1i},t_{2i})  \right]^{\delta_{1i}\delta_{2i}}
\left[-\frac{\partial S(t_{1i},t_{2i})}{\partial t_{1i}} \right]^{\delta_{1i}(1-\delta_{2i})}\left[-\frac{\partial S(t_{1i},t_{2i})}{\partial t_{2i}} \right]^{\delta_{2i}(1-\delta_{1i})}\left[S(t_{1i},t_{2i}) \right]^{(1-\delta_{1i})(1-\delta_{2i})}.
\end{eqnarray}

In the absence of censored observations, the expression above is  reduced to the form,
\begin{eqnarray}\label{dev1}
L&=&\prod_{i =1}^{n}\frac{\partial^2S(t_{1i},t_{2i})}{\partial t_{1i}\partial t_{2i}}  = \prod_{i =1}^{n}f(t_1,t_2).
\end{eqnarray}

For the Clayton copula, the first partial derivatives of $ S(t_{1}, t_{2}) $ with respect to $ t_{1} $ and $ t_{2} $ are given by the following relations,
\begin{equation}\label{dev2}
-\frac{\partial S(t_{1},t_{2})}{\partial t_{1}}=f_1(t_1)S_1(t_1)^{-(\phi+1)}\left[S_1(t_1)^{-\phi}+S_2(t_2)^{-\phi}-1\right]^{-(1+1/\phi)}
\end{equation}
and
\begin{equation}\label{dev12}
-\frac{\partial S(t_{1},t_{2})}{\partial t_{2}}=f_2(t_2)S_2(t_2)^{-(\phi+1)}\left[S_1(t_1)^{-\phi}+S_2(t_2)^{-\phi}-1\right]^{-(1+1/\phi)}.
\end{equation}

\subsection{Bayesian Analysis}
Assuming the proposed model, let $ \bm{\theta}=({\alpha_1},{\beta_1}, {\alpha_2},{\beta_2},{\phi}) $	 be the vector of unknown parameters. Under a Bayesian framework, the joint posterior distribution for the model parameters is obtained by combining the joint prior distribution of the parameters and the likelihood function given by equation (\ref{Lik}) \citep{gelman2013bayesian}. To simulate samples from the joint posterior distribution, we could consider the use of MCMC (Markov Chain Monte Carlo) algorithms implemented in the R2jags package \citep{plummer2003jags} in \textit{R} software, where we  just need to specify the data distribution and the prior distribution for the parameters.

Under a Bayesian approach, we assume independent uniform prior distributions for the parameters $ {\alpha_1},\;{\beta_1}\;{\alpha_2},\;{\beta_2}\; \text{and}\;{\phi}$. That is, we assume $ \alpha_1 \sim Unif(a_1,b_1) $, $ \alpha_2 \sim Unif(a_2,b_2) $, $ \beta_1 \sim Unif(a_3,b_3) $, $ \beta_2\sim Unif(a_4,b_4) $ and $ \phi \sim Unif(a_6,b_6) $, where $ a_k $ and $ b_k $,$  k =1,...,4 $, are known hyperparameters, and $ Unif(a,b) $ denotes a uniform distribution with mean $ (a+b)/2 $ and variance $ (a+b)^2/12 $.  The values of hyperparameters $a$ and $b$ were chosen in order to reflect prior knowledge of experts and better performance of the MCMC algorithm in terms of good convergence. These values were obtained using empirical Bayesian methods \citep{carlin2000} as information on the cure rate obtained from the non-parametrical  Kaplan-Meier estimator for the survival function and information on the correlation obtained from empirical estimators.

\section{Simulation Study}\label{simul}

The simulation study was carried out in order to evaluate the performance of the maximum likelihood (ML) estimation. The coverage probability of the Wald confidence intervals  for the parameters $\alpha_1,\;\alpha_2,\; \beta_1,\;\beta_2,\;\rho_1,\;\rho_2$ and $\phi$, with their corresponding bias and mean squared errors (MSE) were considered. Calculations of the coverage probabilities were carried out for a nominal coverage of 95\%, corresponding to 95 successes in each 100 simulated samples. Since $ \rho_1 $ and $ \rho_2 $ are functions of other parameters, the Wald confidence interval  for these parameters were obtained using the delta method \citep{oehlert1992note}. In this simulation study, the coverage probability is defined as the observed percentage of times that the confidence interval includes the respective parameter. The bias and MSE in the estimation of a parameter $\eta$ are given, respectively, by,
\begin{equation}\label{bias}
\widehat{\text{Bias}}(\widehat{{\eta}})=\frac{1}{N}\sum_{i=1}^{N}\left(\widehat{{\eta}}^{(i)}-{\eta} \right) 
\end{equation}
and
\begin{equation}\label{MSE}
\widehat{\text{MSE}}(\widehat{{\eta}})=\frac{1}{N}\sum_{i=1}^{N}\left(\widehat{{\eta}}^{(i)}-{\eta} \right) ^2,
\end{equation}
where we denote  $ \widehat{{\eta}}$ as each $ \widehat{\alpha_1},\;\widehat{\alpha_2},\;\widehat{\beta_1},\;\widehat{\beta_2},\;\widehat{\rho_1},\;\widehat {\rho_2}\;\text{and}\;\widehat{\phi} $, $\eta$ is the nominal value of the corresponded parameter, and  $ N $ is the number of simulated samples of size $ n $.

To generate bivariate data, we used an adaptation of the algorithm introduced by \cite{balakrishnan2009continuous} and used by \cite{ribeiro2017abordagem} and by \cite{peres2018bivariate}, along with an  algorithm to defective distributions presented by \cite{rocha2017new} and used by \cite{martinez2017defective,martinez2018new}. We generate random samples of size $n=50,75,100,\ldots,500$ in twelve different scenarios presented in Table (\ref{scen}). The steps of the proposed generation algorithm are described below. 
\begin{description}
	\item[Step 1:] Fix values for the parameters: $\alpha_1,\;\alpha_2,\;\beta_1,\;\beta_2$ and $\phi$.
	
	\item[Step 2:] Calculate $ \rho_1 $ and $ \rho_2 $.
	
	\item[Step 3:] Generate $n$ random samples from $ M_{1i}\sim Bernoulli(1-\rho_1)$.
	
	\item[Step 4:] Generate $n$ random samples from $ u_{1i}\sim U(0,1-\rho_1)$.
	
	\item[Step 5:] For  $i=1,...,n$ consider $ t_{1i}^*=\infty $ if $  M_{i1}=0 $ and $ t_{1i}^*=F_1^{-1}(u_{1i}) $ if $  M_{i1}=1 $, where the inverse of the distribution function is given by,
		\begin{equation}
		F_1^{-1}(u_{1i})=-\dfrac{1}{\beta_1}\ln\left[1+\dfrac{\beta_1}{\alpha_1}\ln(1-u_{1i})\right].
		\end{equation}
	
	\item[Step 6:] Generate $n$ random samples from $ u^*_{1i}\sim U(0,max(t_{1i}^*))$, considering only finite values of $t_{1i}^*  $.
	
	\item[Step 7:] Consider $ t_{1i}=min(t_{1i}^*,u^*_{1i}) $.
	
	\item[Step 8.] Pairs of values  $ (t_{1i},\delta_{1i})$ are thus obtained, where $\delta_{1i} =1$ if $ t_{1i} < u^*_{1i} $ and $\delta_{1i} =0$ if $ t_{1i} > u^*_{1i}$.
	
	\item[Step 9:] Generate $n$ random samples from $ M_{2i}\sim Bernoulli(1-\rho_2)$.
	
	\item[Step 10:] Generate $n$ random samples from $ u_{2i}\sim U(0,1-\rho_2)$.
	
	\item[Step 11:] Generate $n$ random samples from $ k_{i}\sim Bernoulli(\phi)$.
	
	\item[Step 12:] Get values from $w_{i}$, considering the following expression,
	\begin{equation}
	w_{i}=\min\left\lbrace u_{1i}^{-(\phi+1)}(u_{1i}^{\phi}+u_{2i}^{\phi})^{-\left(\frac{1+\phi}{\phi} \right) },1-\rho_2\right\rbrace .
	\end{equation}
	This expression is the derivative of (\ref{clayton}) with respect to  $u_{i1}$, when $ S(t_{1i})=u_{i1}$ and $ S(t_{2i})=w_{i}$. 
	
	\item[Step 13:] For  $i=1,...,n$ consider $ K_{i}= M_{1i}$ if $  k_{i}=1 $ and $ K_{i}= M_{2i}$ if $  k_{i}=0 $.
	
	\item[Step 14:] For  $i=1,...,n$ consider $ t_{2i}^*=\infty $ if $  K_{i}=0 $ and $ t_{2i}^*=F_2^{-1}(w_i) $ if $  K_{i}=1 $, where the inverse of the distribution function is given by,
	\begin{equation}
	F_2^{-1}(u_{1i})=-\dfrac{1}{\beta_2}\ln\left[1+\dfrac{\beta_2}{\alpha_2}\ln(1-w_{i})\right].
	\end{equation}
	
	\item[Step 15:] Generate $n$ random samples from $ u^*_{2i}\sim U(0,max(t_{2i}^*))$, considering only finite values of $t_{2i}^*  $.
	
	\item[Step 16:] Consider $ t_{2i}=min(t_{2i}^*,u^*_{2i}) $.
	
	\item[Step 17.] Pairs of values  $ (t_{2i},\delta_{2i})$ are thus obtained, where $\delta_{2i} =1$ if $ t_{2i} < u^*_{2i} $ and $\delta_{2i} =0$ if $ t_{2i} > u^*_{2i}$.
\end{description}

\begin{table}[!h]
	\caption{Nominal values assumed for each scenario considered in the simulation study.}
	\label{scen}
	\setlength{\tabcolsep}{12pt}
	\renewcommand{\arraystretch}{1.2}
	\centering
	\begin{adjustbox}{max width=\textwidth,max totalheight=\textheight,keepaspectratio}
	\begin{tabular}{@{}cccccccccccccc@{}}
		\toprule
		&          & \multicolumn{12}{c}{Scenarios}                                                                                 \\ \cmidrule(l){3-14} 
		&          & 1   & 2   & 3   & \multicolumn{1}{c|}{4}   & 5   & 6   & 7   & \multicolumn{1}{c|}{8}   & 9    & 10   & 11   & 12  \\ \midrule
		\multirow{5}{*}{\rotatebox{90}{Parameter}} & $ \phi  $    & \multicolumn{4}{c|}{1.0}                   & \multicolumn{4}{c|}{3.0}                   & \multicolumn{4}{c}{10.0} \\
		& $ \alpha_1 $ & 1.0 & 0.5 & 1.0 & \multicolumn{1}{c|}{0.5} & 1.0 & 0.5 & 1.0 & \multicolumn{1}{c|}{0.5} & 1.0  & 0.5  & 1.0  & 0.5 \\
		& $ \alpha_2 $ & 1.0 & 0.5 & 0.5 & \multicolumn{1}{c|}{1.0} & 1.0 & 0.5 & 0.5 & \multicolumn{1}{c|}{1.0} & 1.0  & 0.5  & 0.5  & 1.0 \\
		& $ \beta_1 $  & 0.8 & 1.5 & 0.8 & \multicolumn{1}{c|}{1.5} & 0.8 & 1.5 & 0.8 & \multicolumn{1}{c|}{1.5} & 0.8  & 1.5  & 0.8  & 1.5 \\
		& $ \beta_2 $  & 0.8 & 1.5 & 1.5 & \multicolumn{1}{c|}{0.8} & 0.8 & 1.5 & 1.5 & \multicolumn{1}{c|}{0.8} & 0.8  & 1.5  & 1.5  & 0.8 \\ \bottomrule
	\end{tabular}\end{adjustbox}
\end{table}

In the presence of a cure rate, it was considered nominal values for the parameters such that the samples generated have low and high percentage of cure rate in scenarios with parameter values $ \alpha_i=1.0 $ and $ \beta_i=0.8, $ where the cure rate parameter is given by $ \rho_i\approx0.2865 $, and  the scenarios parameter values $ \alpha_i=0.5 $ and $ \beta_i=1.5 $ where we have cure rate parameter given by $ \rho_i\approx0.7165$ $ (i=1,2) $. From scenarios 1 to 4 (combinations of the fixed parameter values) in Table (\ref{scen}), it was considered $ \phi=1.0 $, that is, $ \tau_k(\phi)=0.3333 $ and $ \tau_s(\phi)=0.4790 $, which corresponds to a moderate correlation between $T_1$ and $T_2$. from 5 to 8 (combinations of the fixed parameter values) given in Table (\ref{scen}), it was considered $ \phi=3.0 $, so $ \tau_k(\phi)=0.6000 $ and $ \tau_k(\phi)=0.7864 $, representing a high correlation between $T_1$ and $T_2$. Finally, in the scenarios from 9 to 12 (combinations of the fixed parameter values), it was considered very high correlation between $T_1$ and $T_2$, with $ \phi=10.0 $, which leads to $ \tau_k(\phi)=0.8333 $ and $ \tau_s(\phi)=0.9583 $ (see Table \ref{scen}).

The ML estimates and corresponding standard errors for each simulated sample were computed using the \textit{maxLik} package in \textit{R} \citep{maxlik}, and the Nelder-Mead maximization method, considering 95\% nominal confidence intervals for the parameters. It was obtained in each scenario (Table (\ref{scen})) the ML estimates of the parameters, the coverage probability of the confidence intervals, bias and MSE for each parameter of interest $ \rho_1 $, $ \rho_2 $ and $\phi$, as well as the percentage of samples resulting in the presence of monotone likelihood functions (error informed by \textit{maxLik}).

\subsection{Results}

This section presents simulations results, for each scenario presented in Table (\ref{scen}). It was observed that the percentage of censored  data generated in the proposed simulation algorithm (Section \ref{simul}) was about 5\% higher than the respective percentage of the nominal cure rate ($ \rho_1$ and $ \rho_2$) considered in the generating samples. Moreover, for each simulated sample, it was calculated the Kendall's correlation $ \tau_k $ by the Clayton copula approach and the Spearman correlation $ \tau_s $ by numerical methods. Also, a re-parametrization of the parameter $ \phi $ was considered in order to obtain flexible results for the coverage probability.

\begin{figure}[H]
\centering
\setkeys{Gin}{height = 4.2cm,width=0.24\linewidth} %
		\includegraphics[]{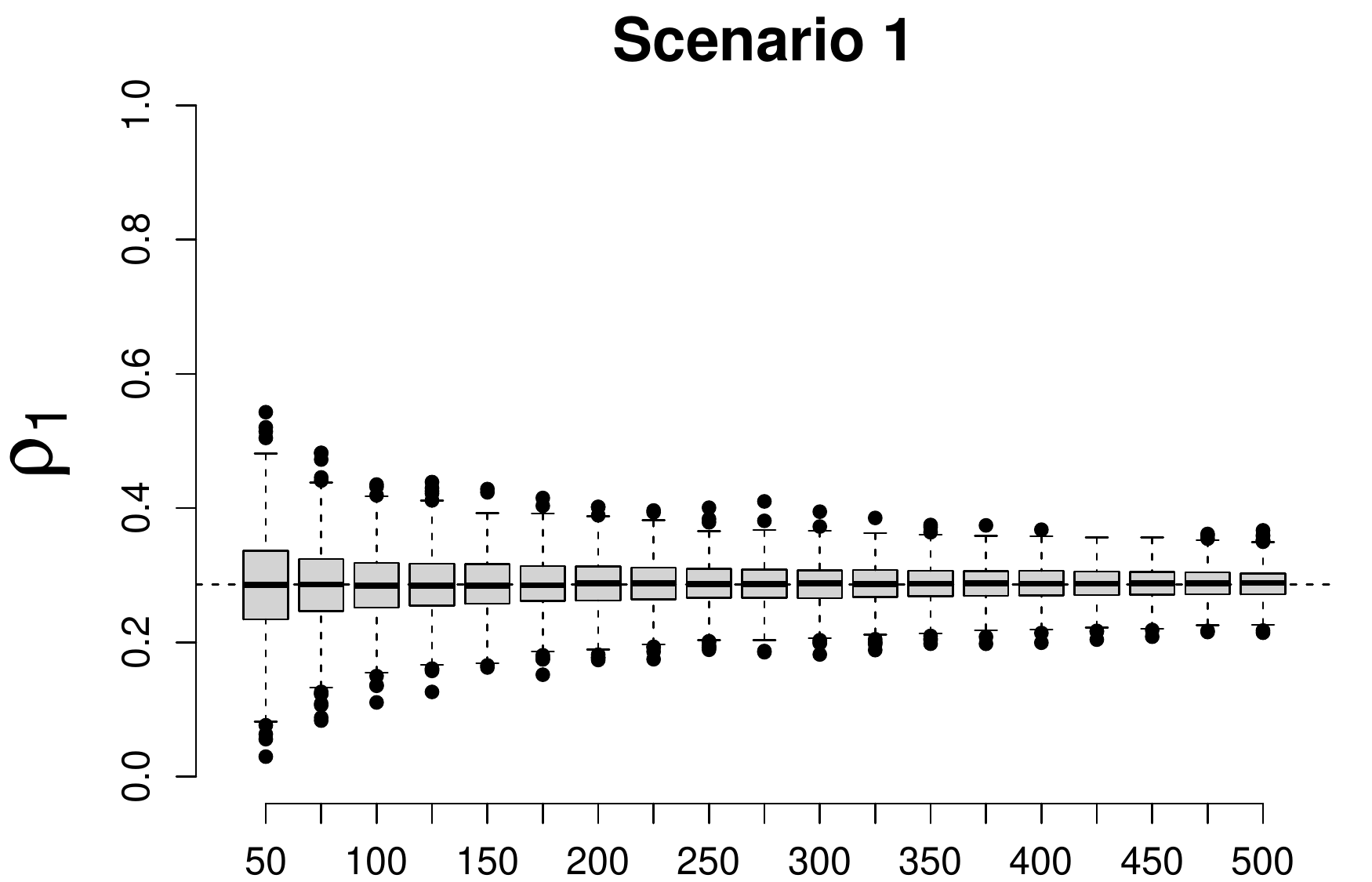}
		\includegraphics[]{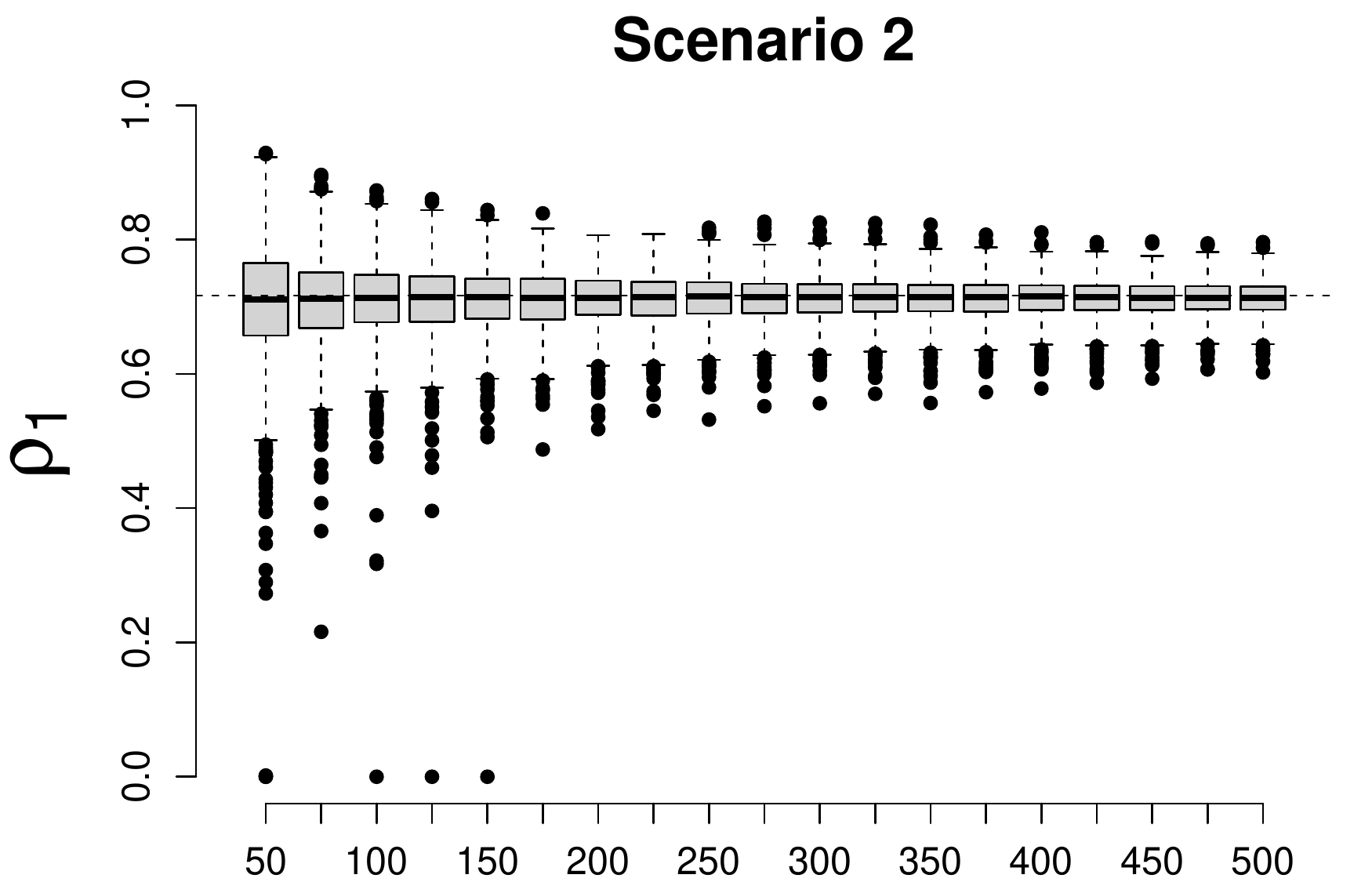}
		\includegraphics[]{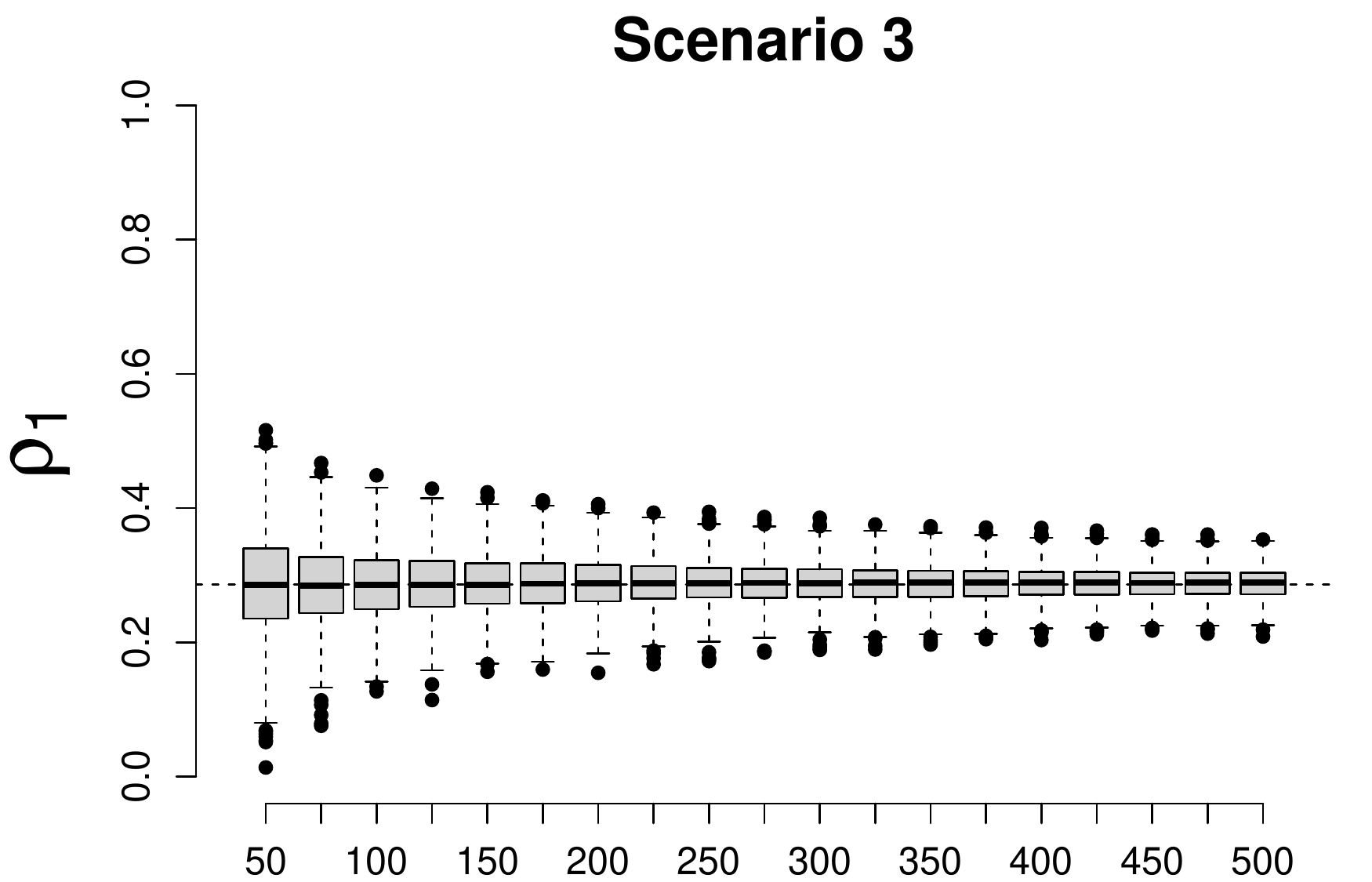}
		\includegraphics[]{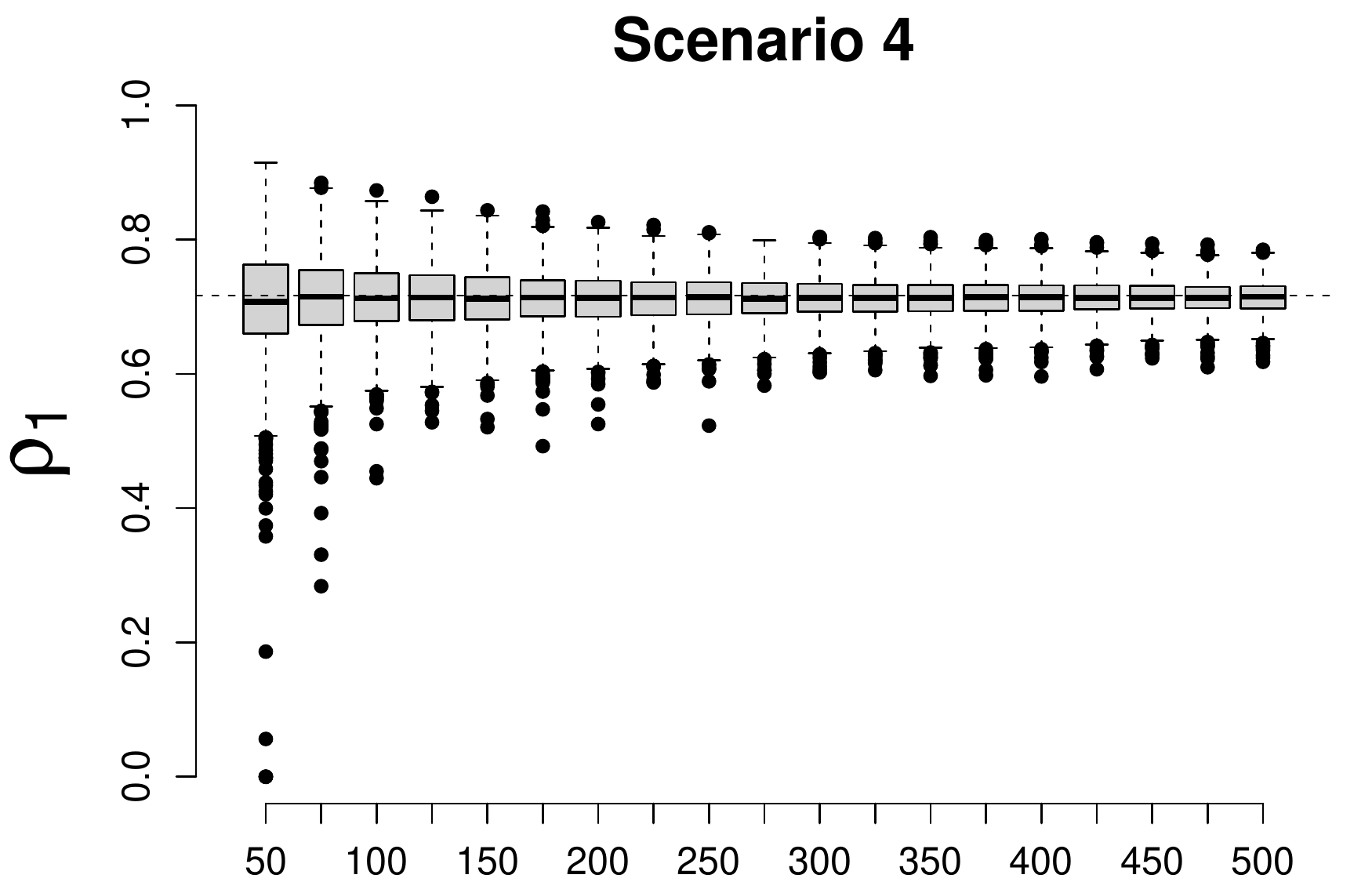}
		\includegraphics[]{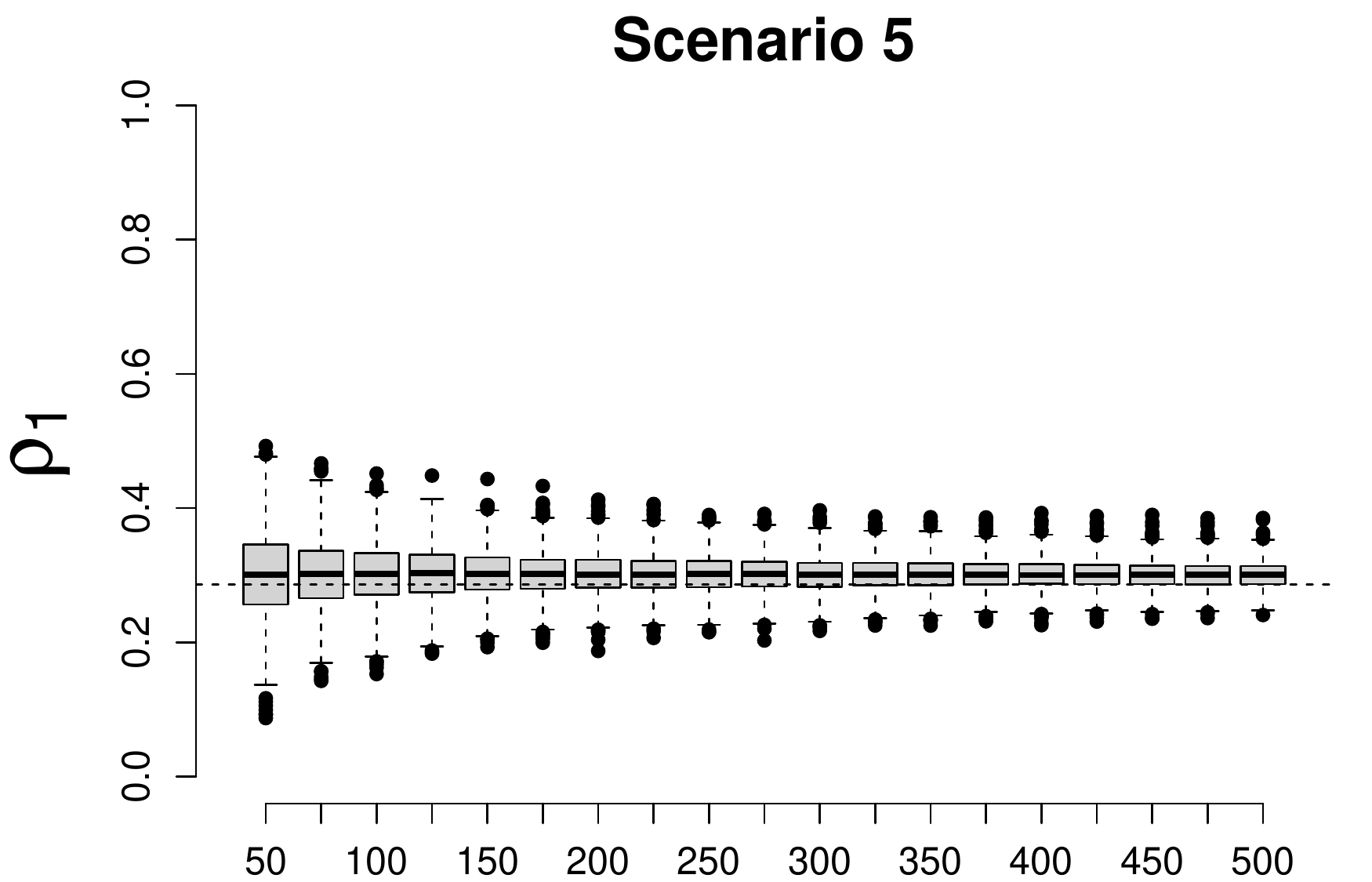}
		\includegraphics[]{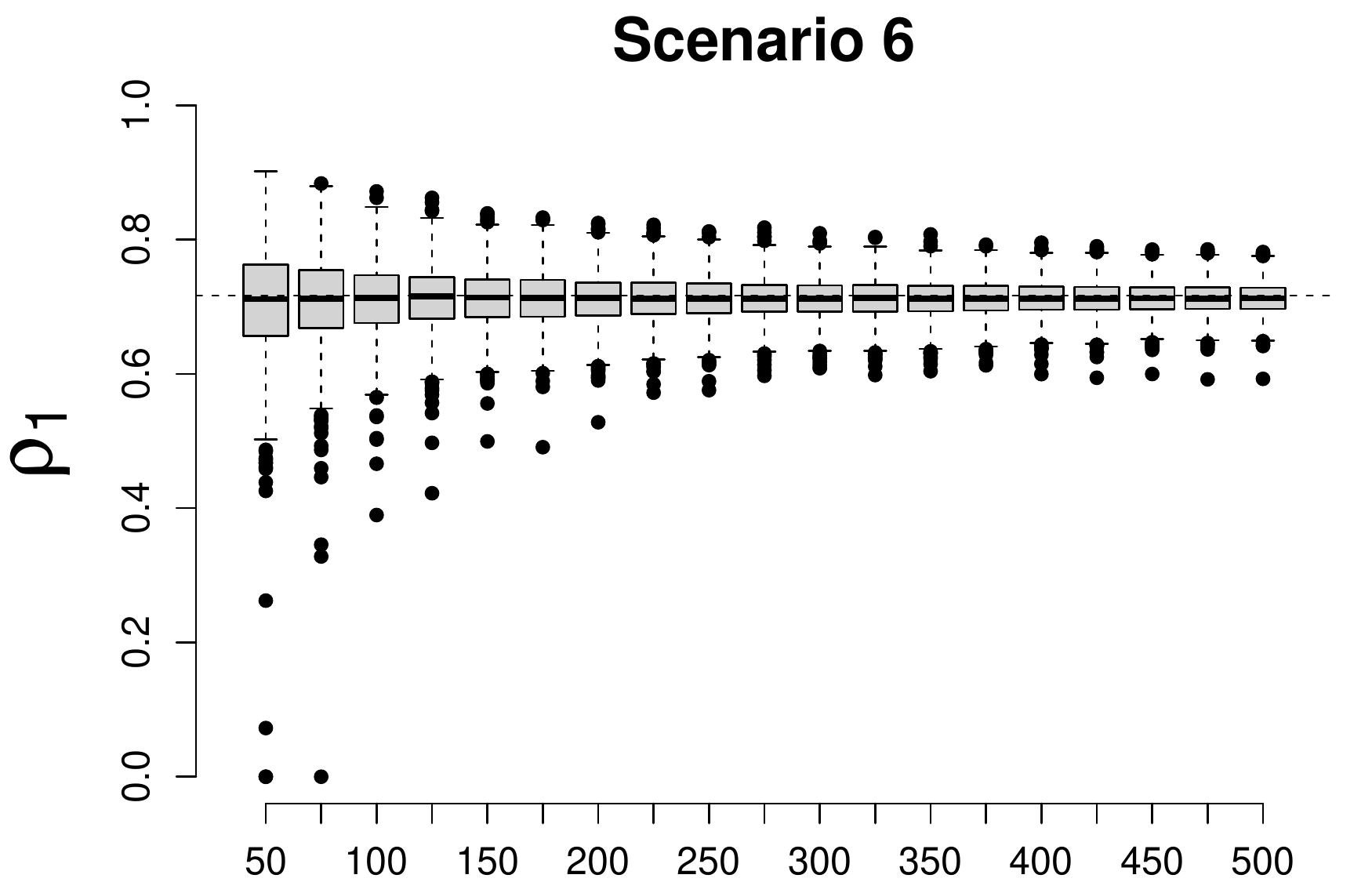}
		\includegraphics[]{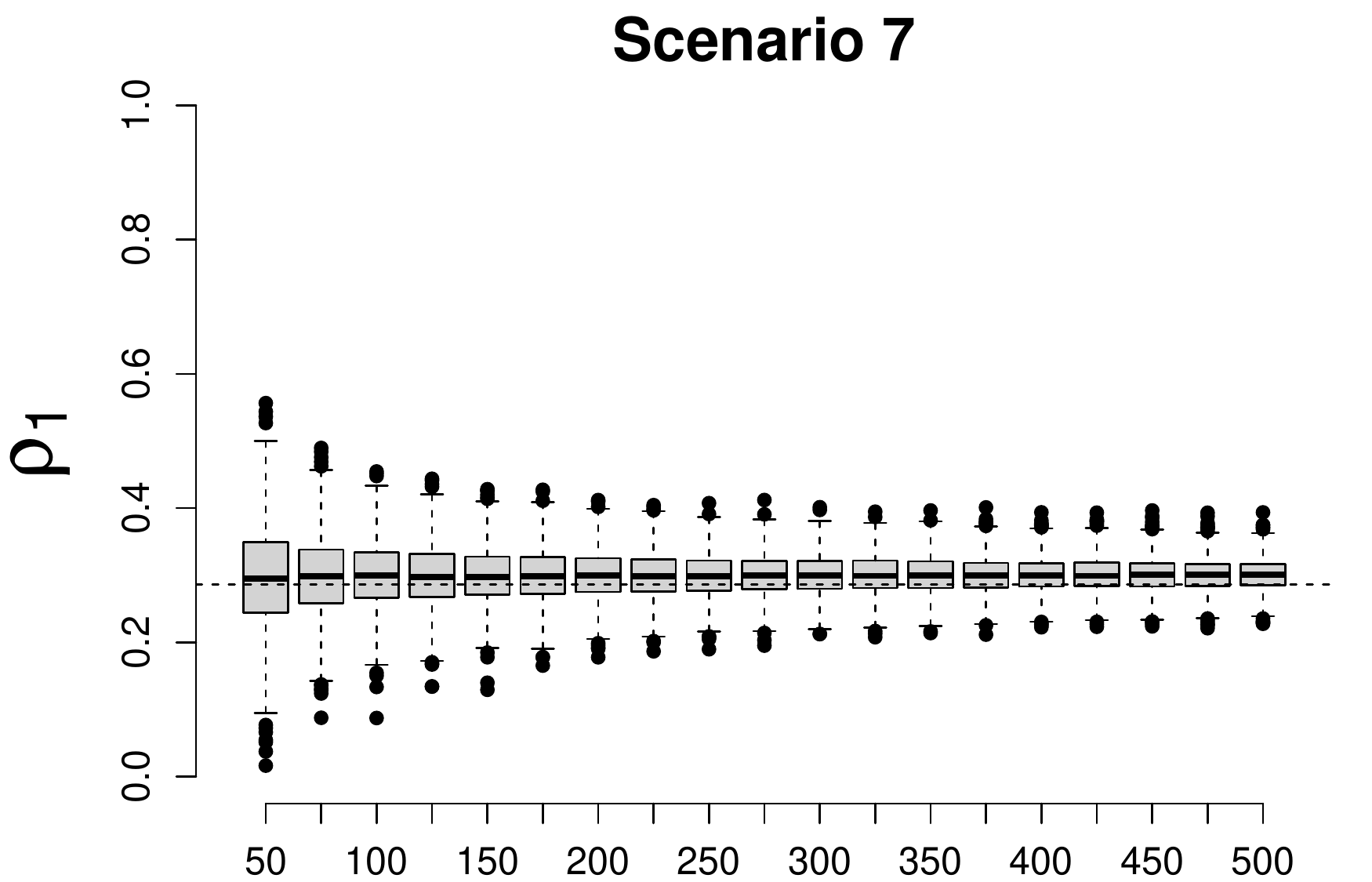}
		\includegraphics[]{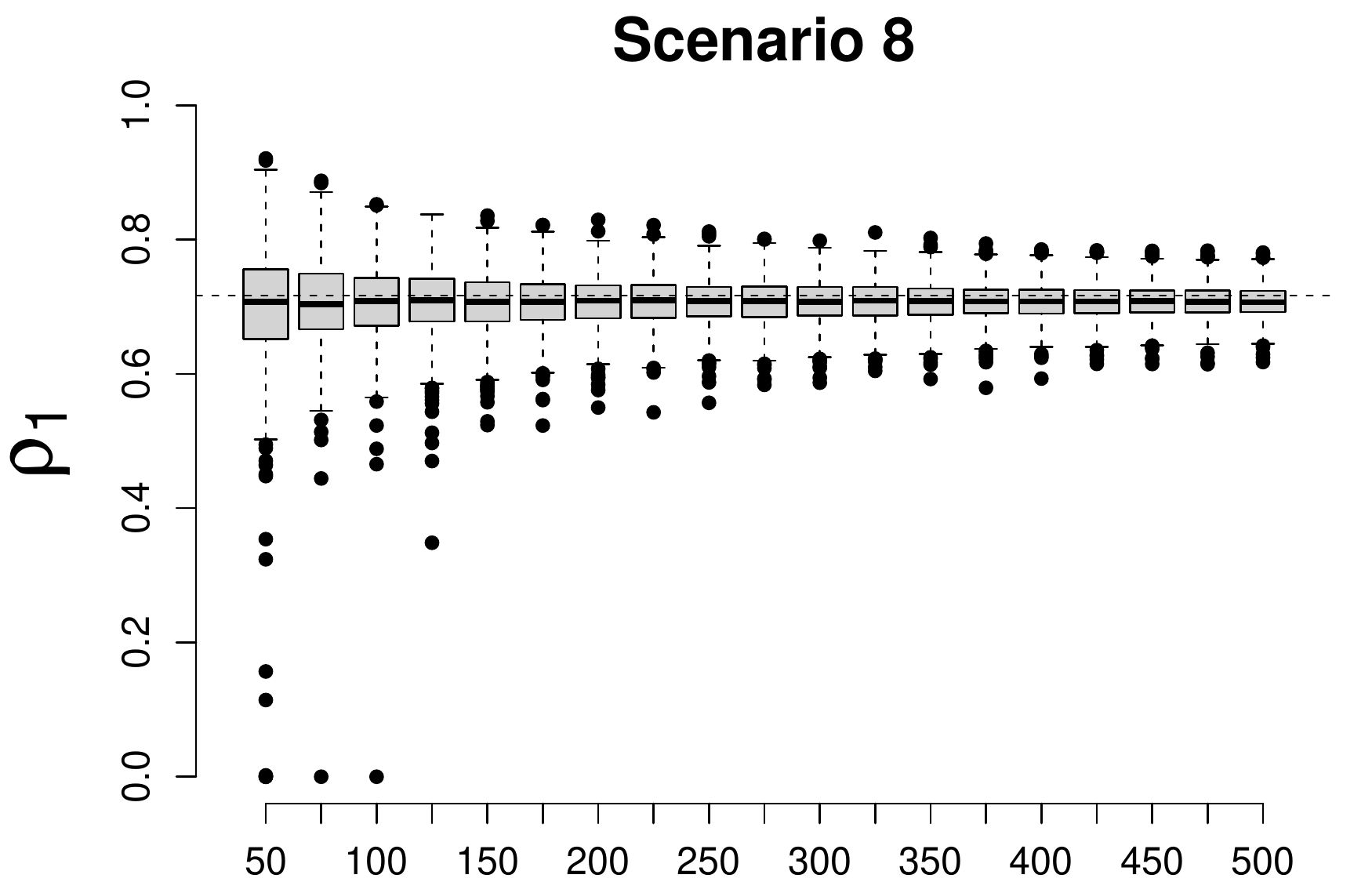}
		\includegraphics[]{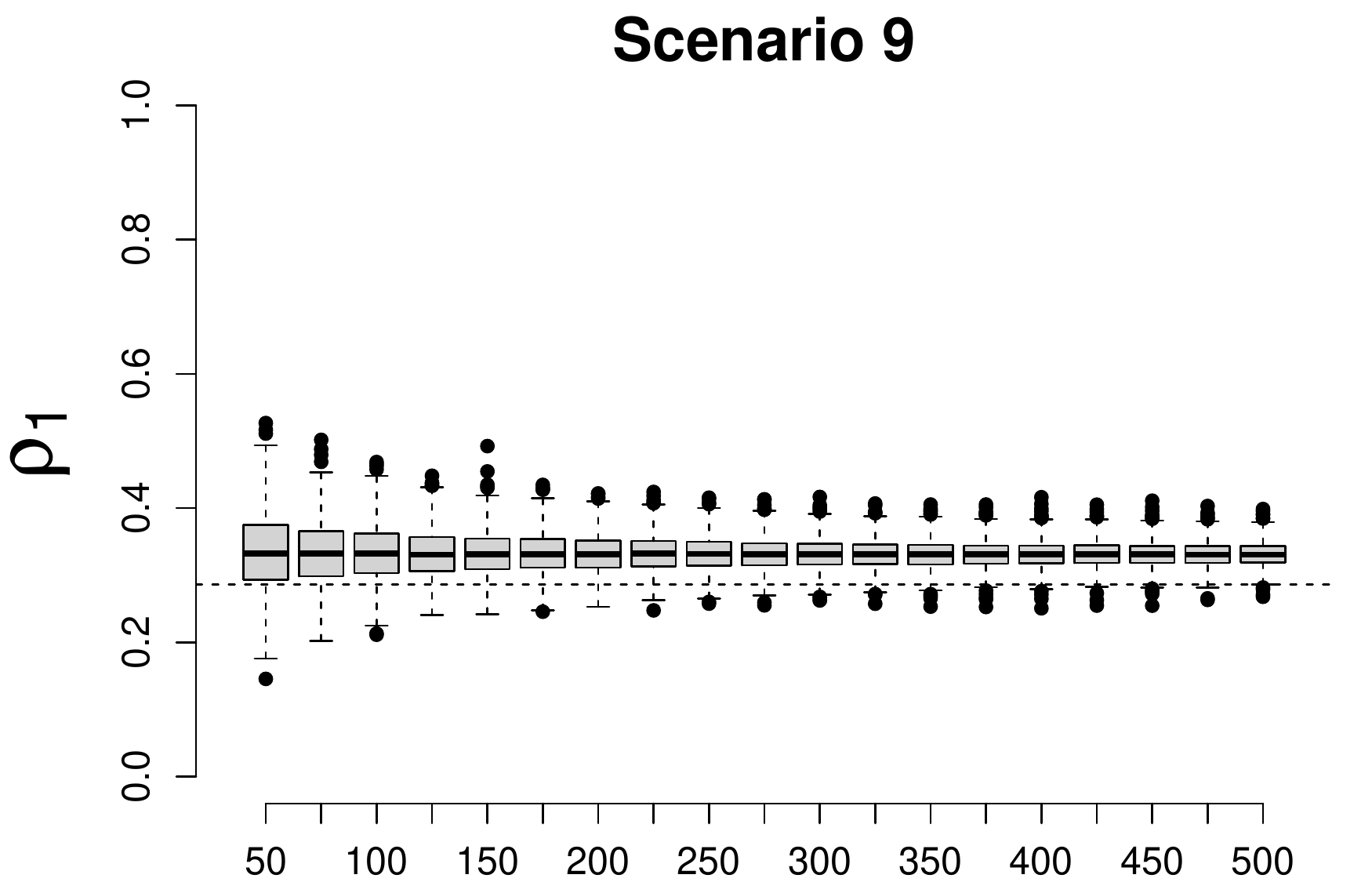}
		\includegraphics[]{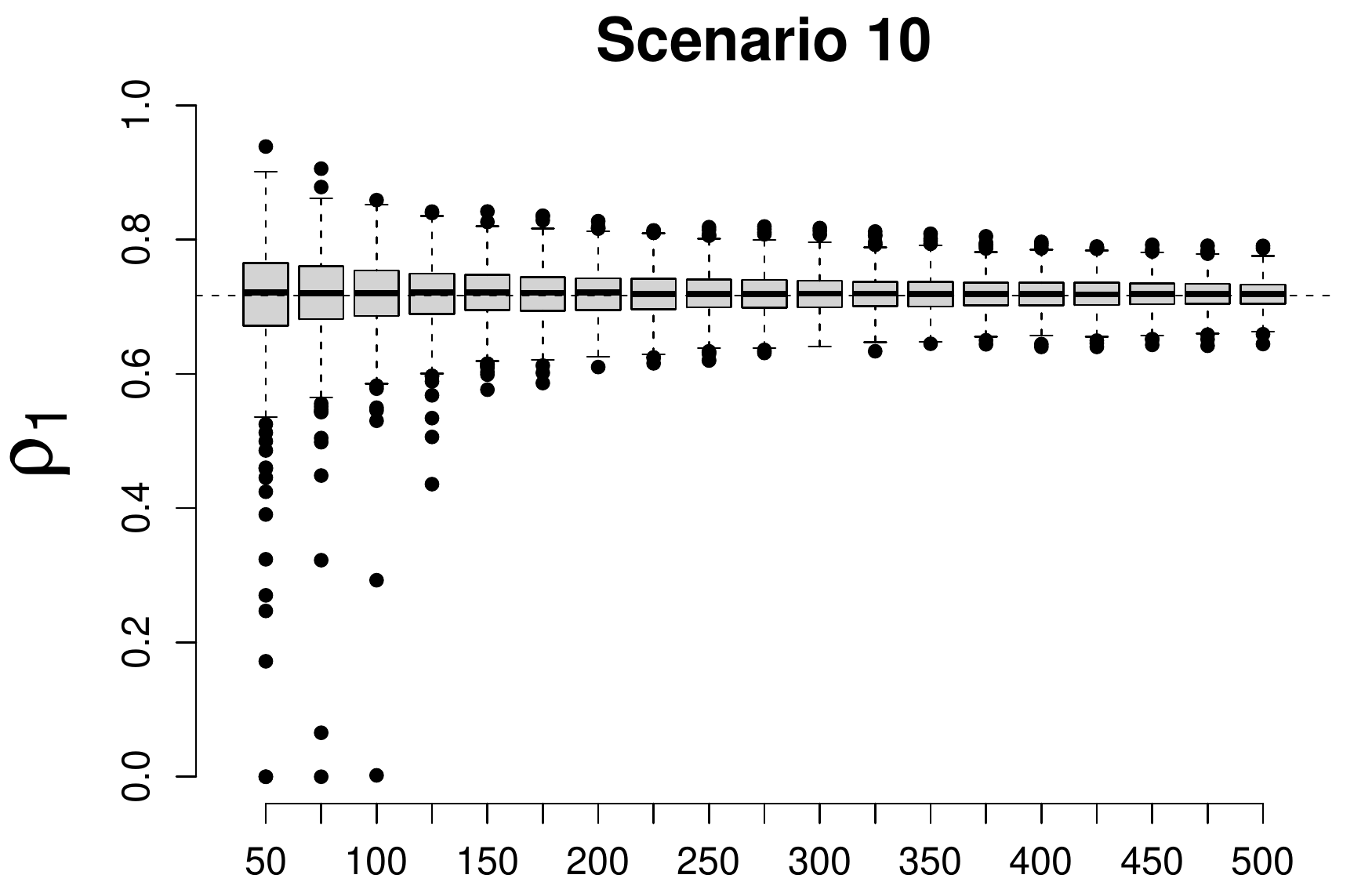}
		\includegraphics[]{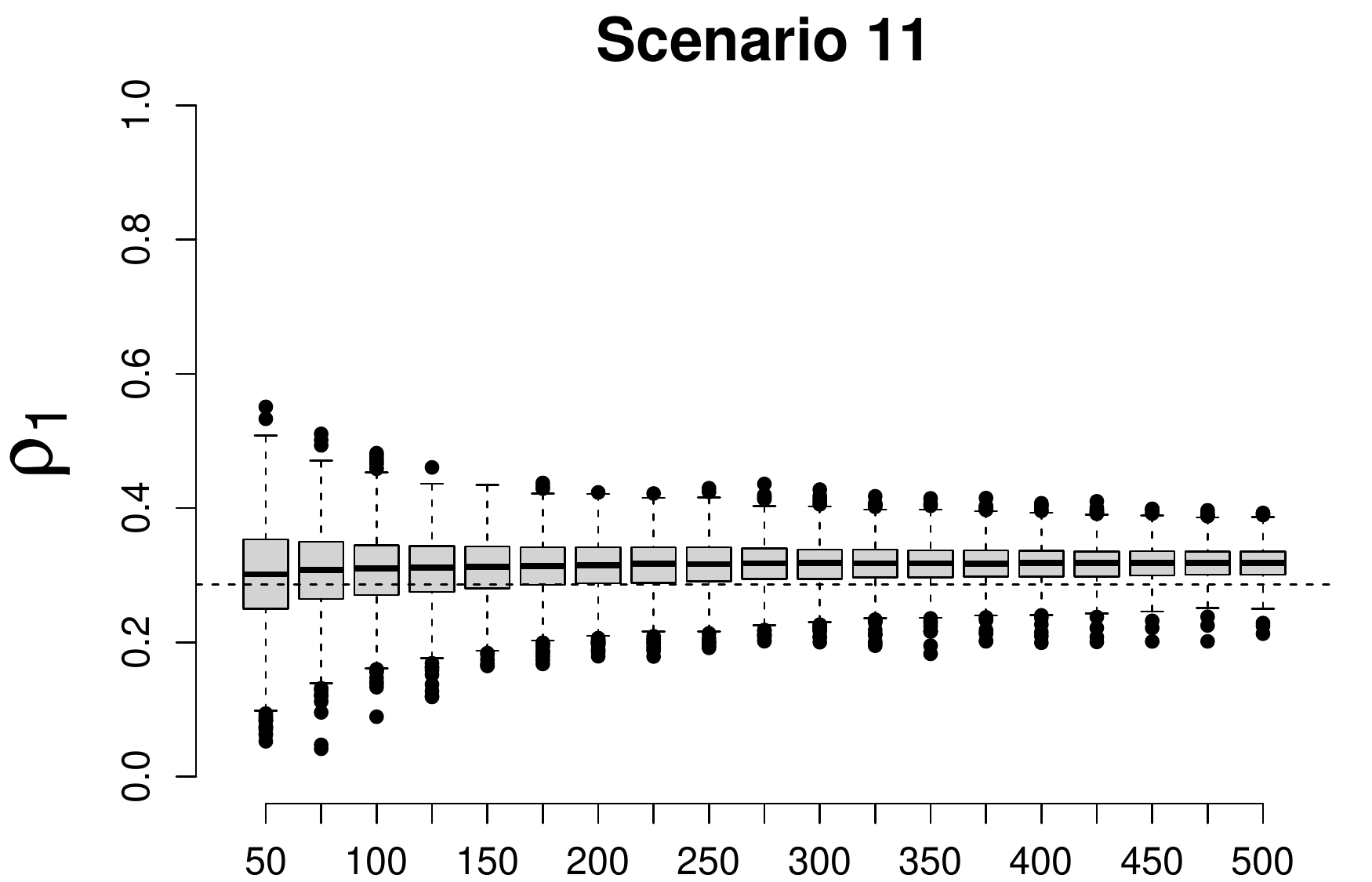}
		\includegraphics[]{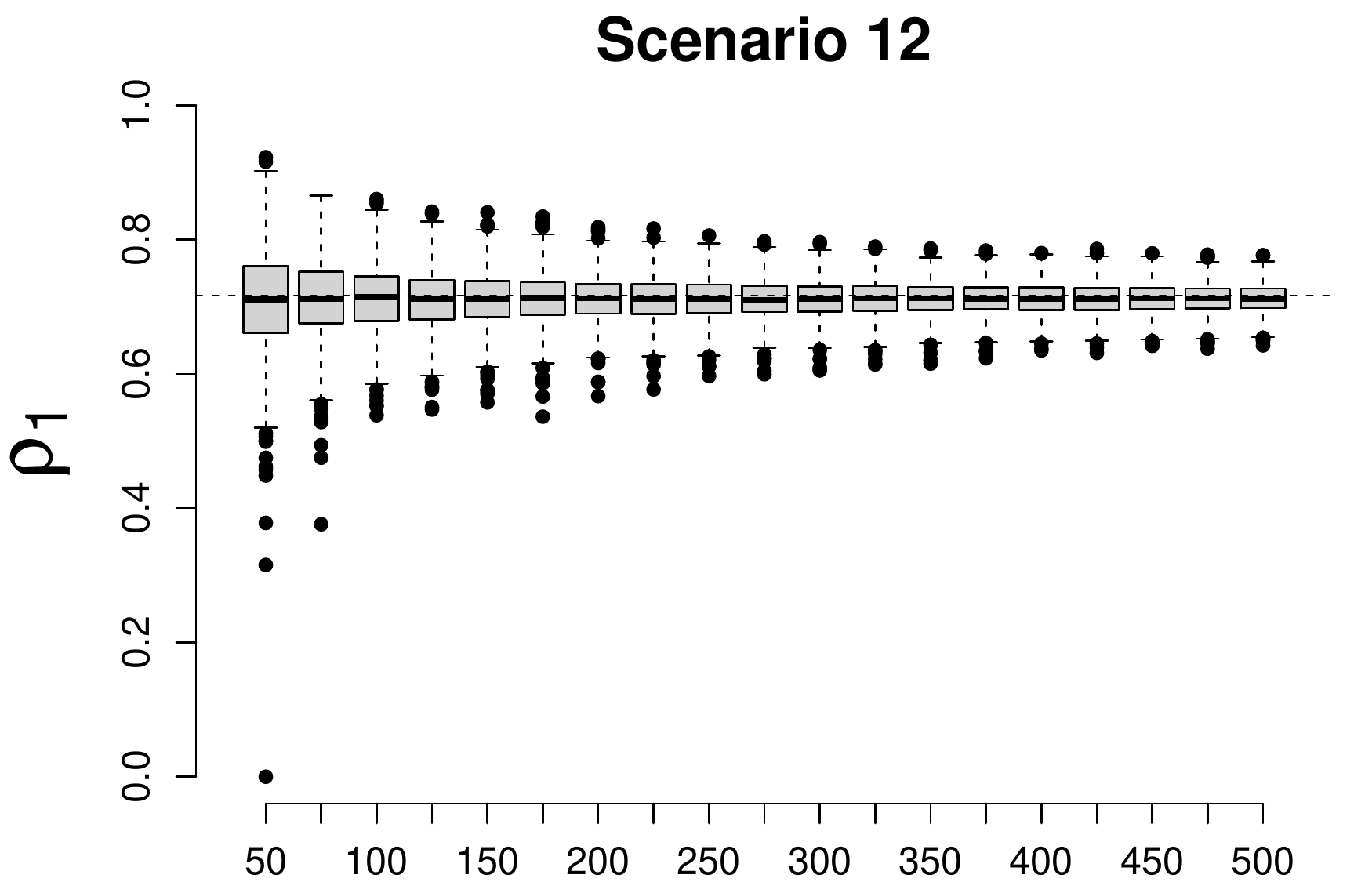}
		\caption{Box-plots of the maximum likelihood estimates for $ \rho_1 $ in each considered scenario considering different sample sizes.}
		\label{figeta1}
\end{figure}

Figure (\ref{figeta1}) shows the box-plots of the ML estimates of the parameter $ \rho_1$ in all scenarios considering different sample sizes (50 to 500), which enables us to observe the variability of these estimates. In each graph of Figure \ref{figeta1}, horizontal dotted line refers to the nominal values of the parameter $ \rho_1$. It is possible to see that the estimated values for $\rho_1$ are closer to the nominal vales, and the sampling variability decreases as the sample size increases as expected.


\begin{figure}[H]
\centering
\setkeys{Gin}{height = 4.2cm,width=0.24\linewidth} %
		\includegraphics[scale=0.3]{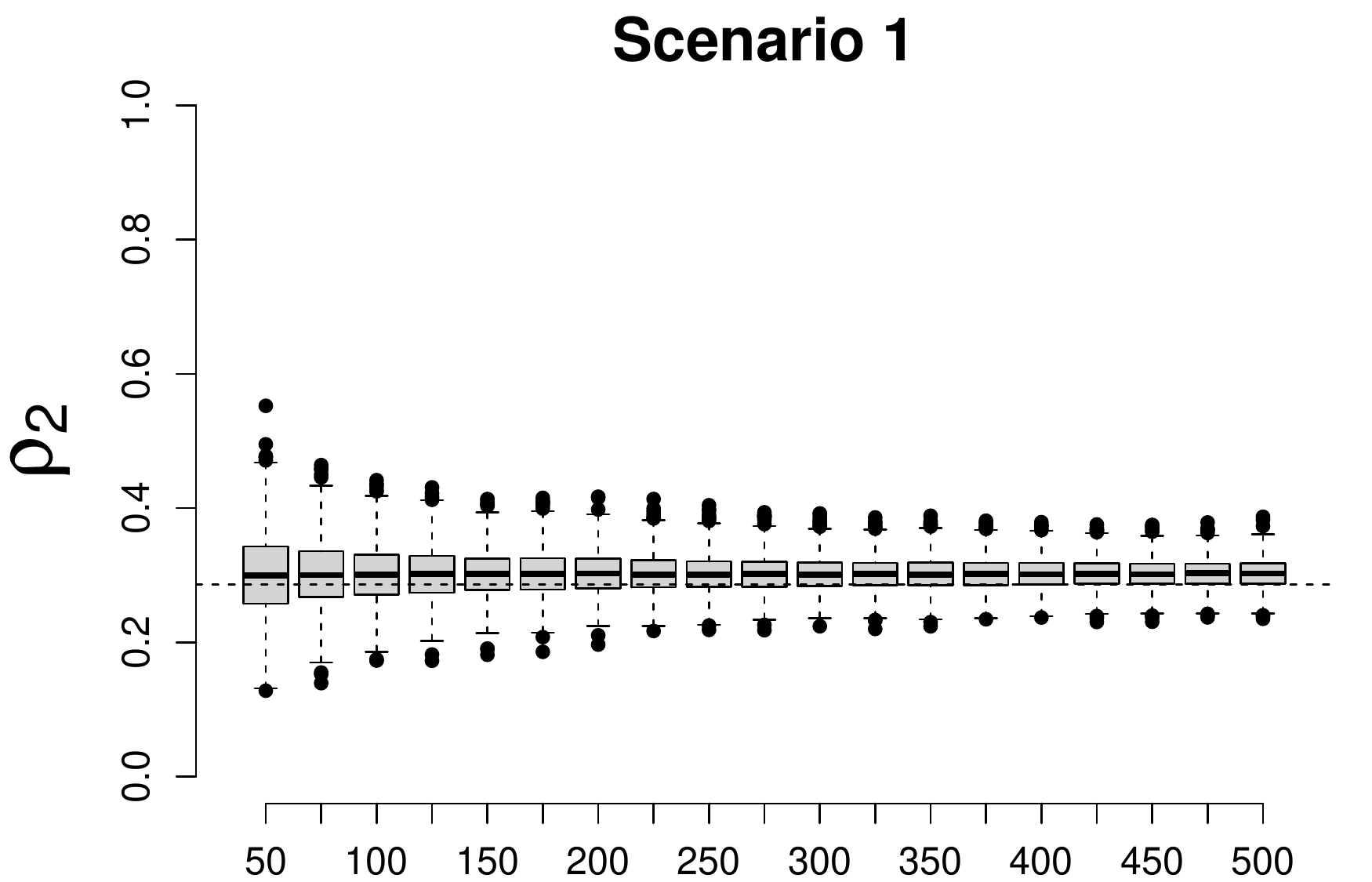}
		\includegraphics[scale=0.3]{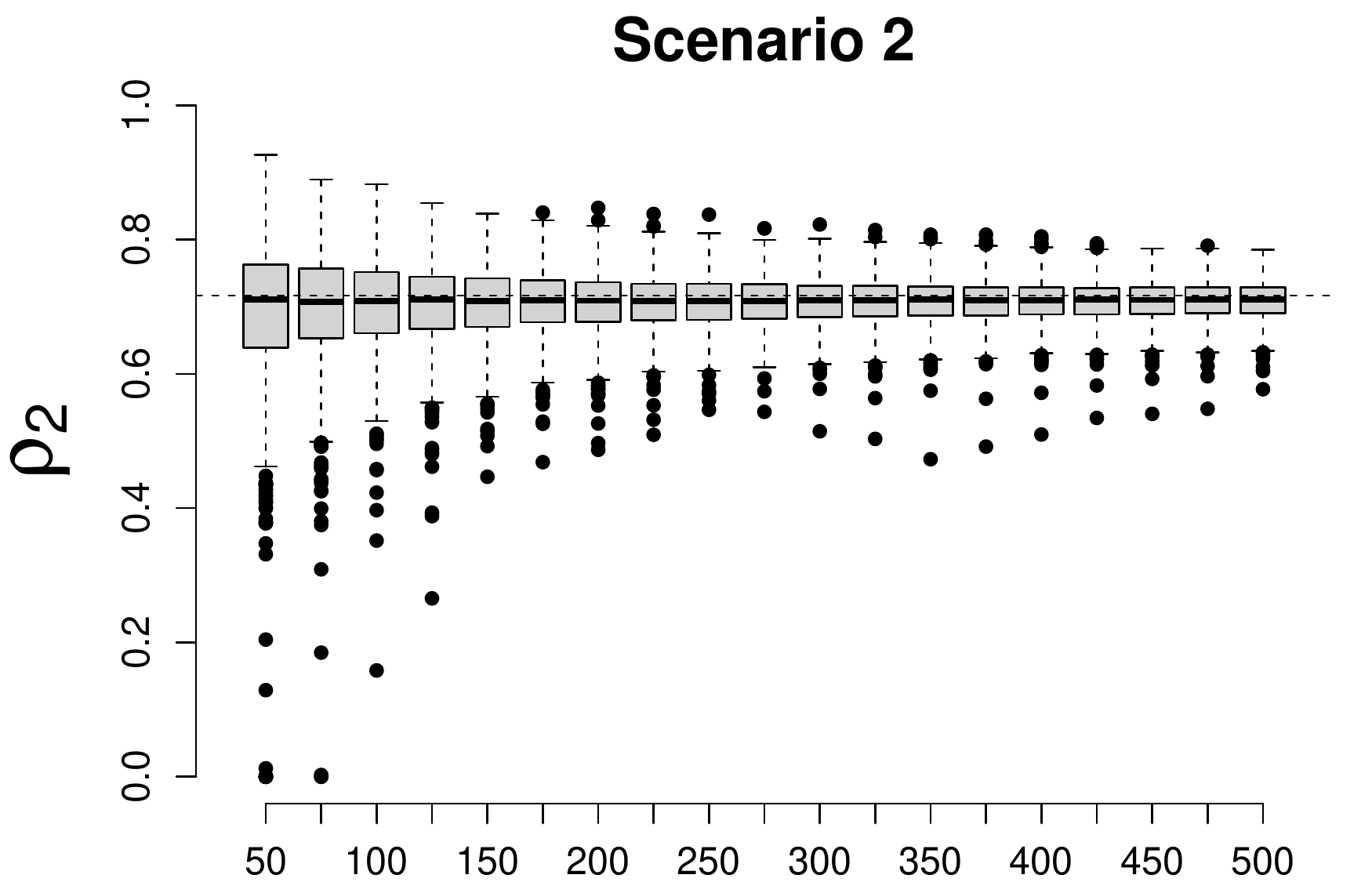}
		\includegraphics[scale=0.3]{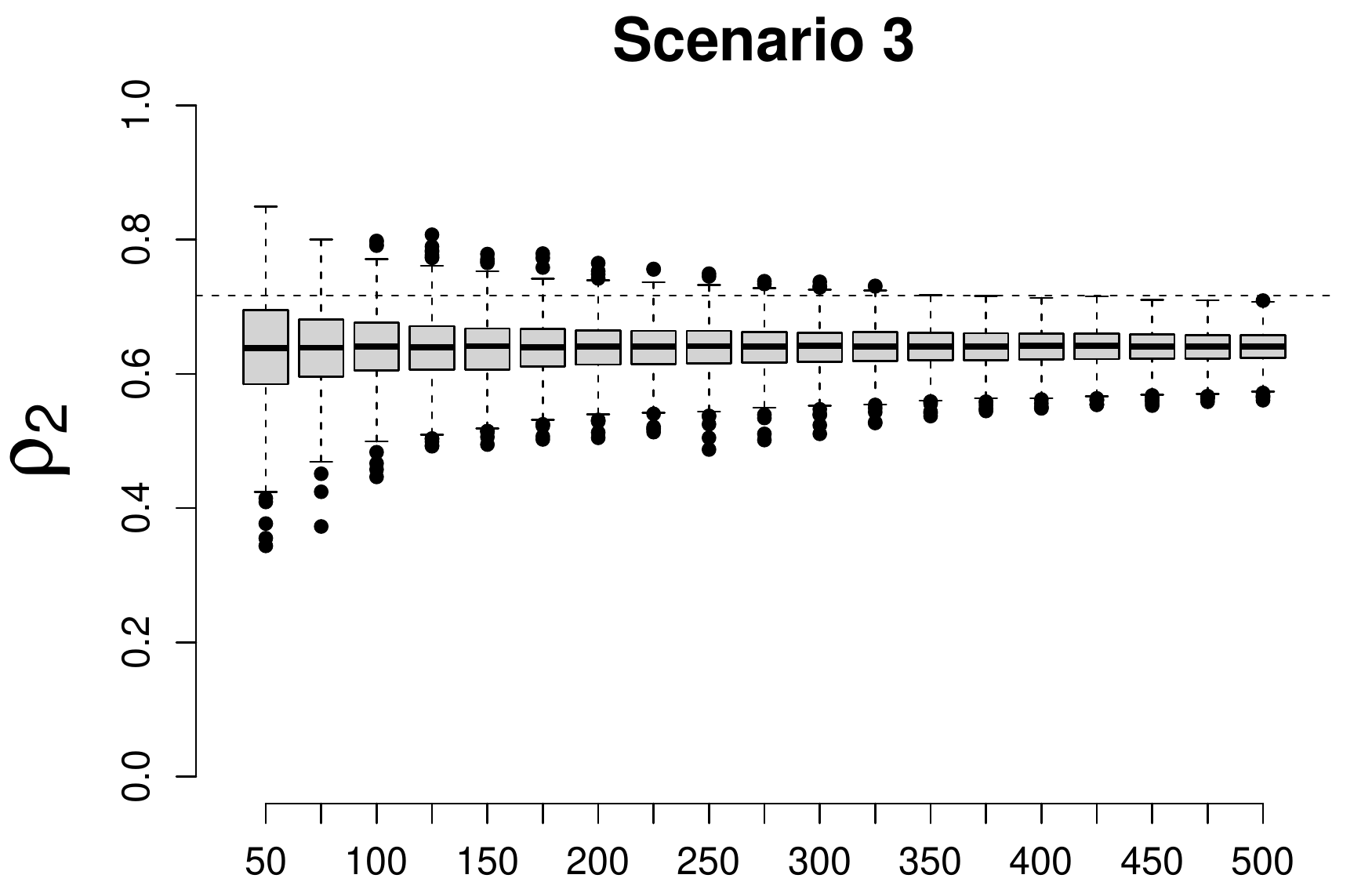}
		\includegraphics[scale=0.3]{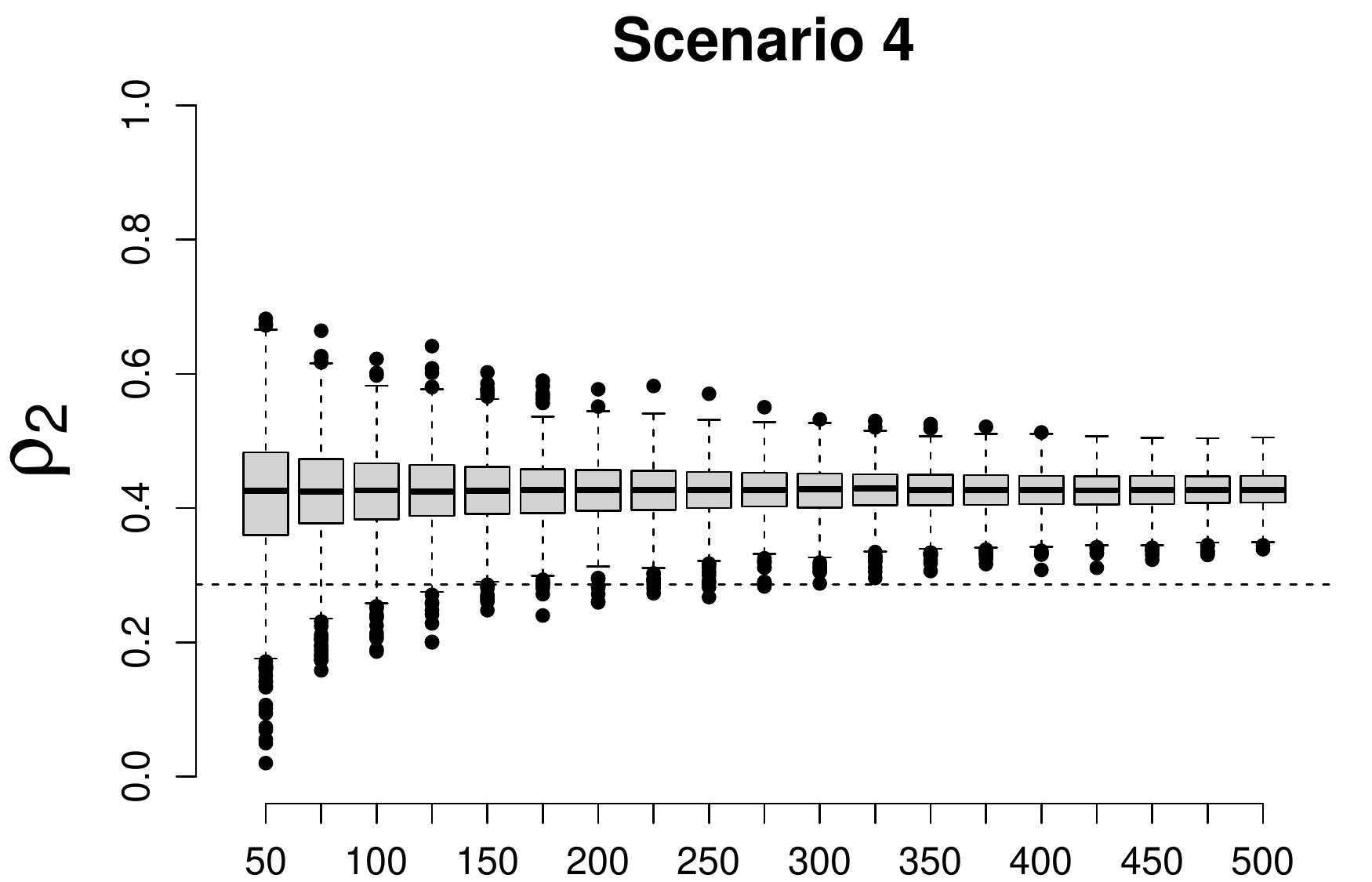}
		\includegraphics[scale=0.3]{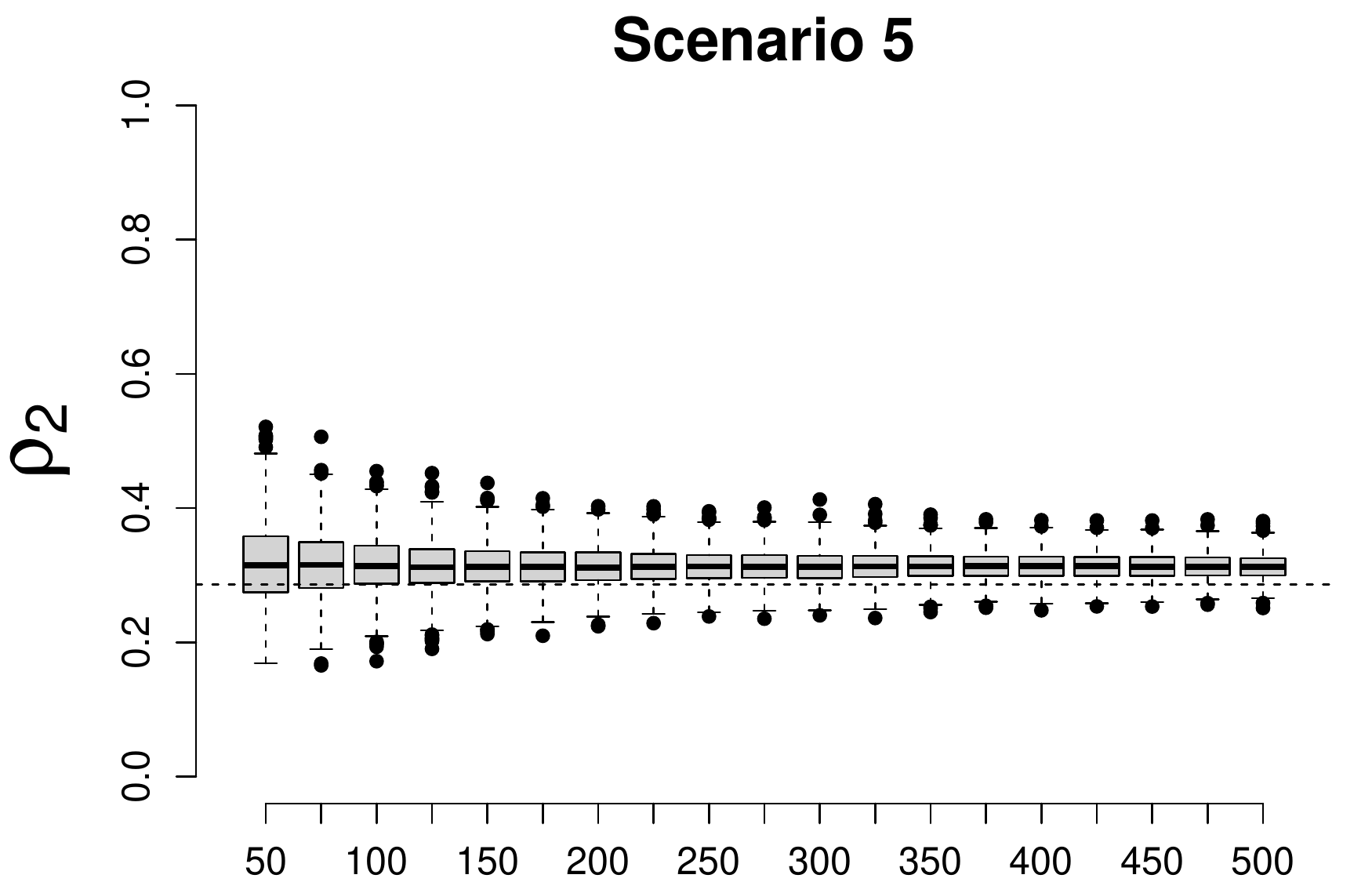}
		\includegraphics[scale=0.3]{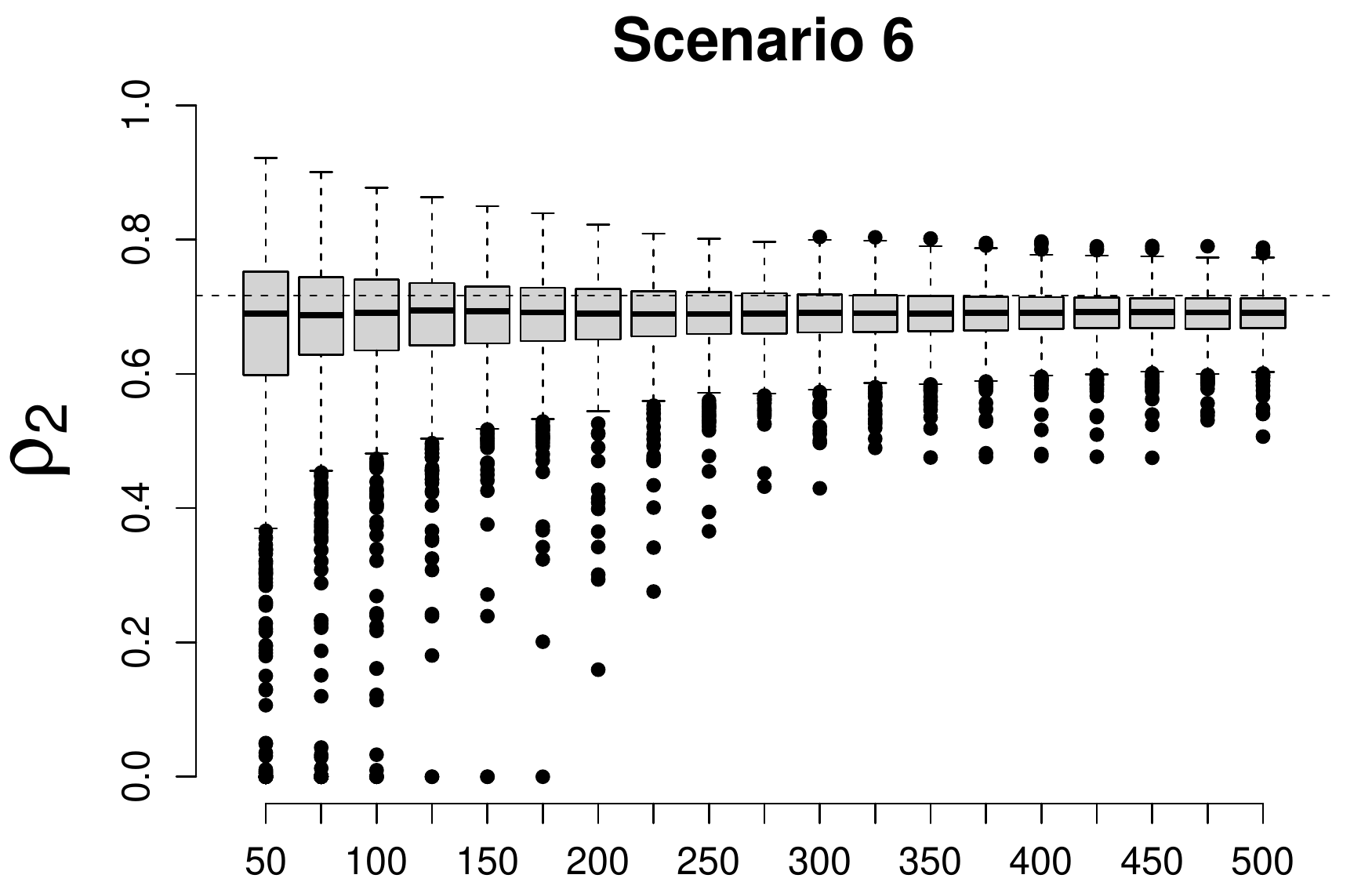}
		\includegraphics[scale=0.3]{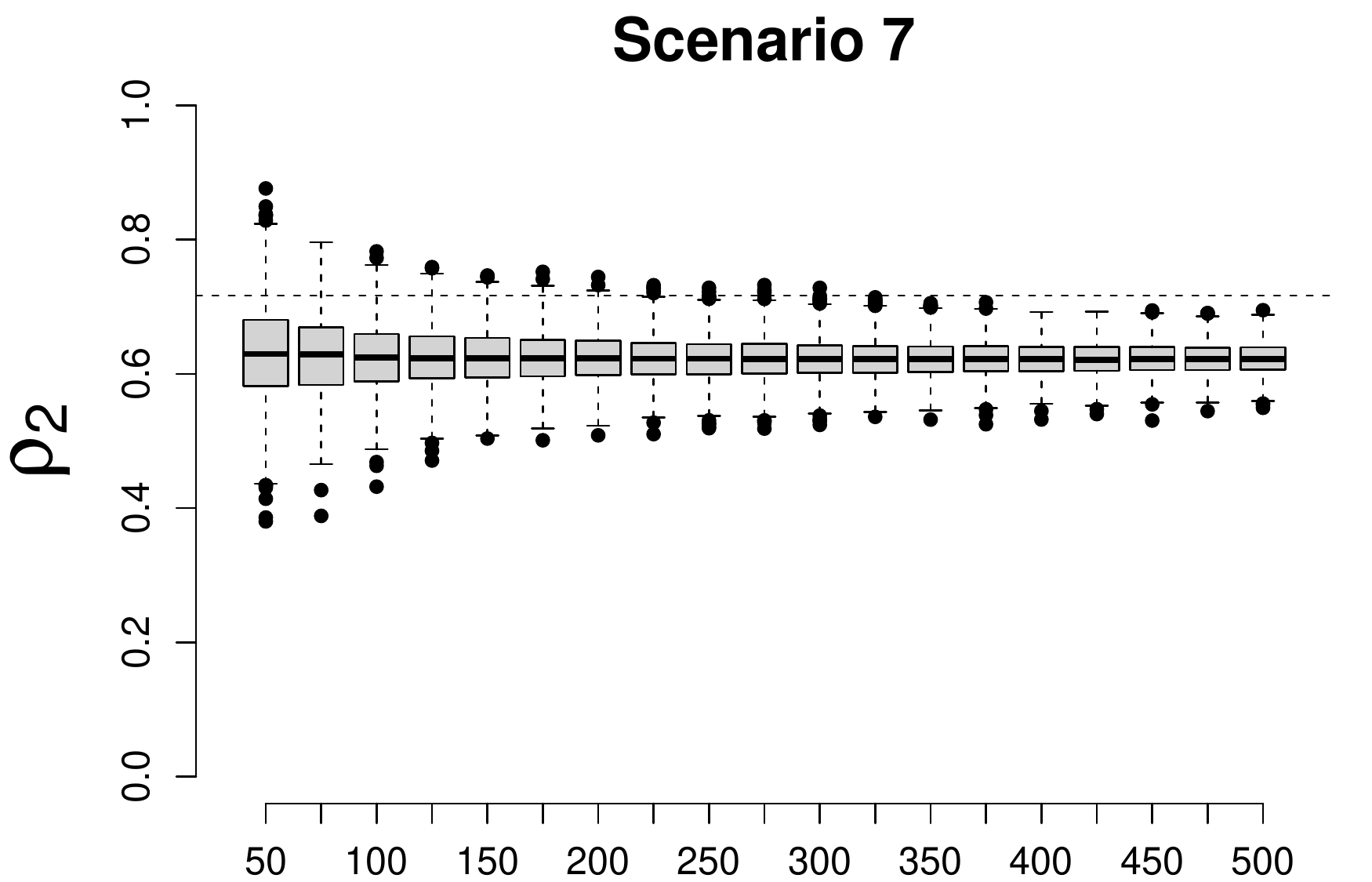}
		\includegraphics[scale=0.3]{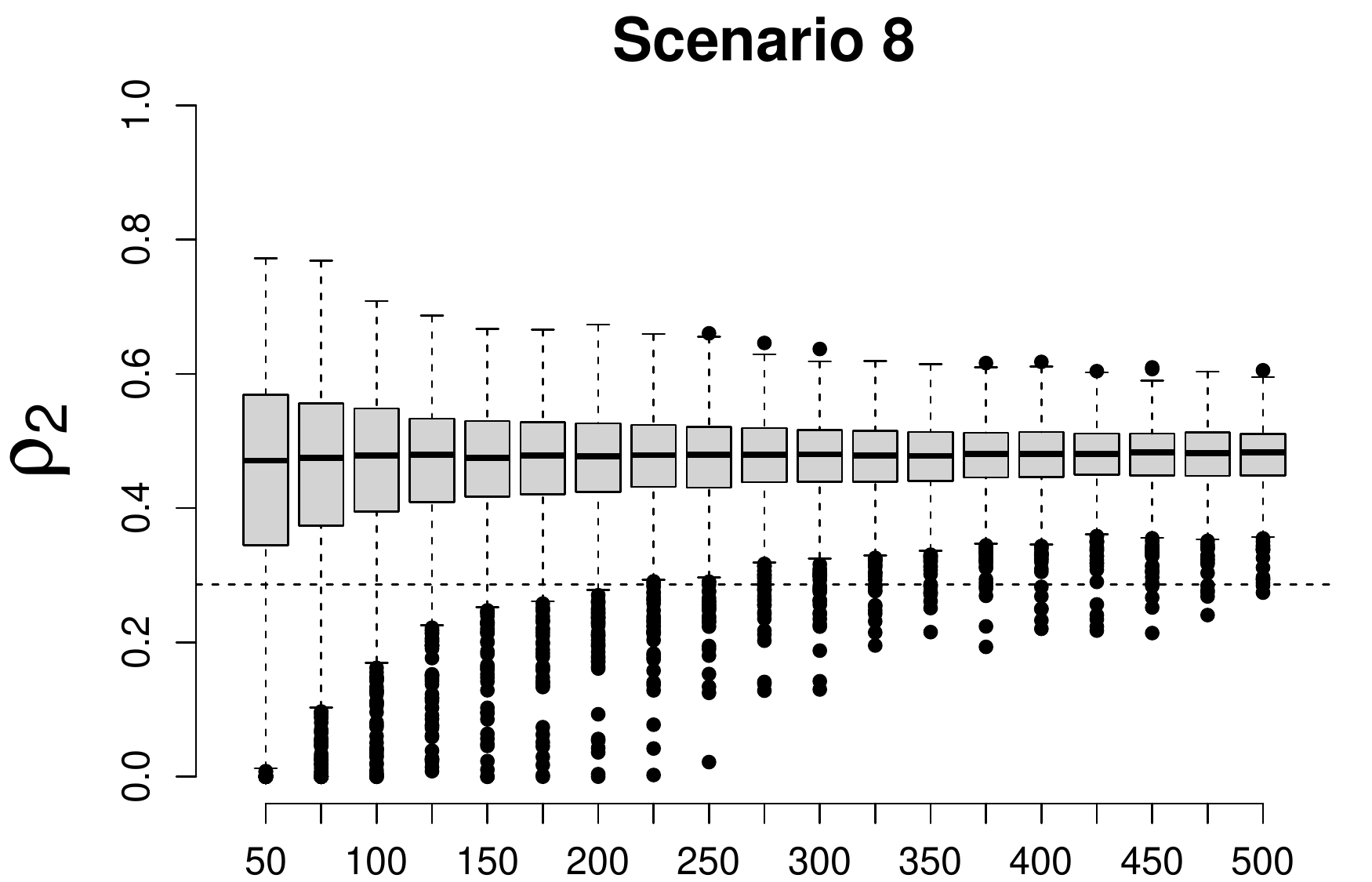}
		\includegraphics[scale=0.3]{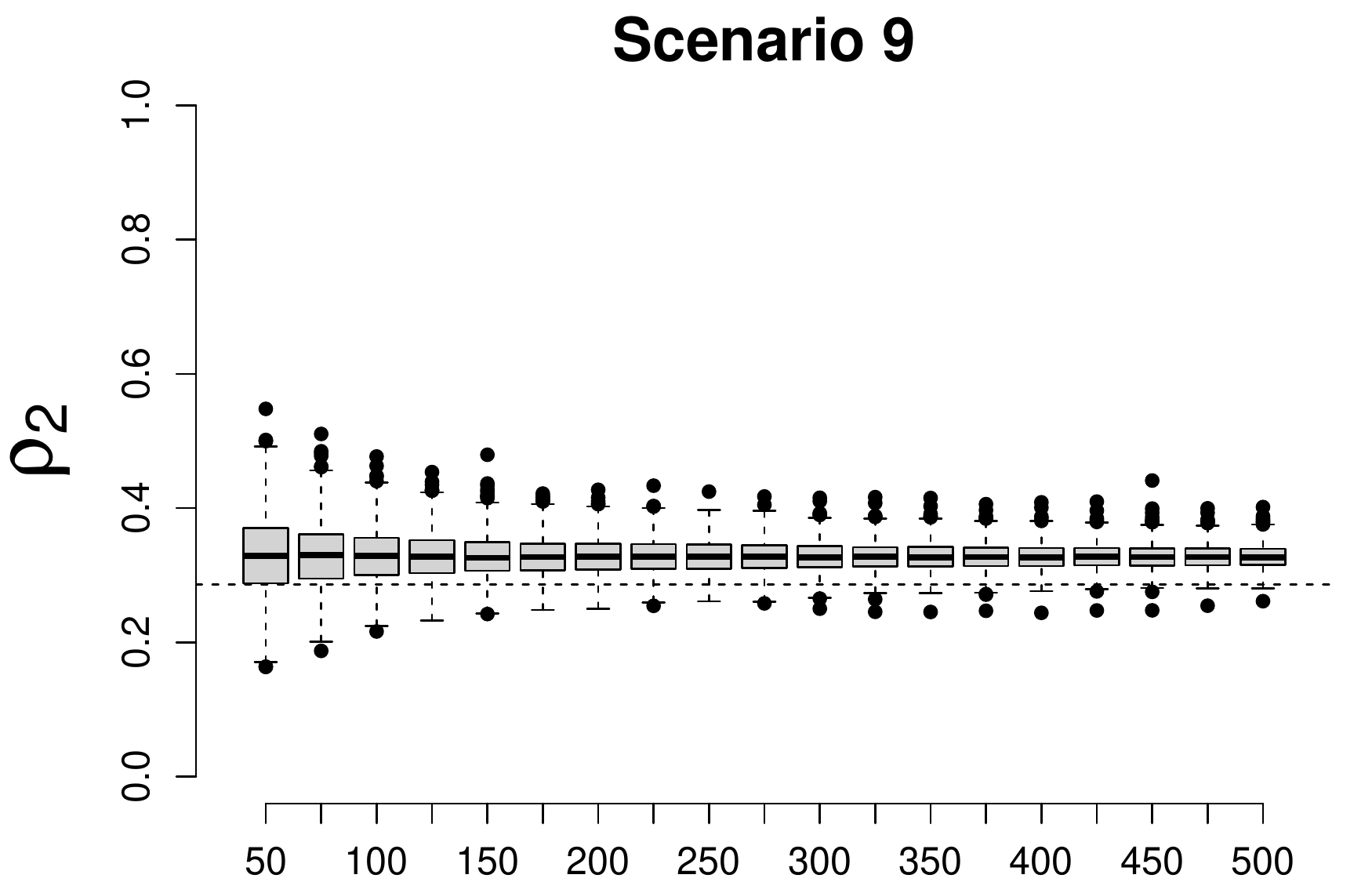}
		\includegraphics[scale=0.3]{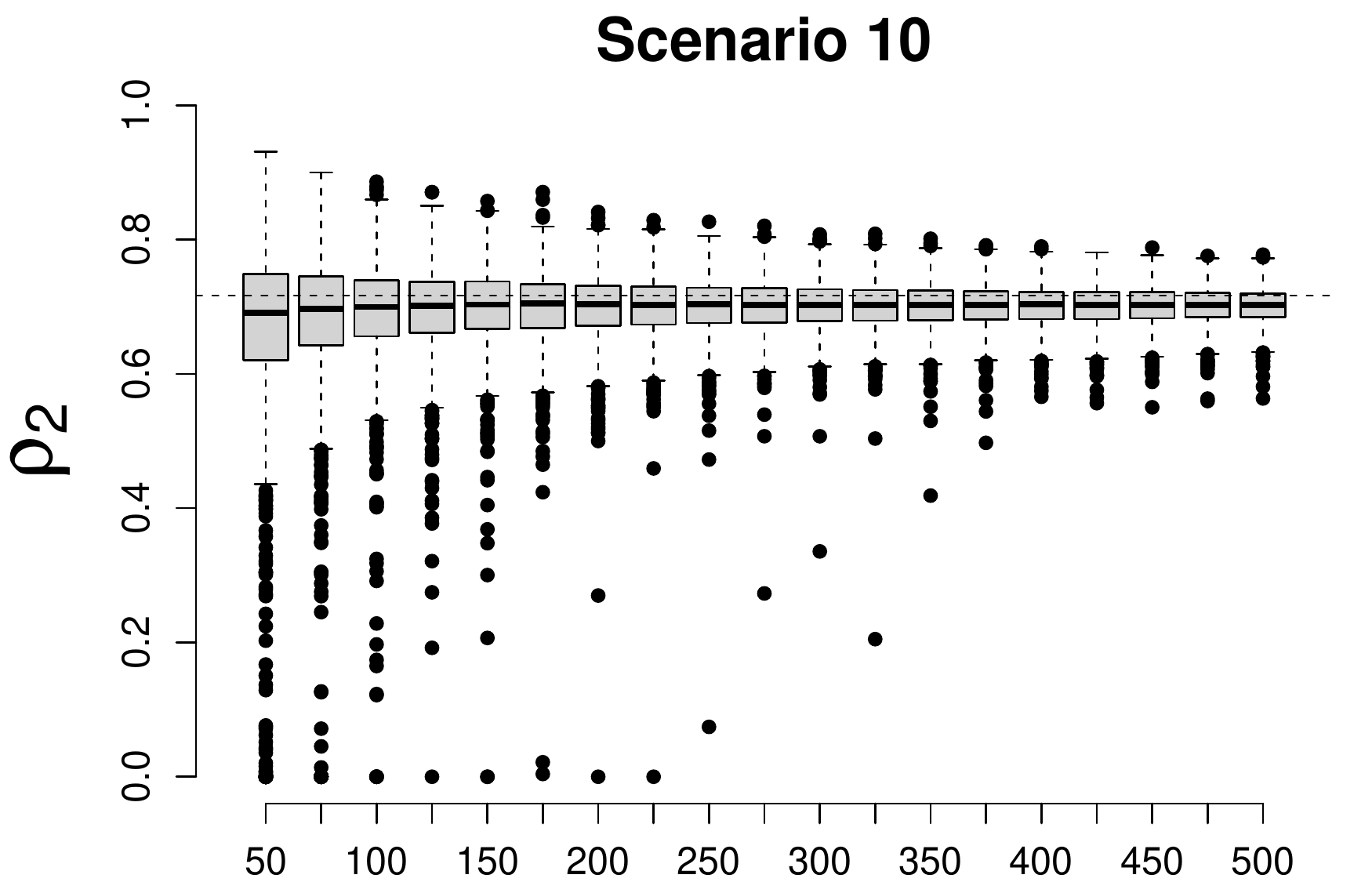}
		\includegraphics[scale=0.3]{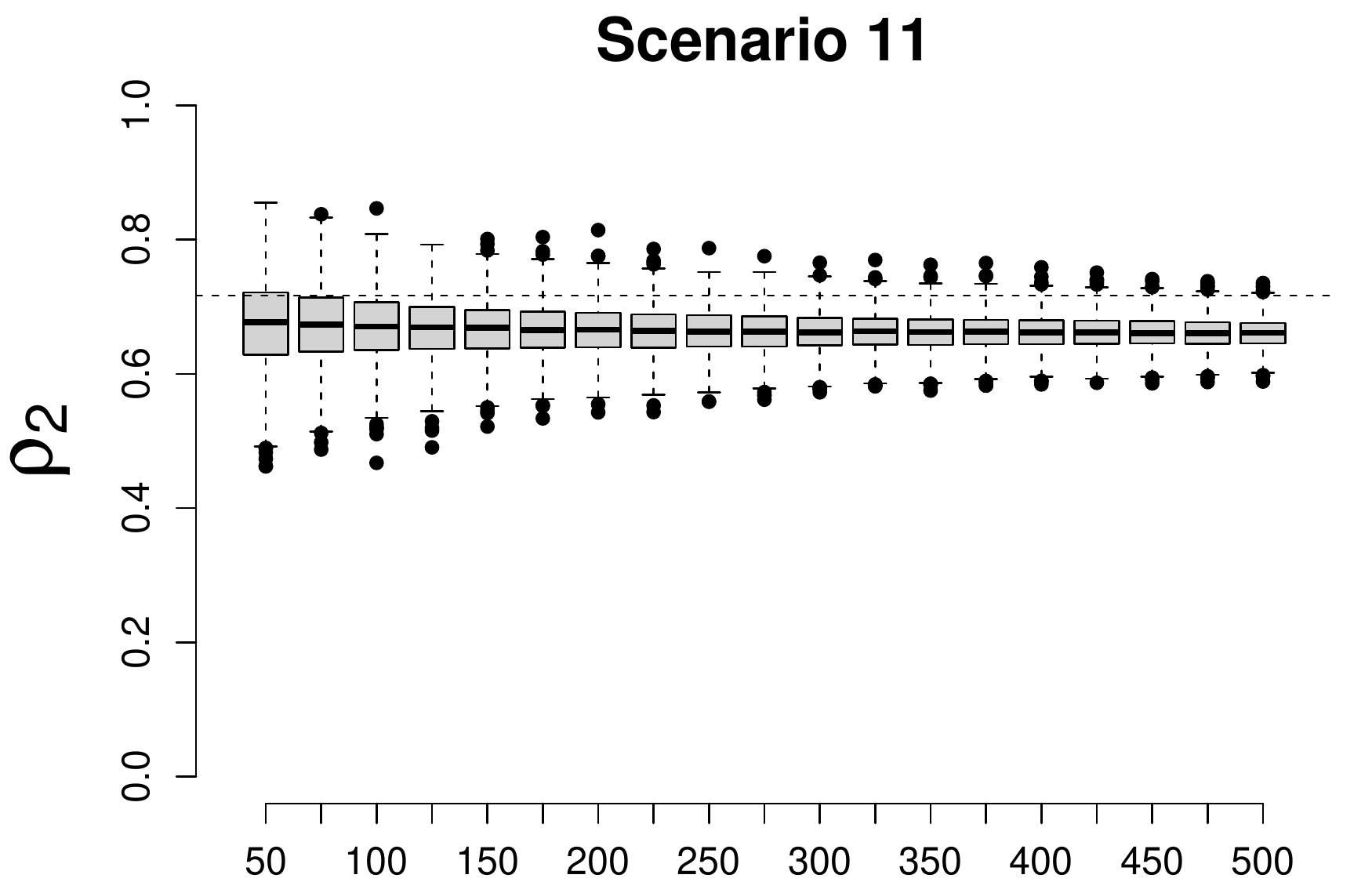}
		\includegraphics[scale=0.3]{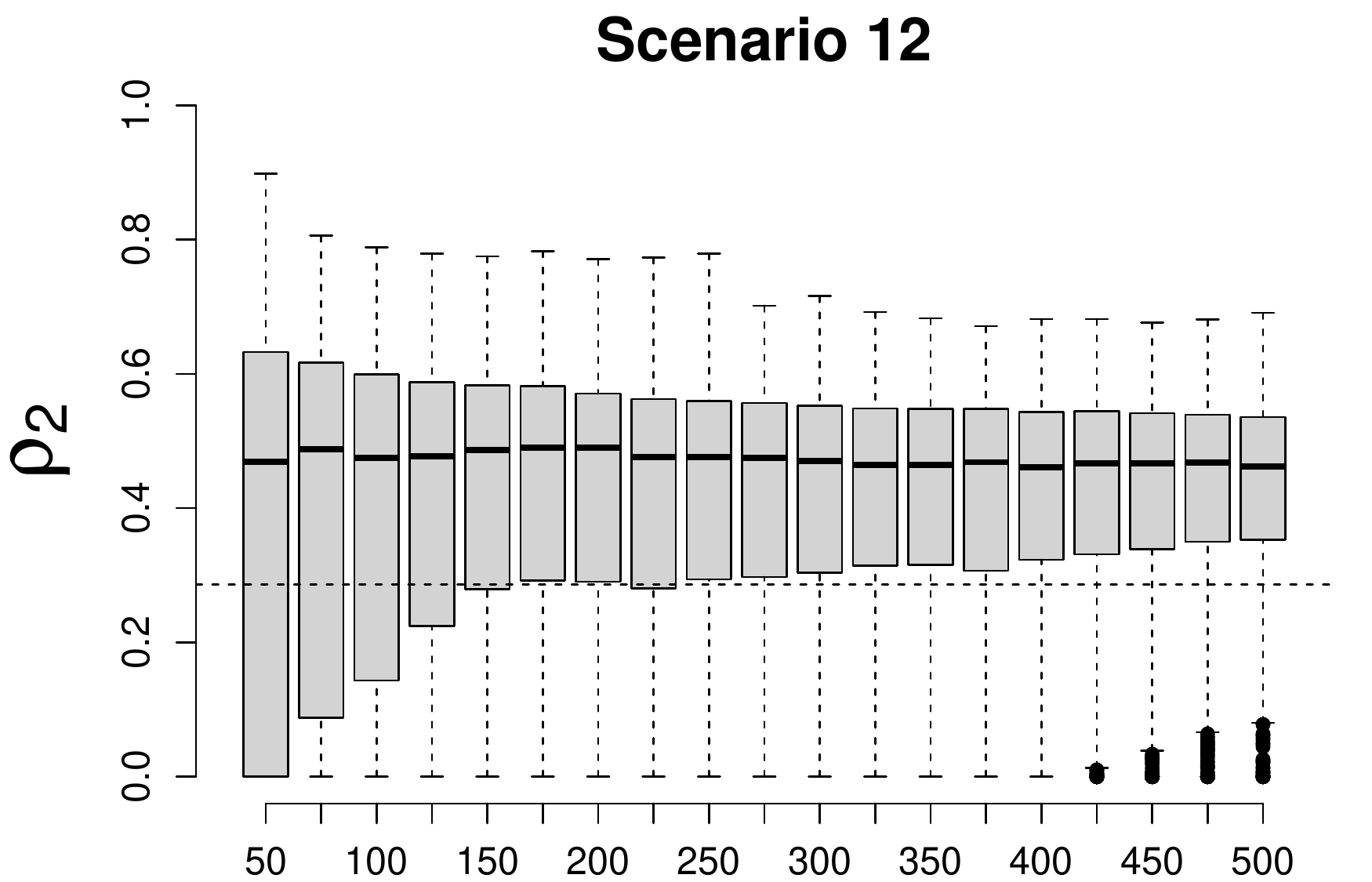}
		\caption{Box-plots of the maximum likelihood estimates for $ \rho_2 $ in each considered scenario considering different sample sizes.}
		\label{figeta2}
\end{figure}

Figure (\ref{figeta2}) shows the box-plots of the ML estimates of the parameter $ \rho_2$  in all considered scenarios, considering samples of size 50 to 500 in increments of 25. Comparing the results from Figures \ref{figeta1} and \ref{figeta2}, we observe the presence of a higher bias for the estimates of $ \rho_2$ than for the estimates of $ \rho_1 $, given that the estimated and the nominal values of $ \rho_1$ and $ \rho_2$ are not close to each other in scenarios 3, 4, 7, 8, 11 and 12. The higher biases and variability of the ML estimates for the parameter $ \rho_2$ are observed in scenarios where we have high correlation between $ T_1 $ and $ T_2 $.  Note that the estimates with lower biases are seen in the estimation of the parameter $ \rho_1 $ instead of the parameter $ \rho_2 $. This is probably due to the correlation between $ T_1 $ and $ T_2 $ included in the simulation process.

\begin{figure}[!ht]
\centering
	\setkeys{Gin}{height = 4.2cm,width=0.24\linewidth} %
		\includegraphics[scale=0.3]{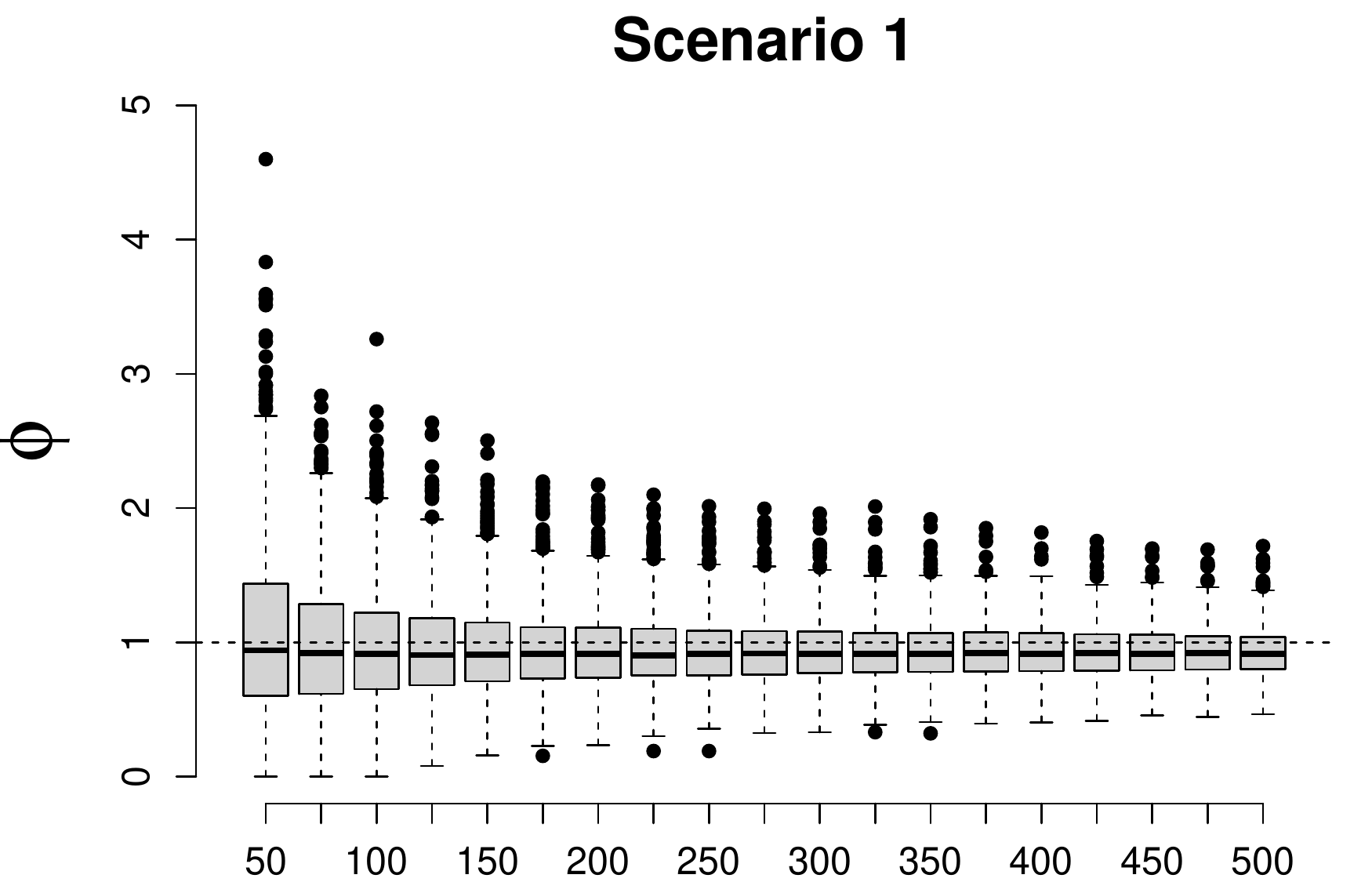}
		\includegraphics[scale=0.3]{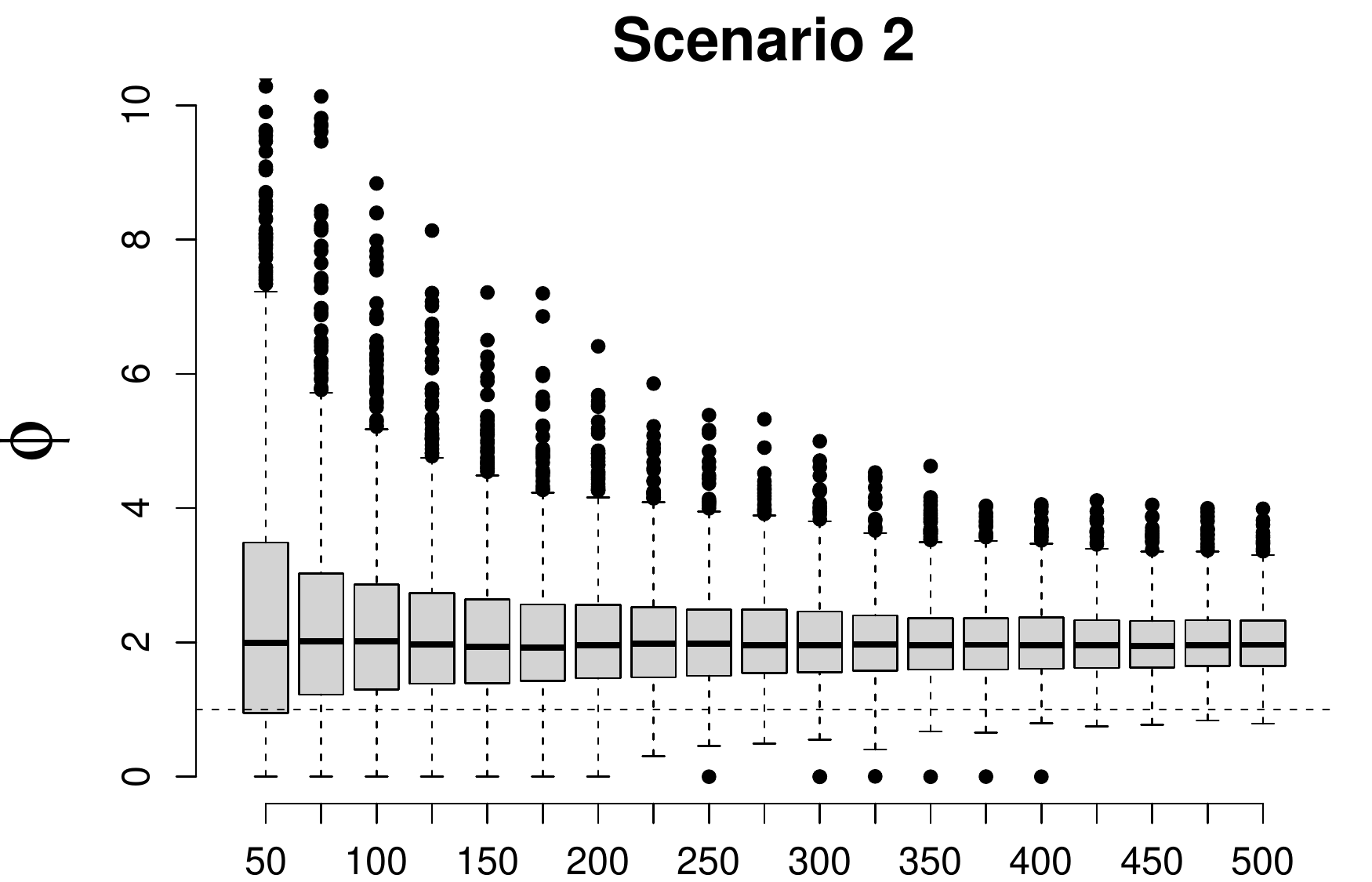}
		\includegraphics[scale=0.3]{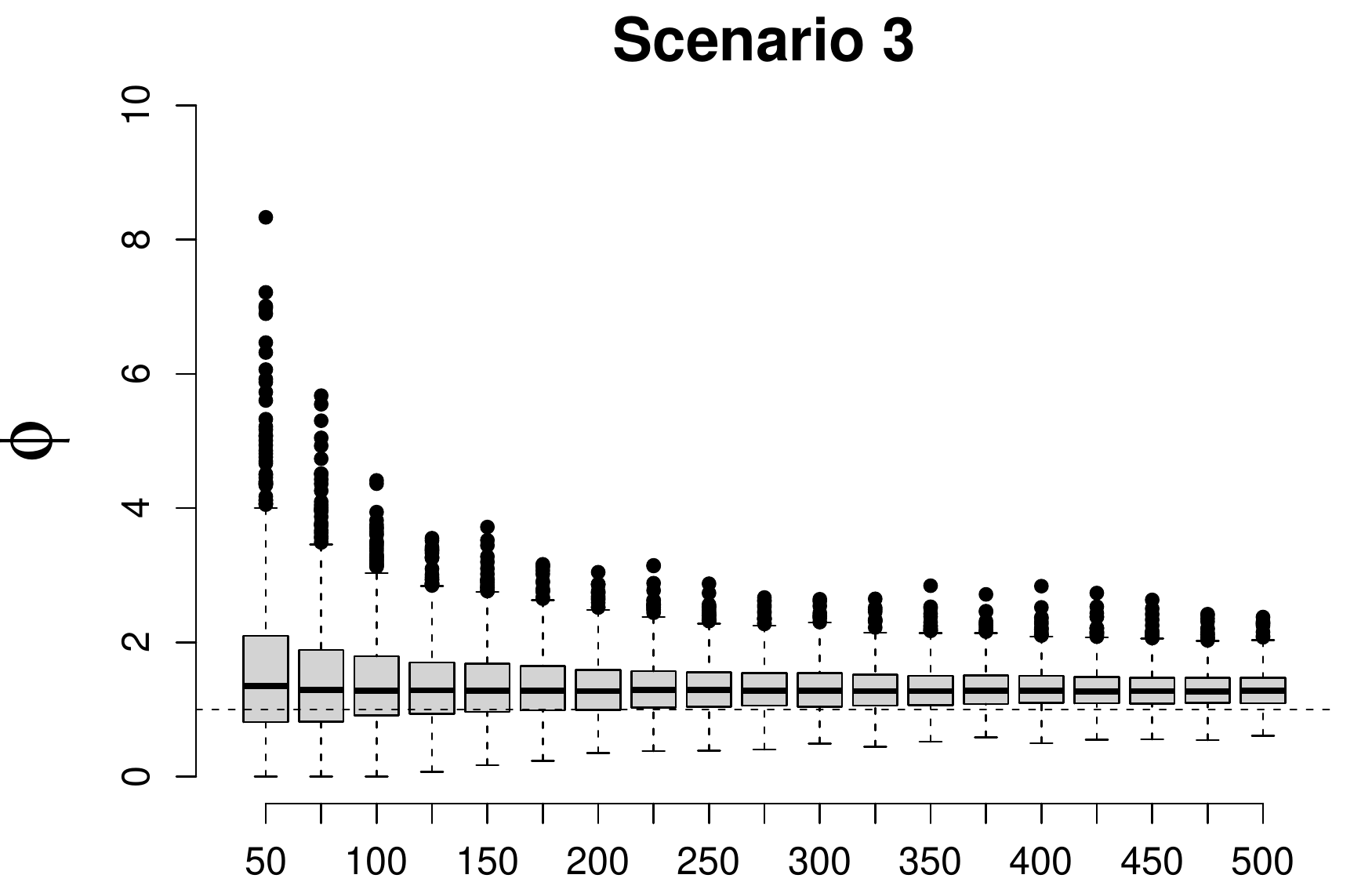}
		\includegraphics[scale=0.3]{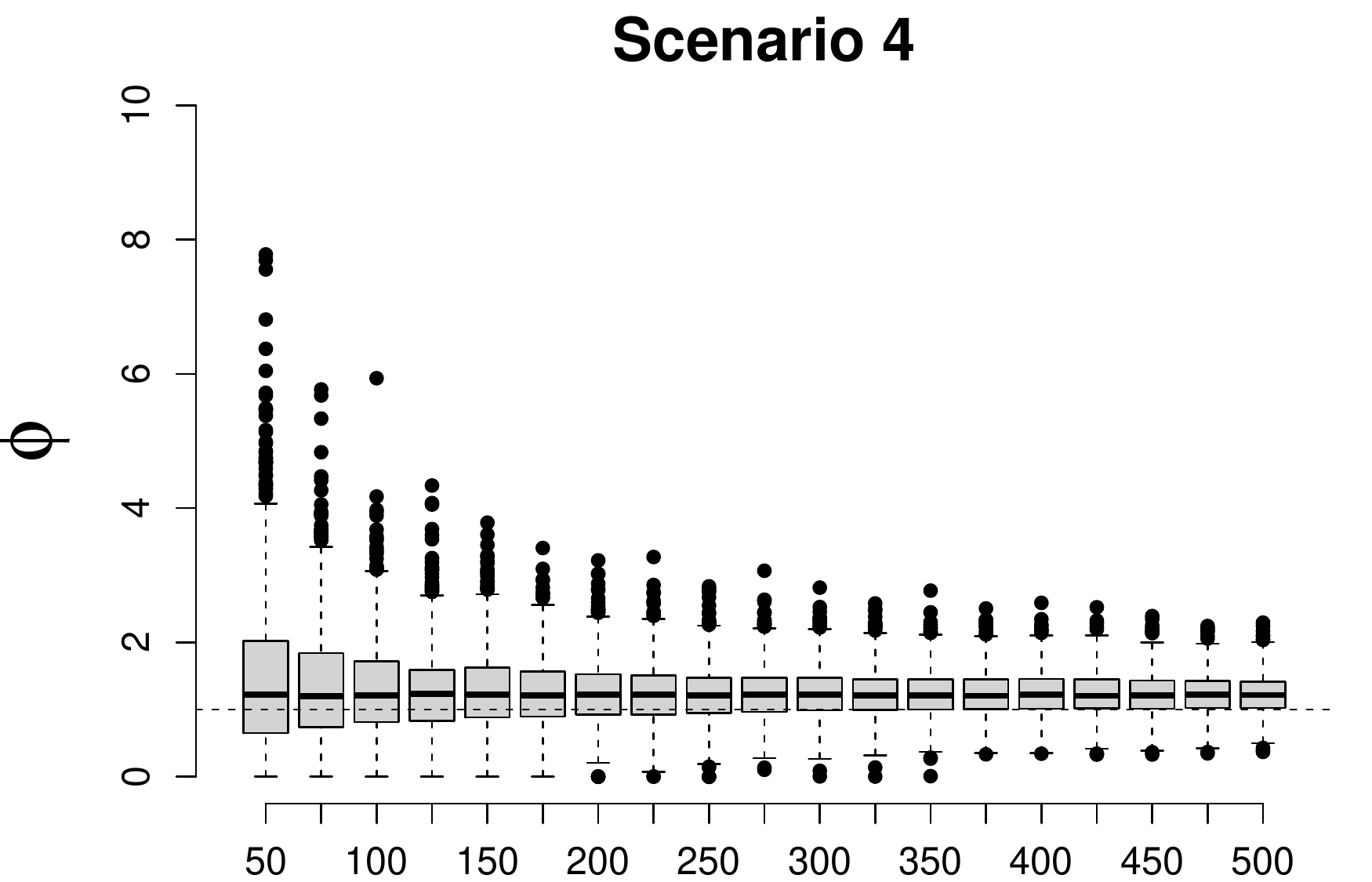}
		\includegraphics[scale=0.3]{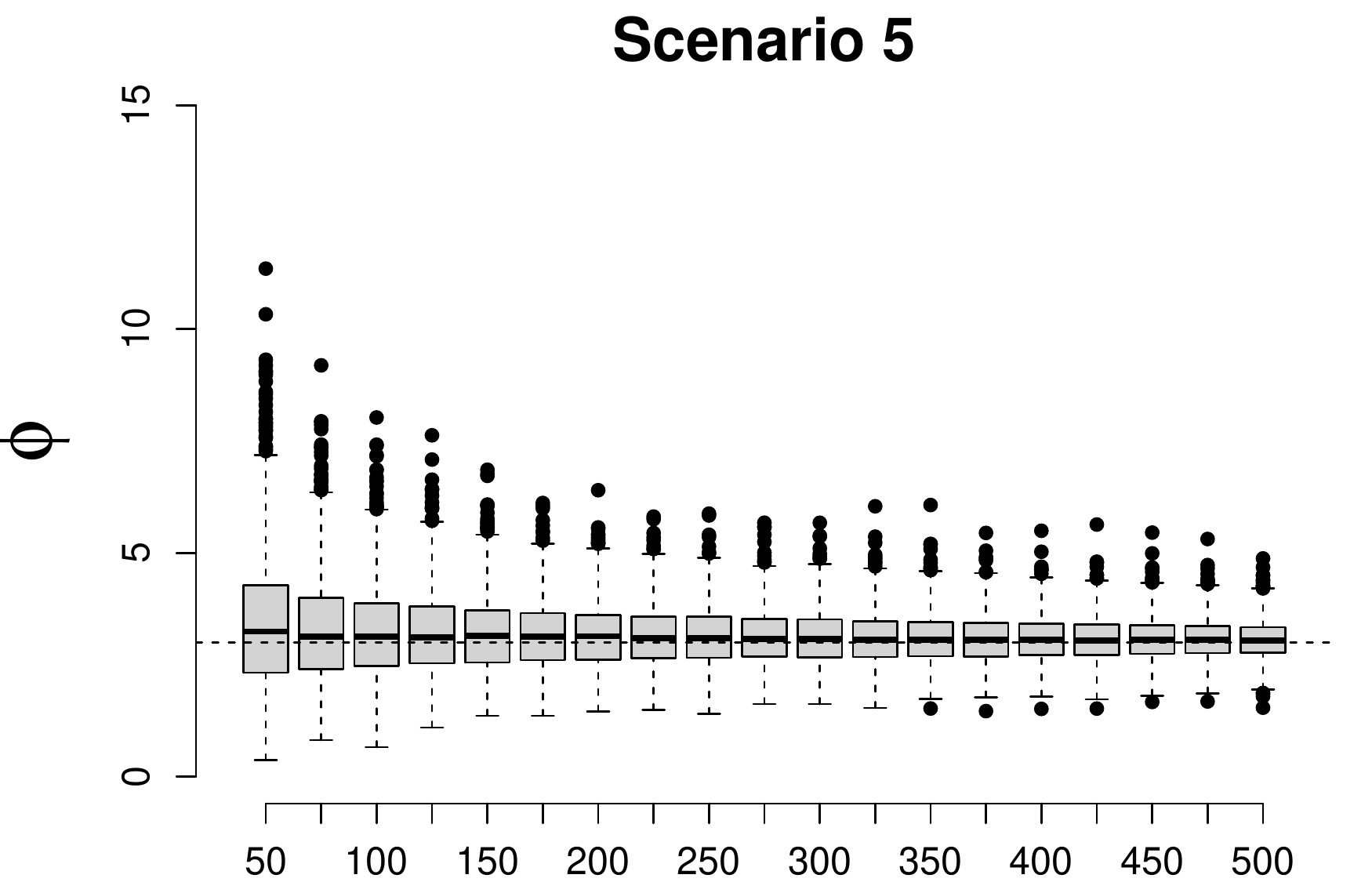}
		\includegraphics[scale=0.3]{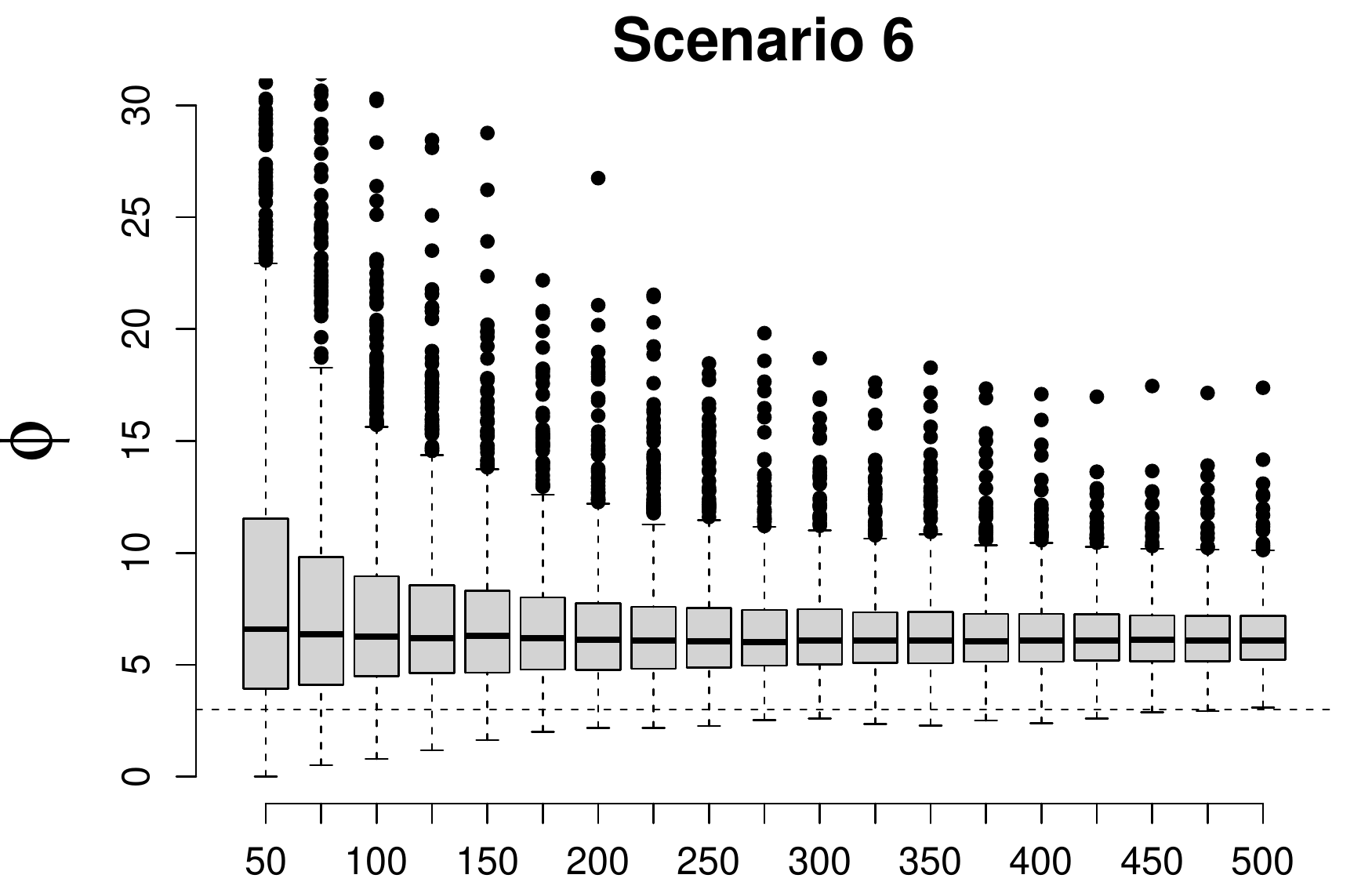}
		\includegraphics[scale=0.3]{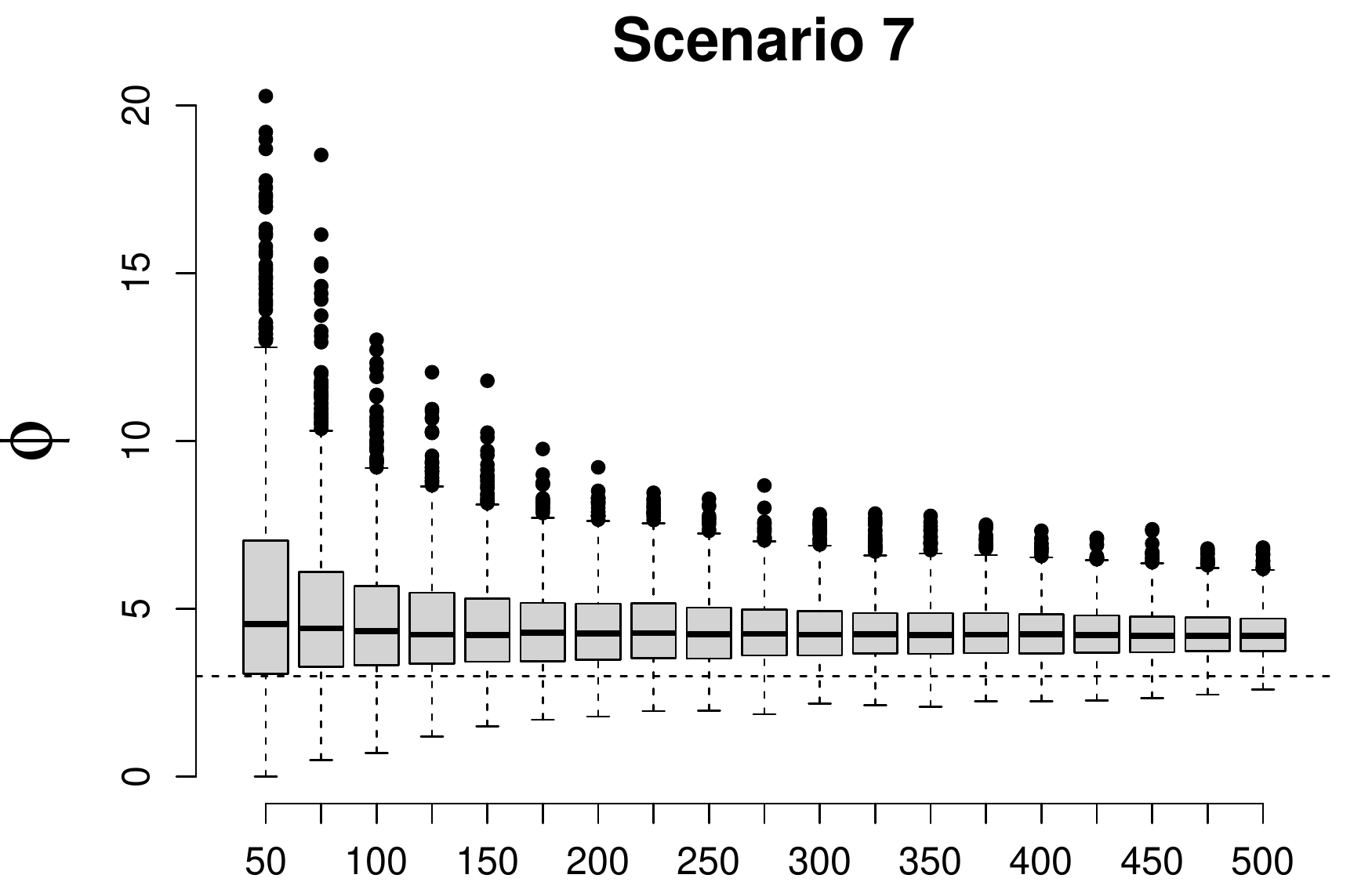}
		\includegraphics[scale=0.3]{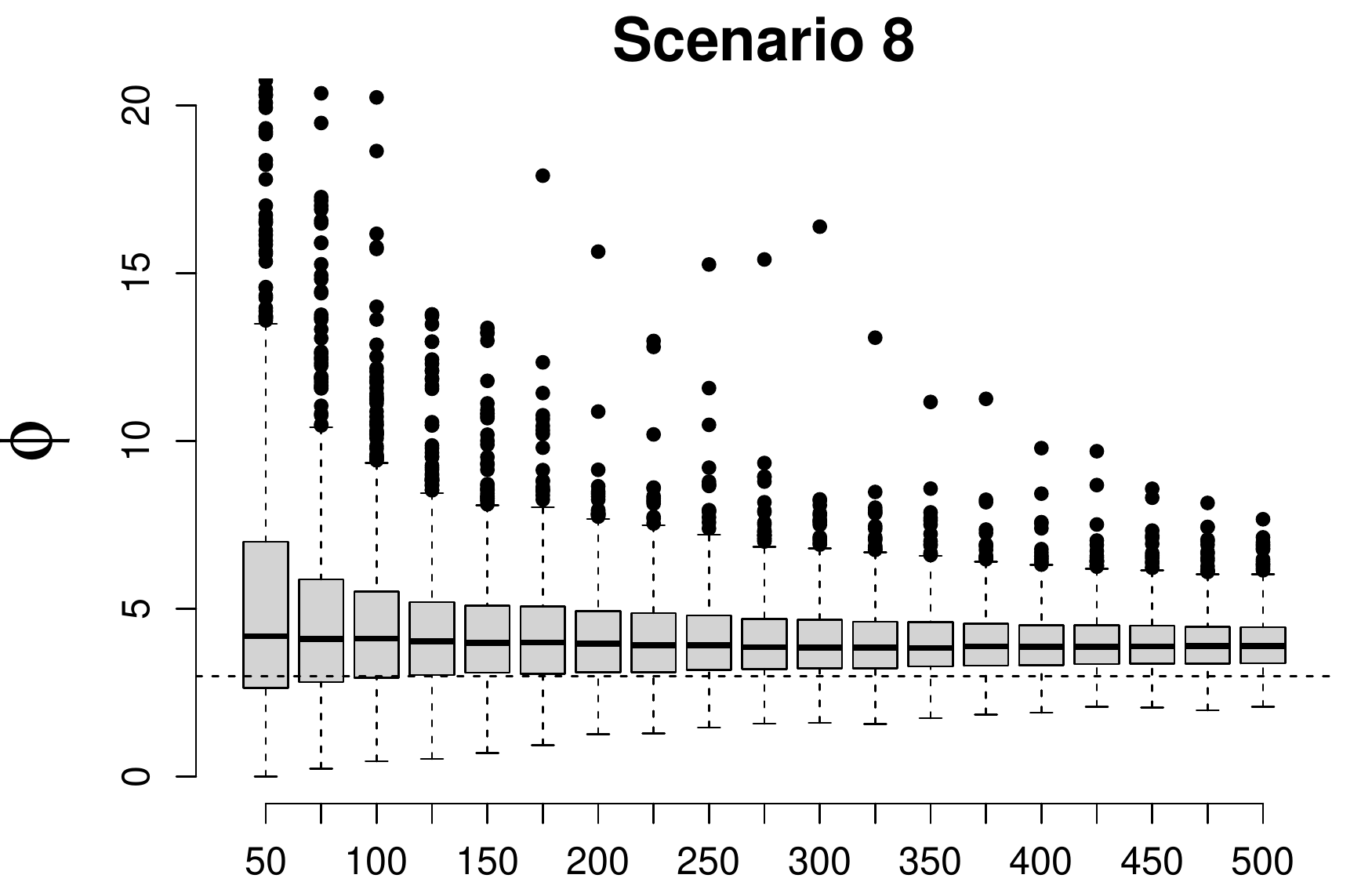}
		\includegraphics[scale=0.3]{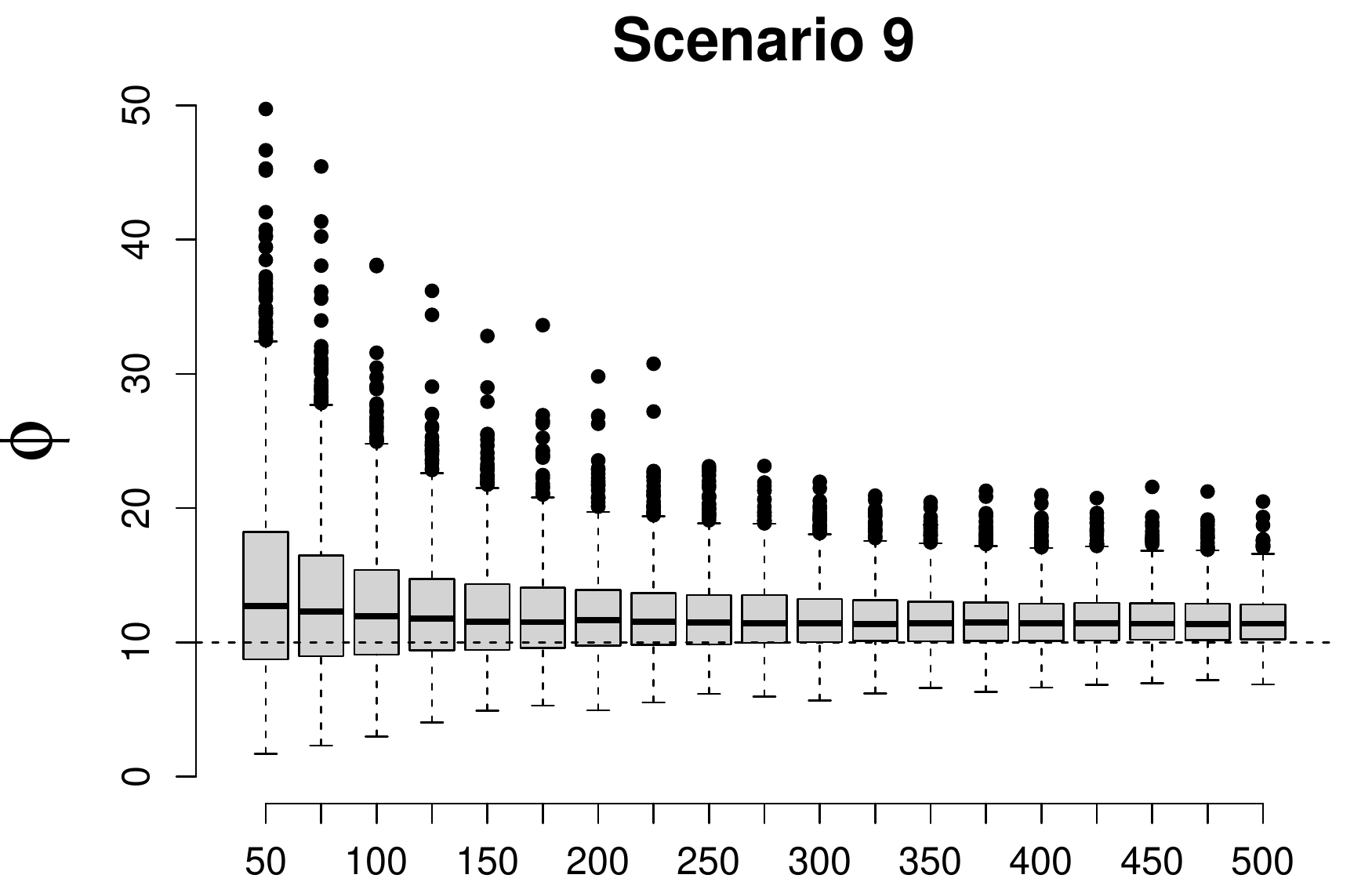}
		\includegraphics[scale=0.3]{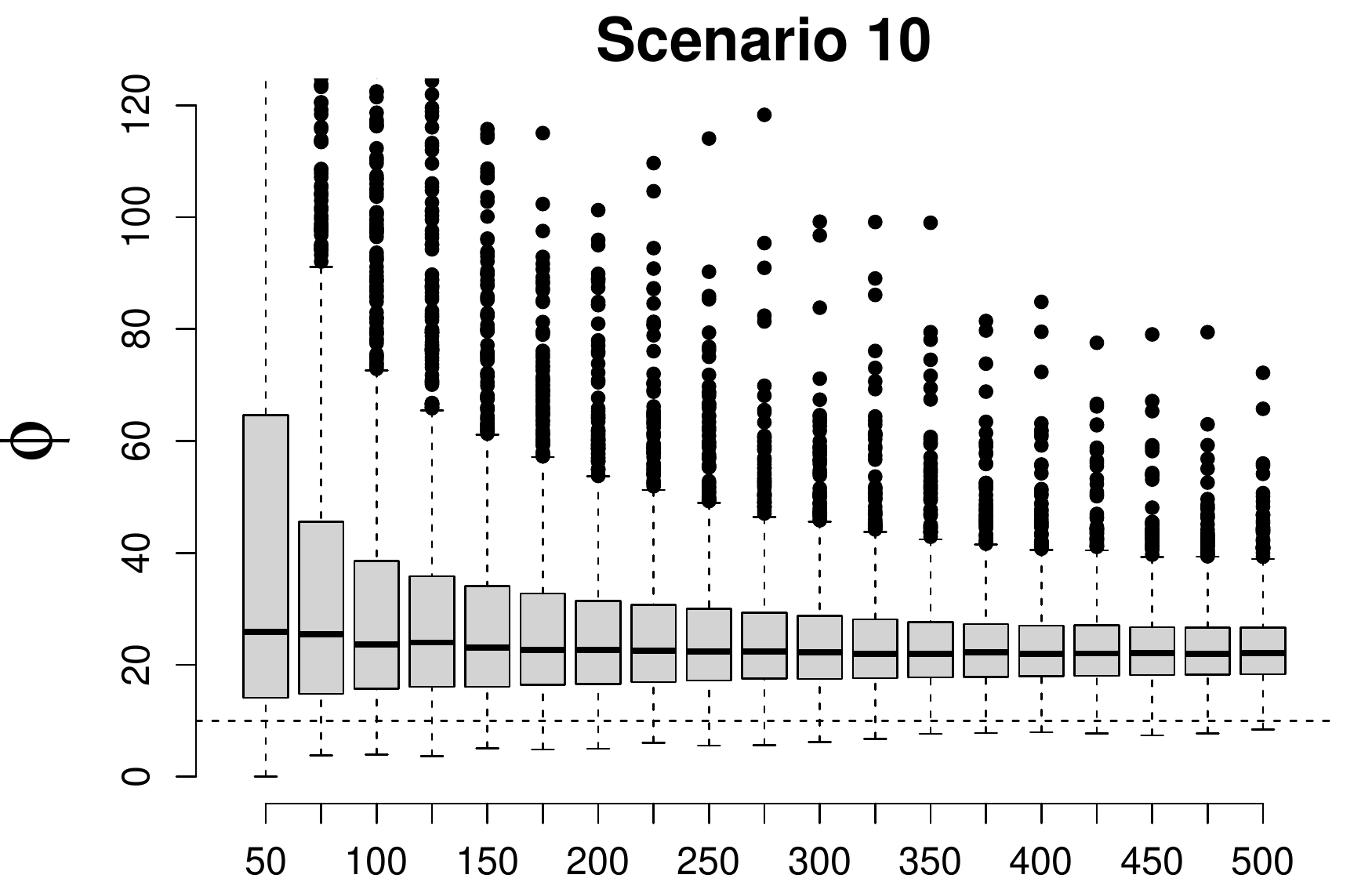}
		\includegraphics[scale=0.3]{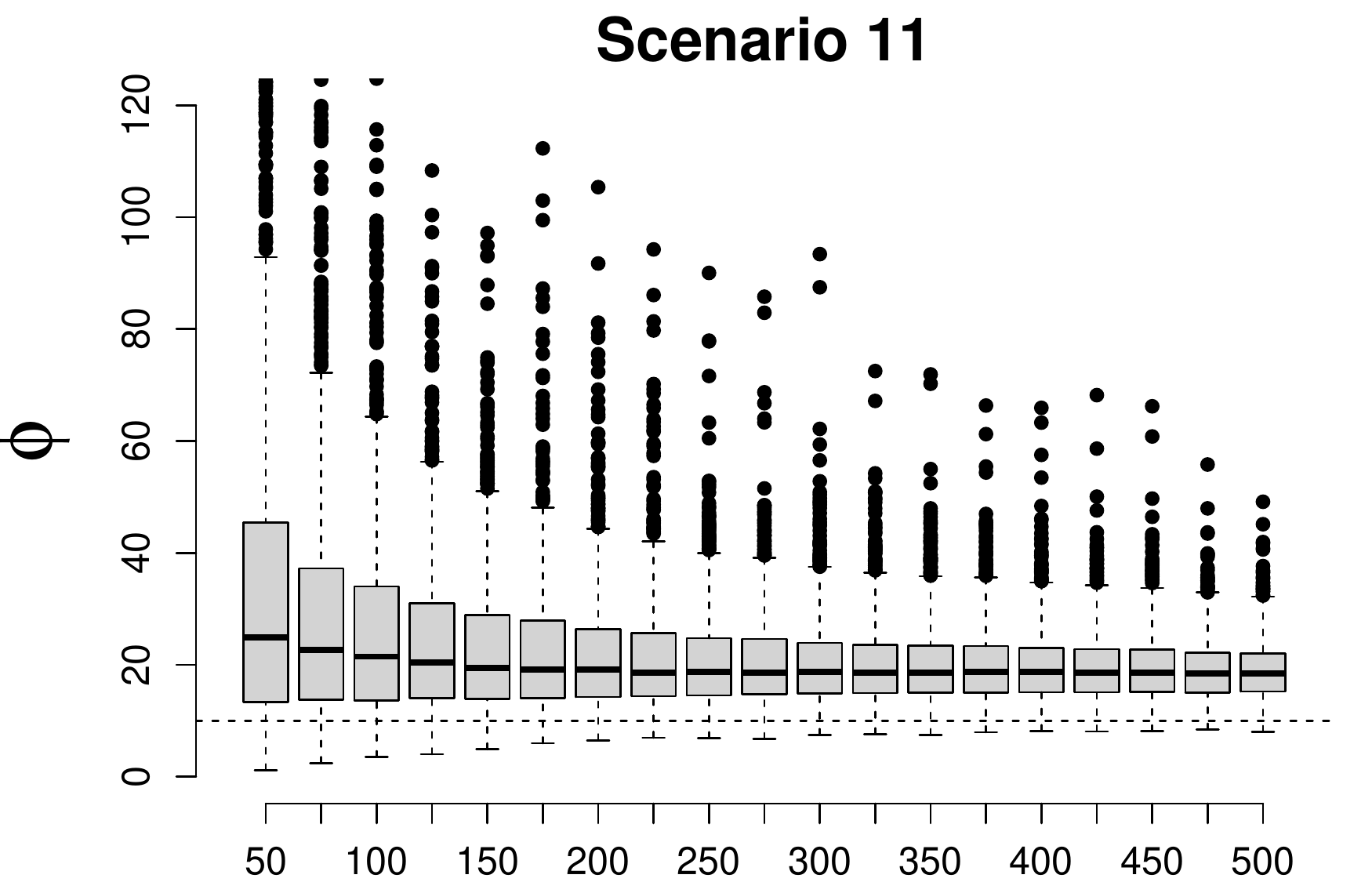}
		\includegraphics[scale=0.3]{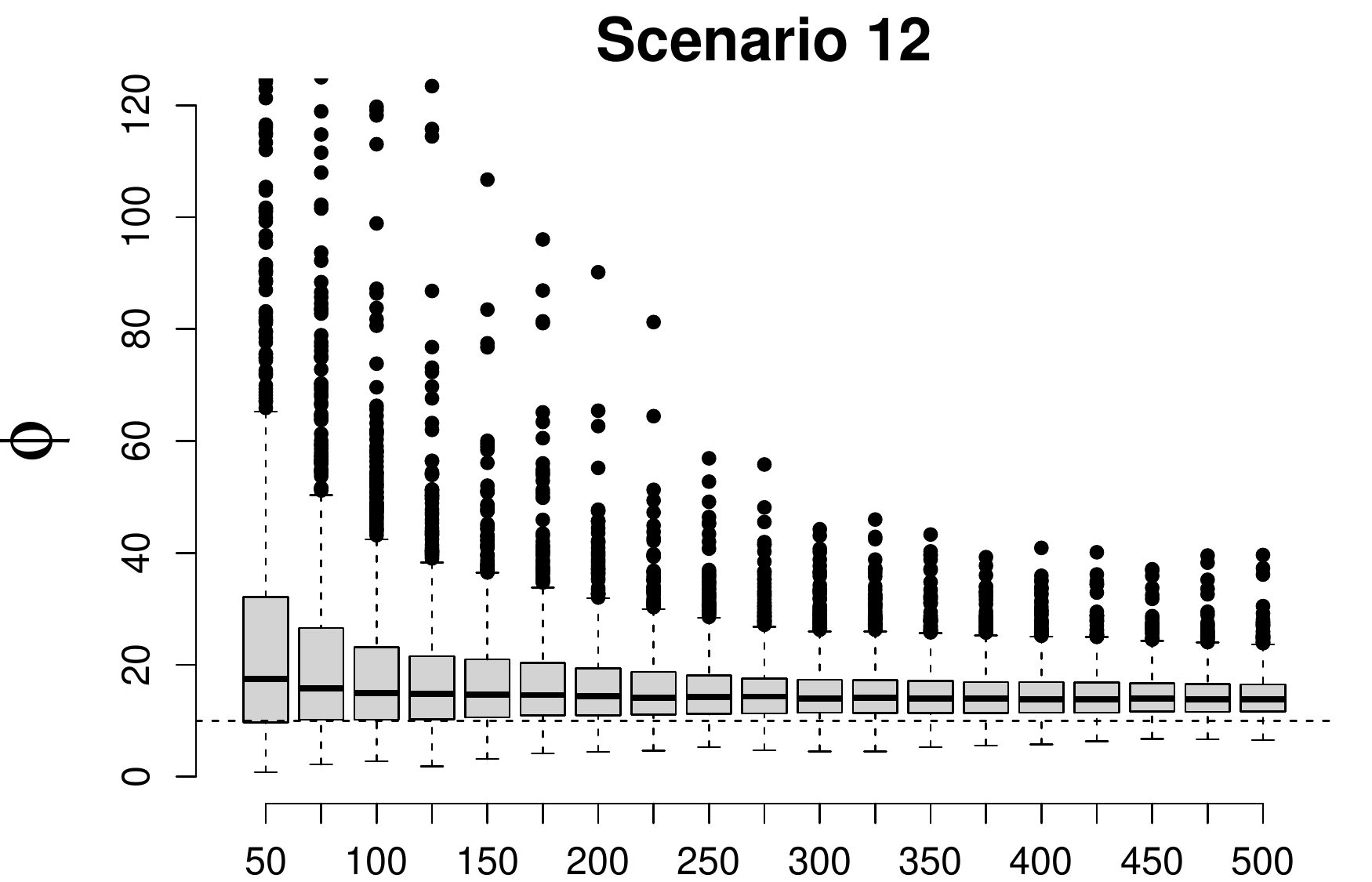}
		\caption{Box-plots of the estimates for $ \phi $, considering the maximum likelihood estimation method  in different scenarios and different sample sizes.}
		\label{figrho}
\end{figure}

The box-plots for the estimates of $\phi$ are presented in the Figure \ref{figrho}.  From these graphs, it is possible to note a great variability of the ML estimates of $ \phi $, and this variability increases as the correlation between $ T_1 $ and $ T_2 $ increases. Morever, it is also possible to observe relatively small interquartile ranges, indicating that most of the estimates obtained are highly concentrated in the central portion of the respective distributions, even in the presence of biases observed in the scenarios with higher cure rate. In addition, it is observed an expressive  presence of bias in scenarios with higher cure rates, so that the medians of the estimates are slightly above the expected nominal values. In general, we could conclude that the model is adequate in these scenarios when the sample size is at least of  100 individuals.

\begin{figure}[H]
	\begin{center}
		\includegraphics[scale=0.3]{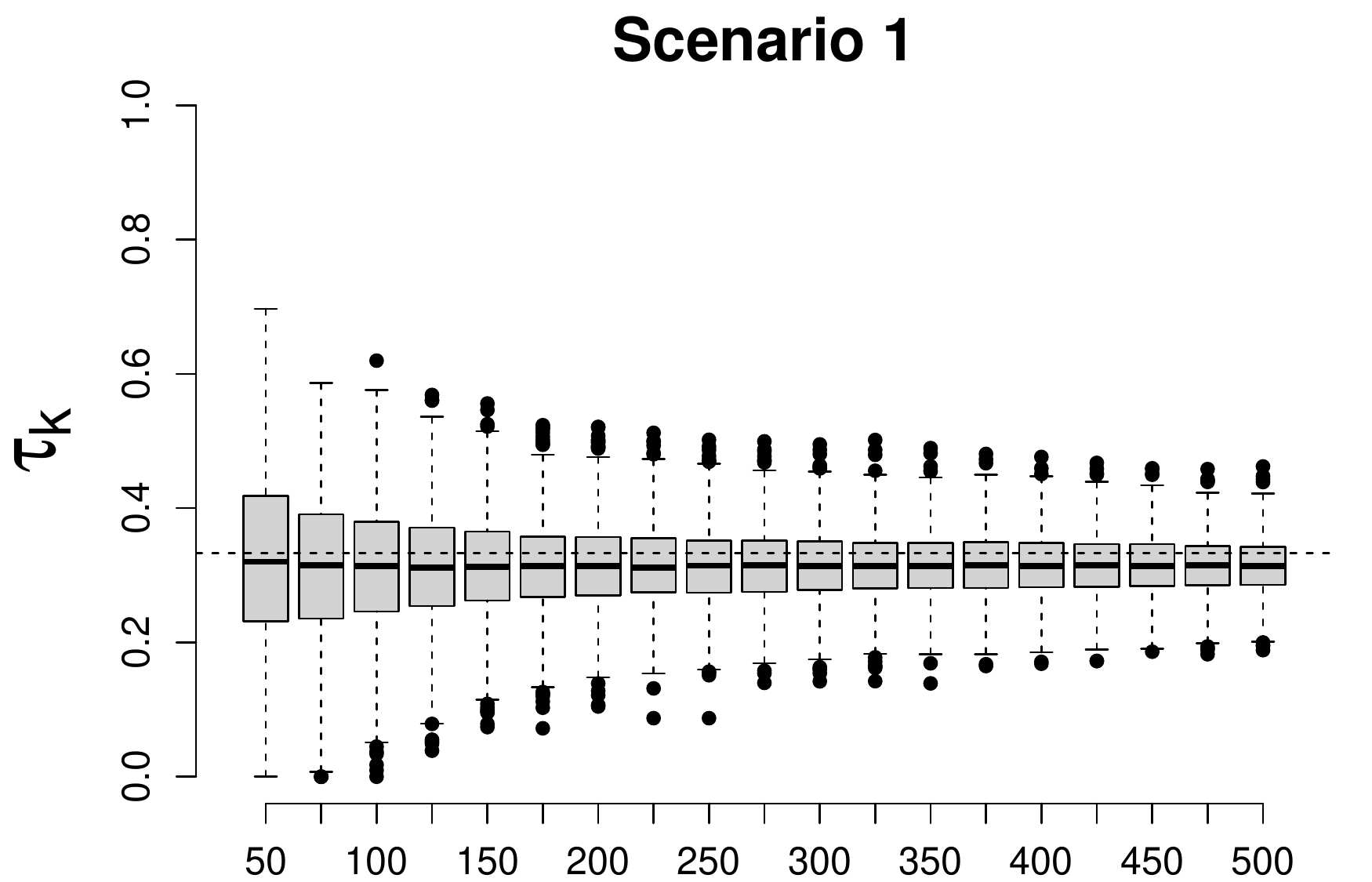}
		\includegraphics[scale=0.3]{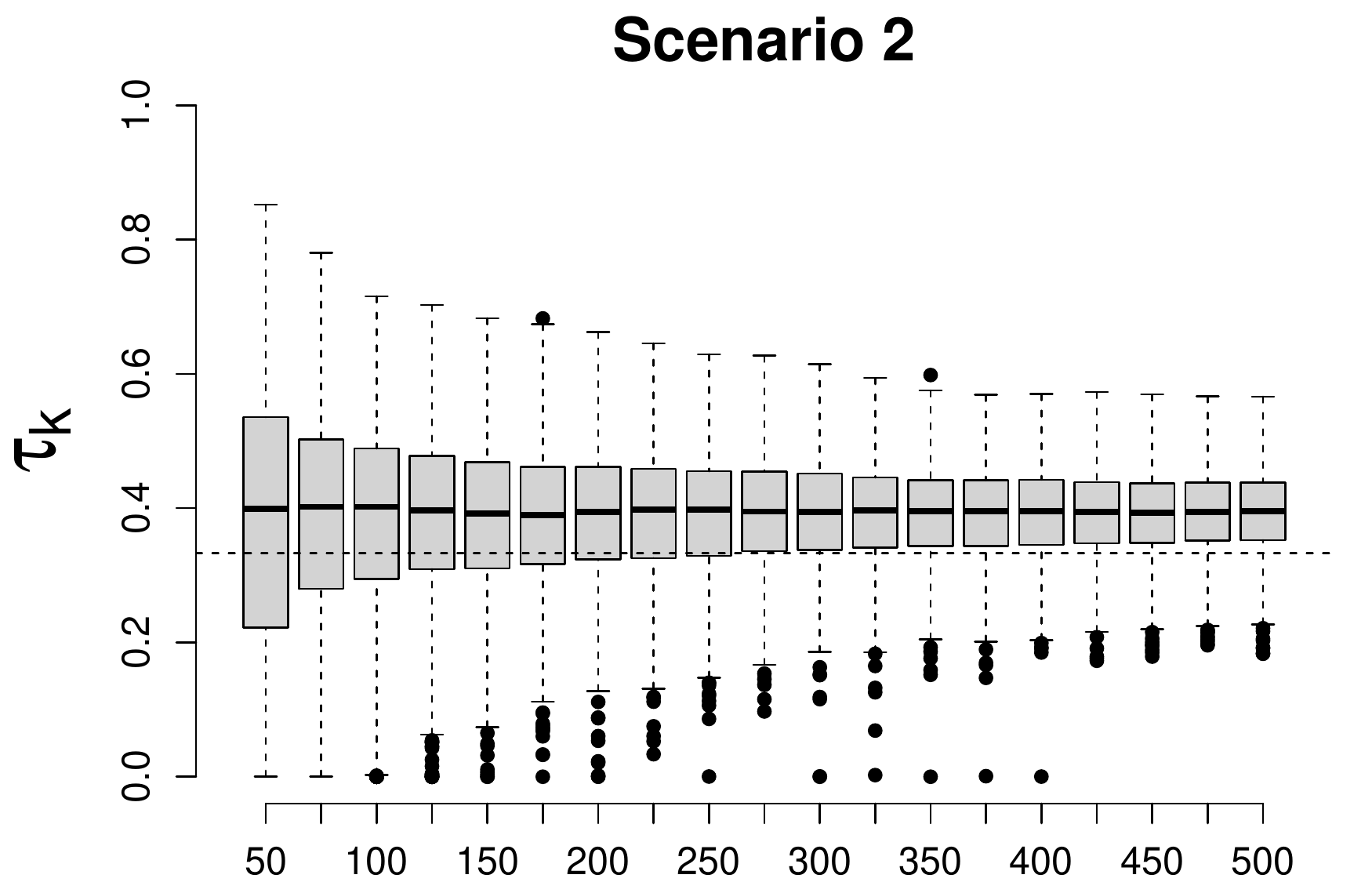}
		\includegraphics[scale=0.3]{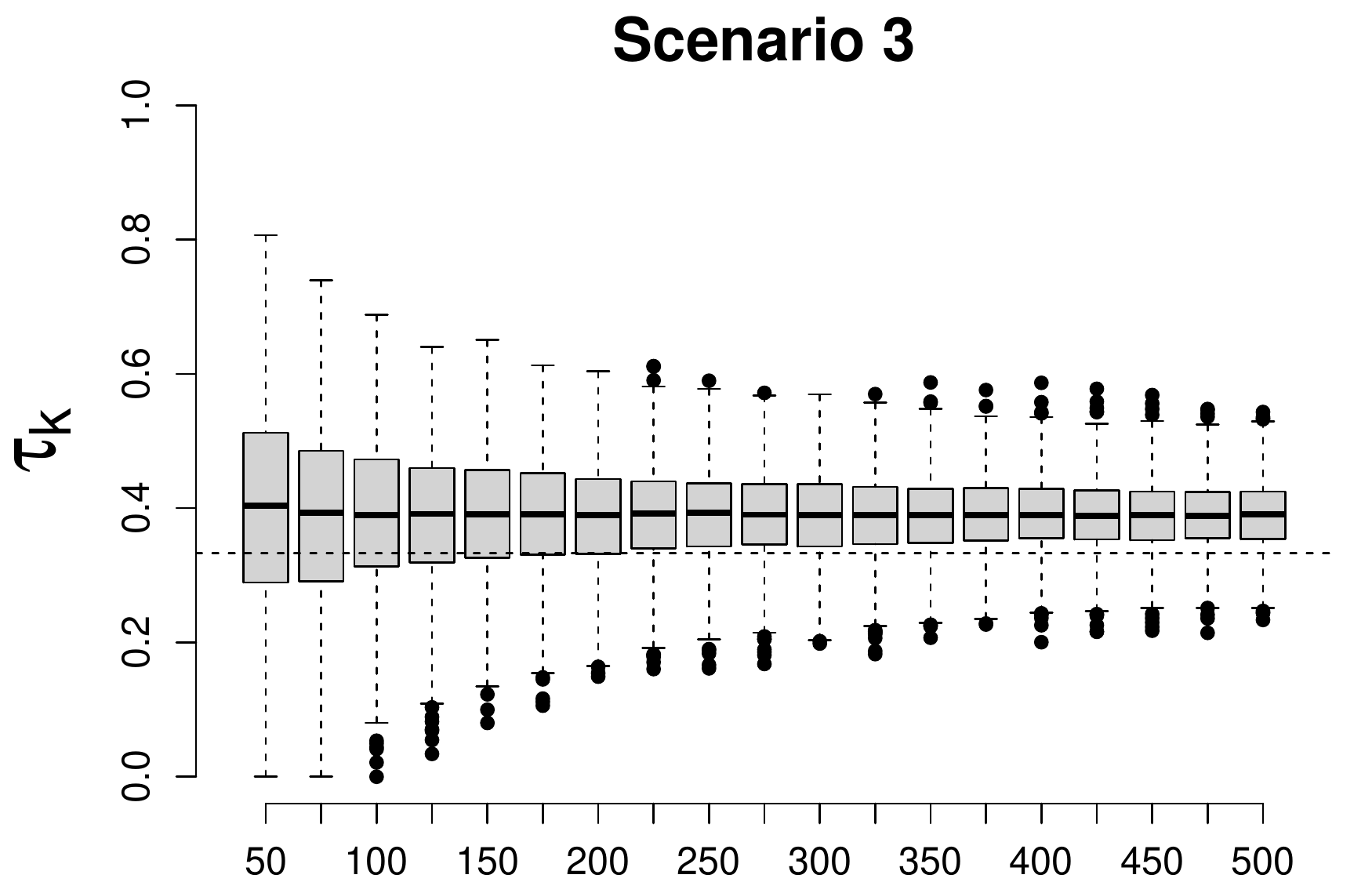}
		\includegraphics[scale=0.3]{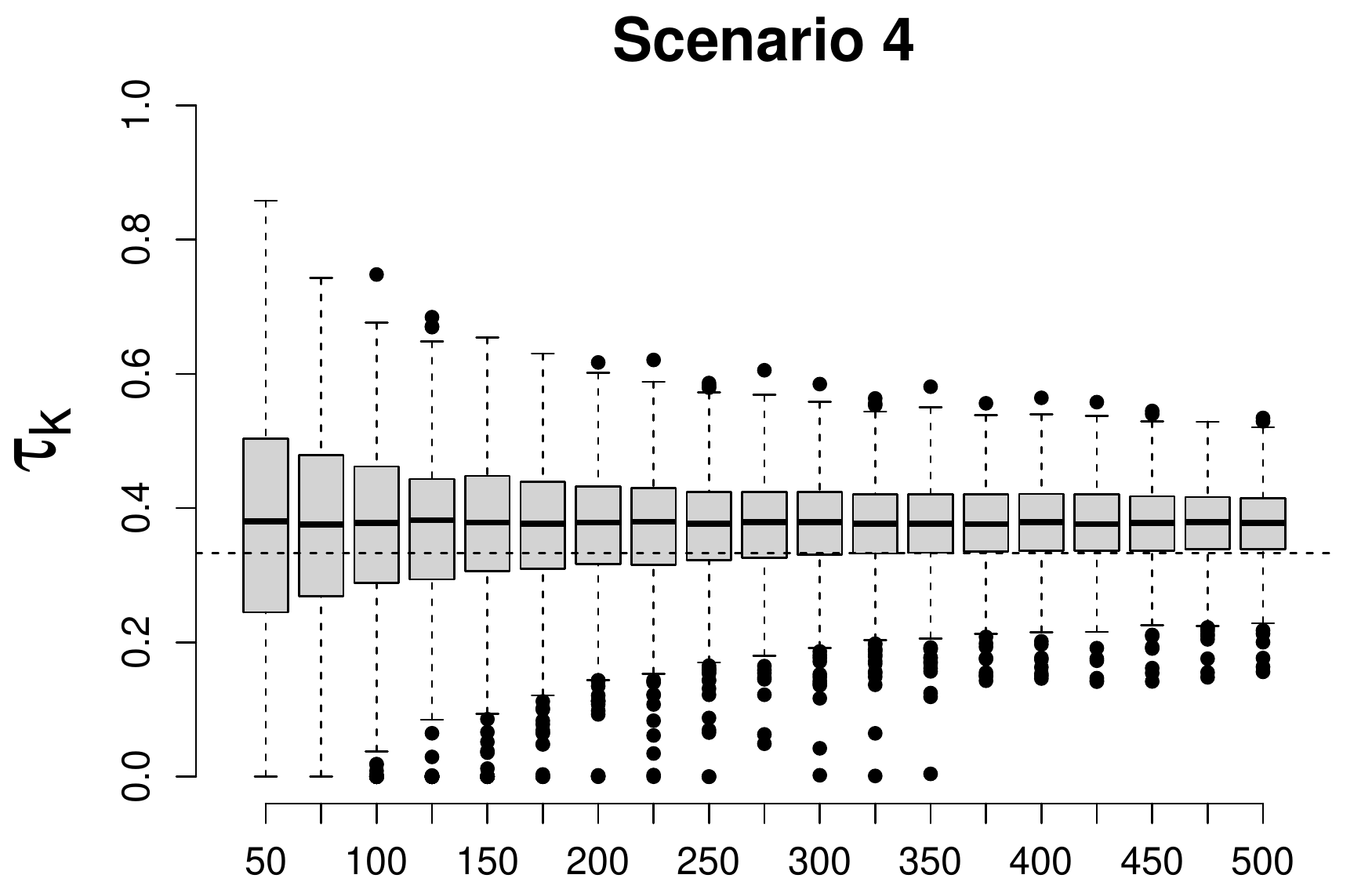}
		\includegraphics[scale=0.3]{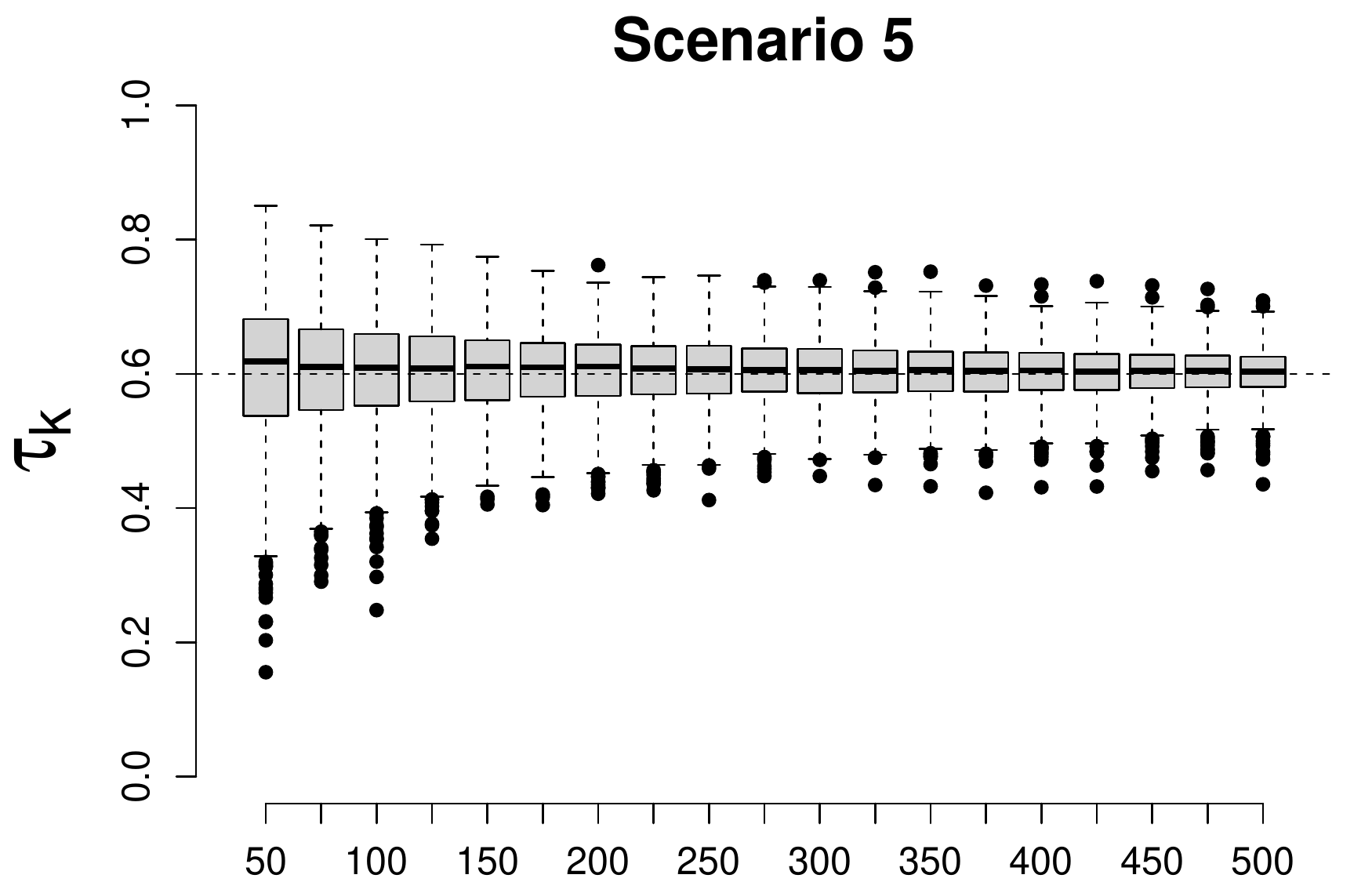}
		\includegraphics[scale=0.3]{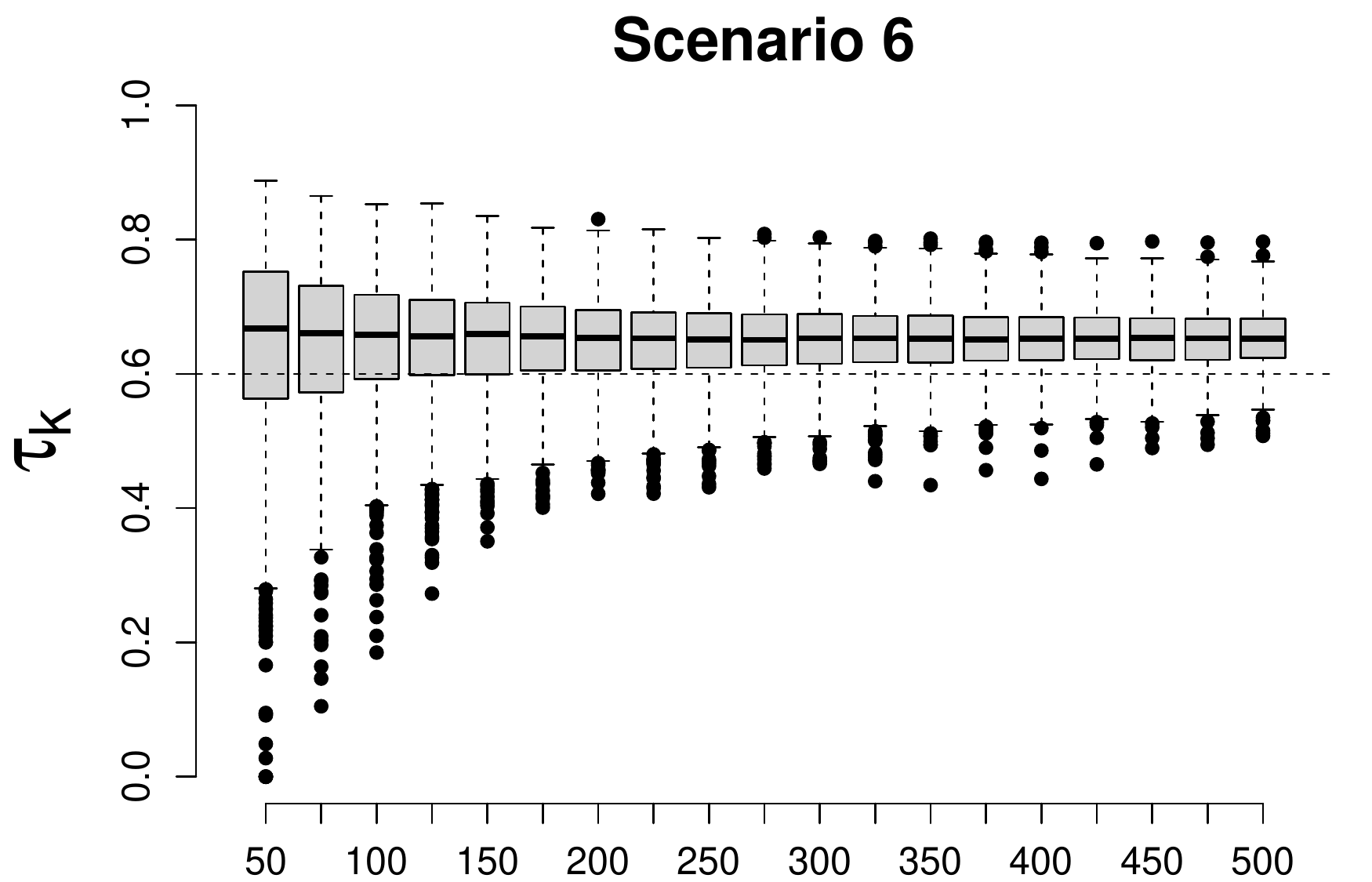}
		\includegraphics[scale=0.3]{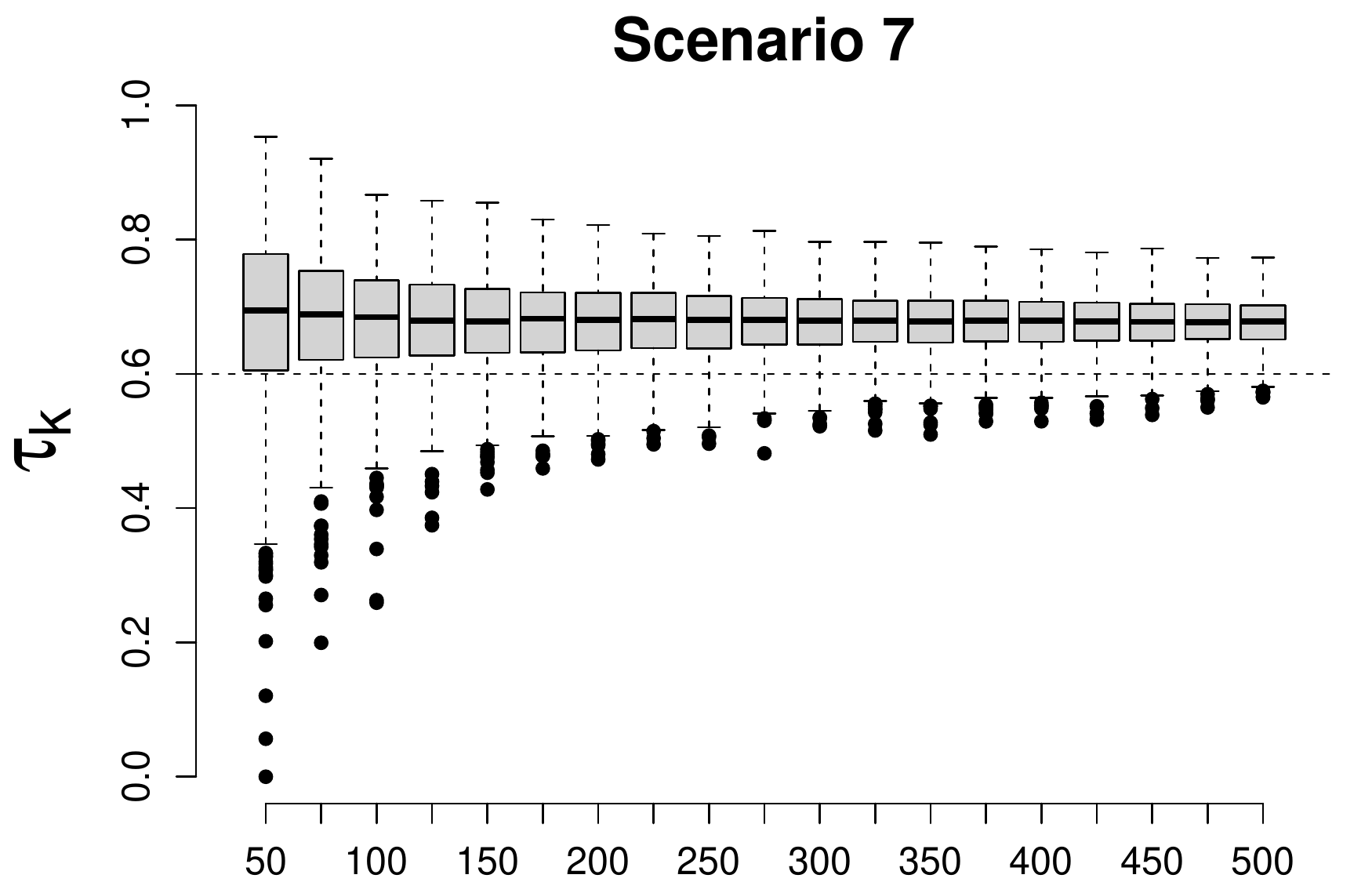}
		\includegraphics[scale=0.3]{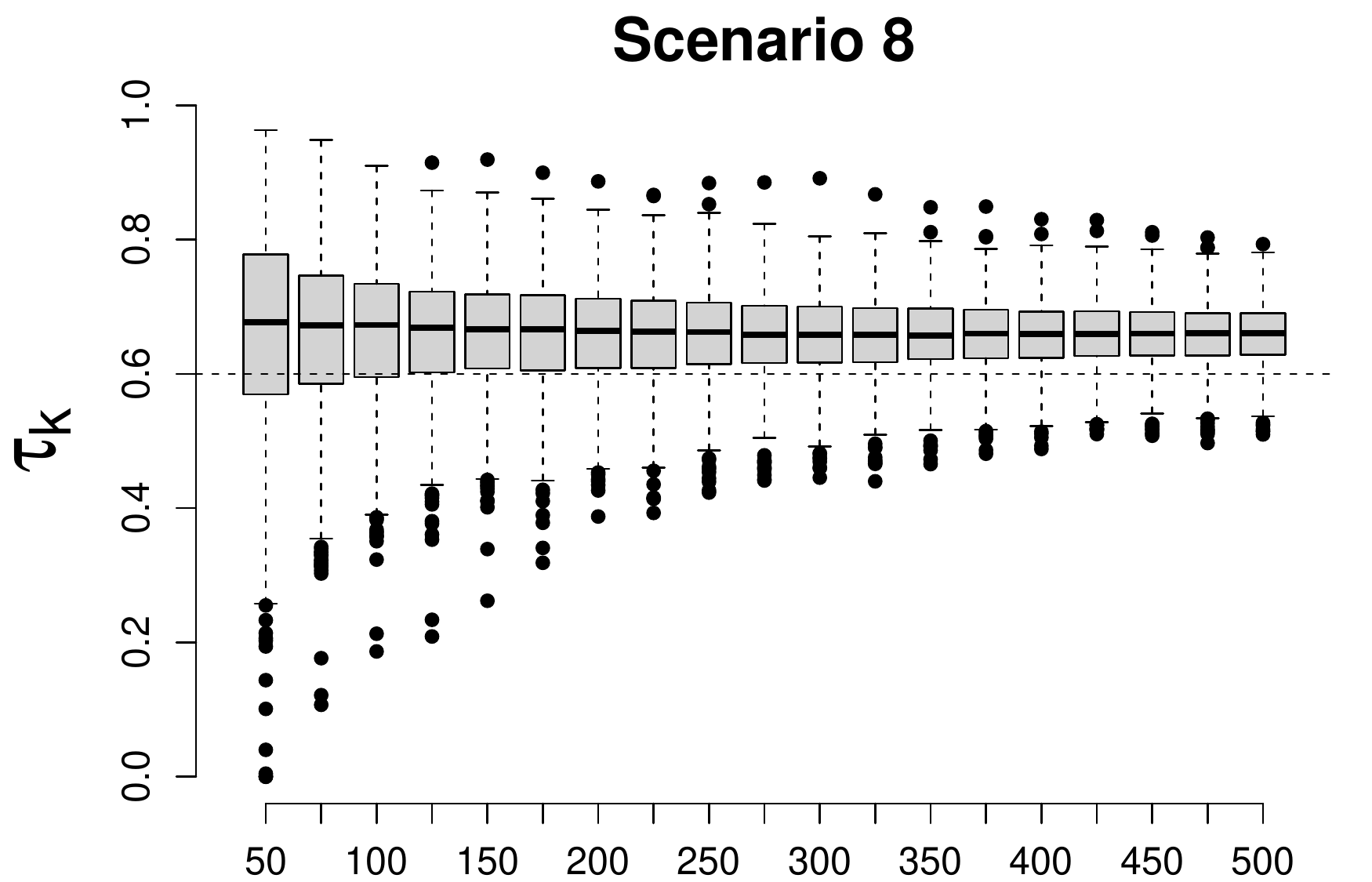}
		\includegraphics[scale=0.3]{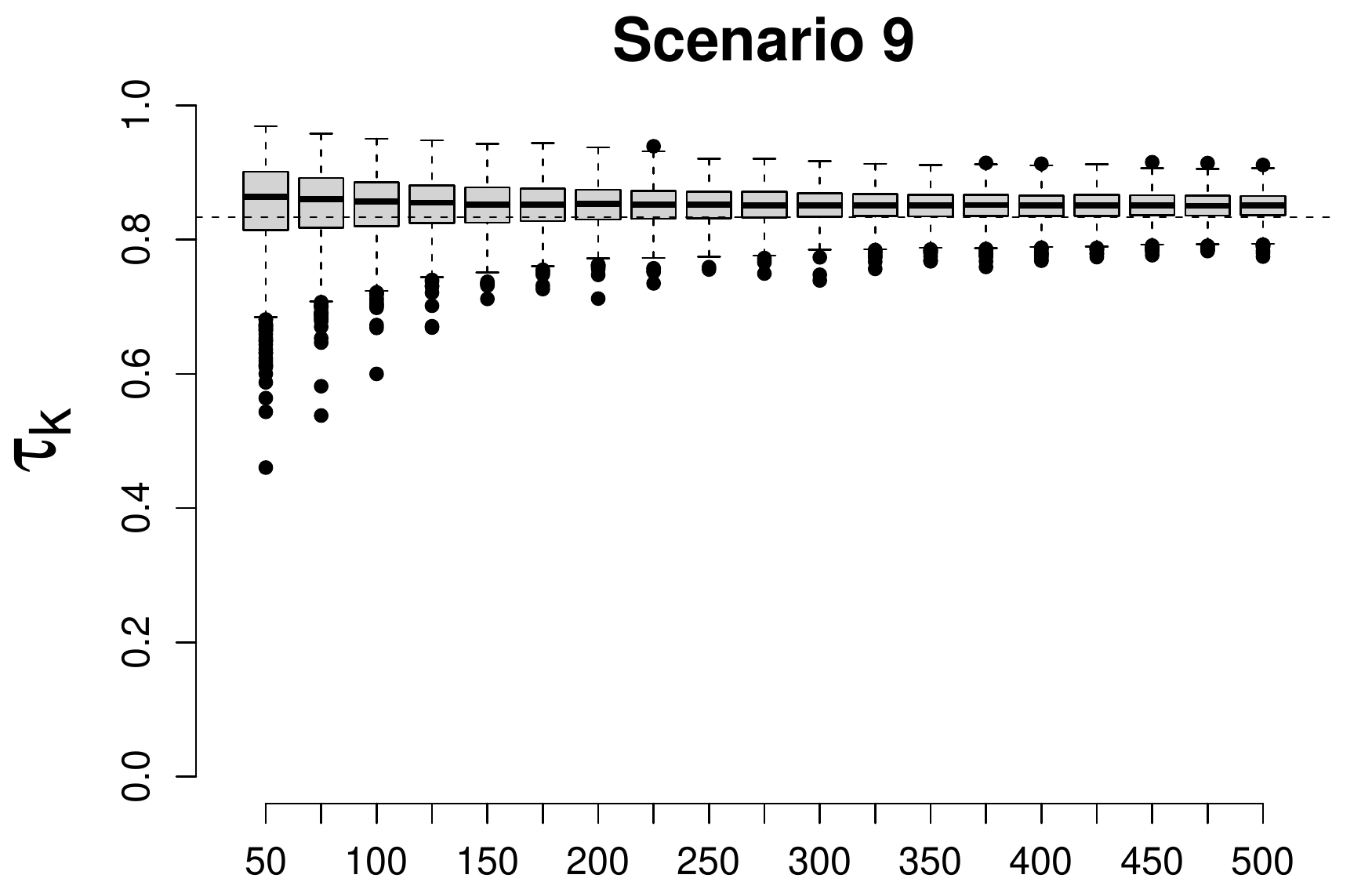}
		\includegraphics[scale=0.3]{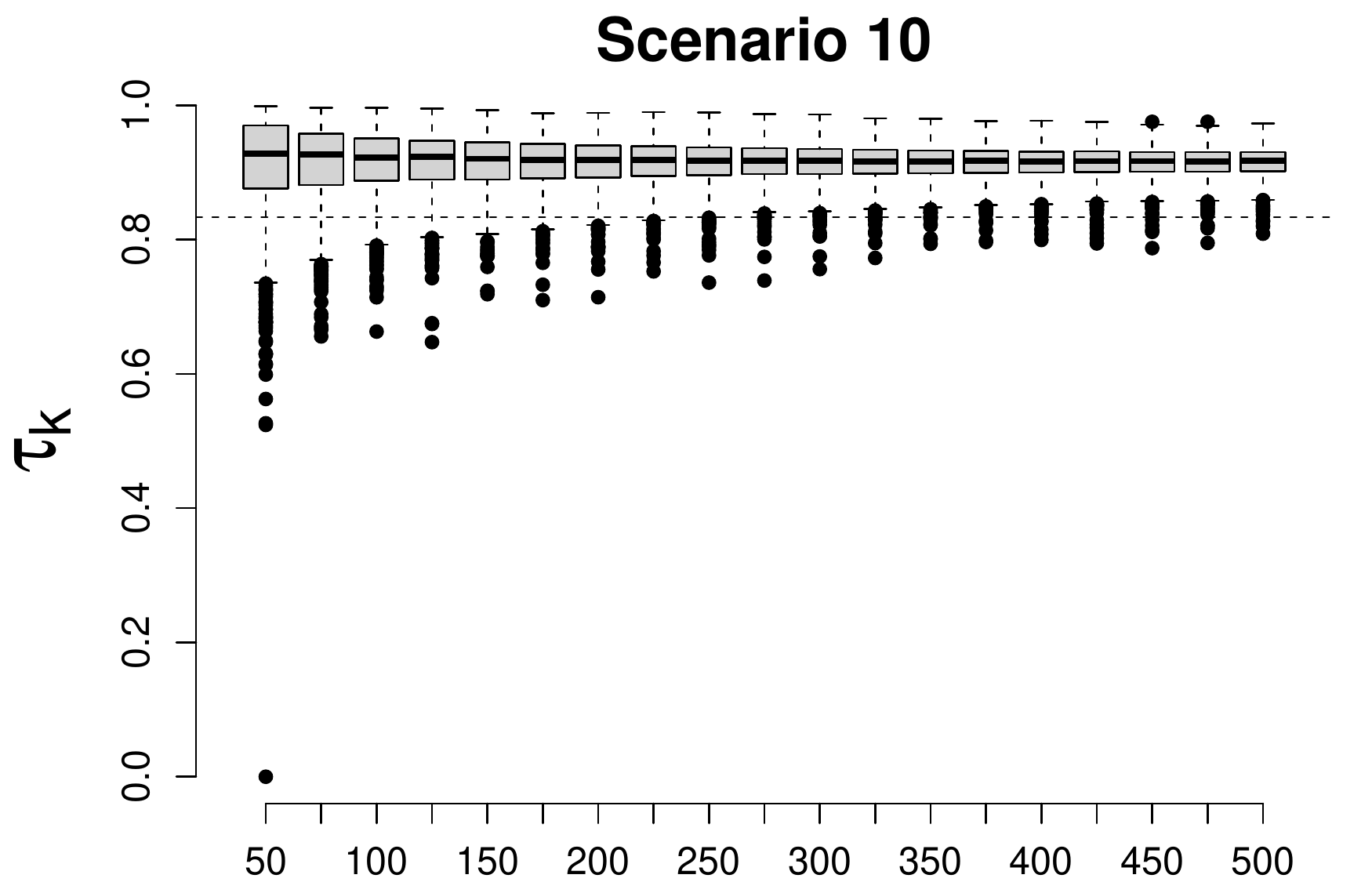}
		\includegraphics[scale=0.3]{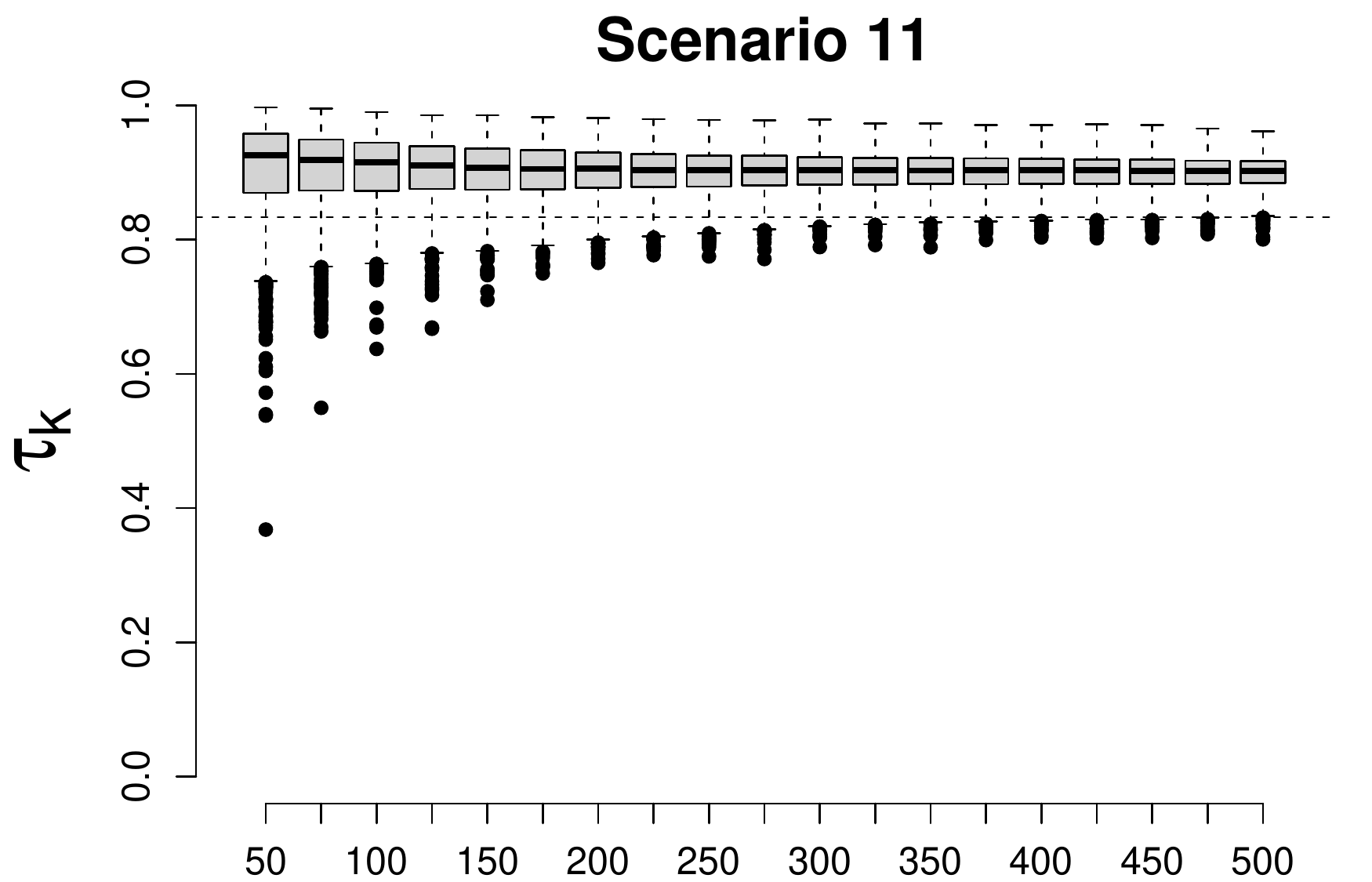}
		\includegraphics[scale=0.3]{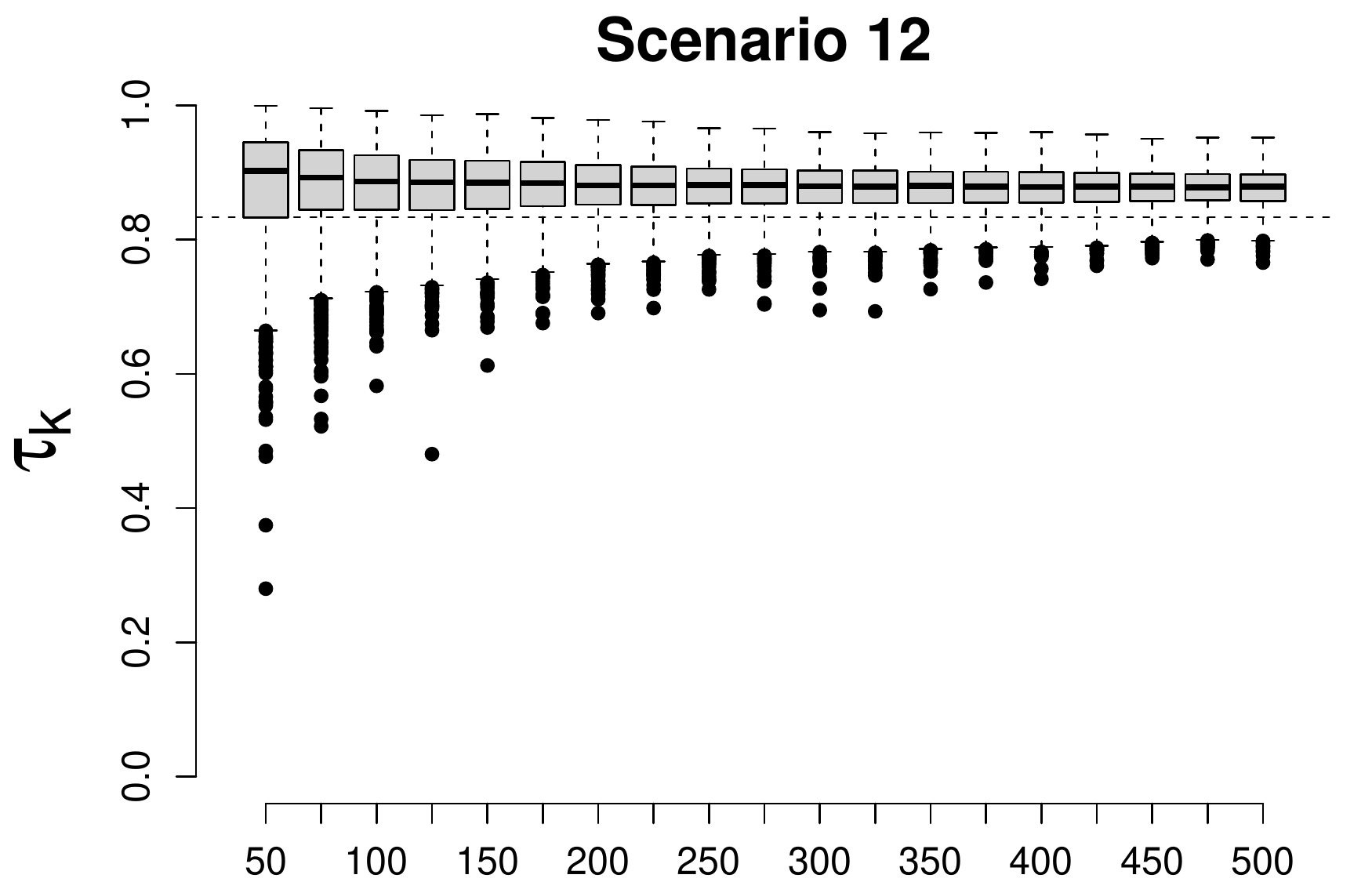}
		\caption{Box-plots of the estimates for $ \tau_k $, considering the maximum likelihood estimation method  in different scenarios and different sample sizes.}
		\label{figkendal}
	\end{center}
\end{figure}

\begin{figure}[H]
	\begin{center}
		\includegraphics[scale=0.3]{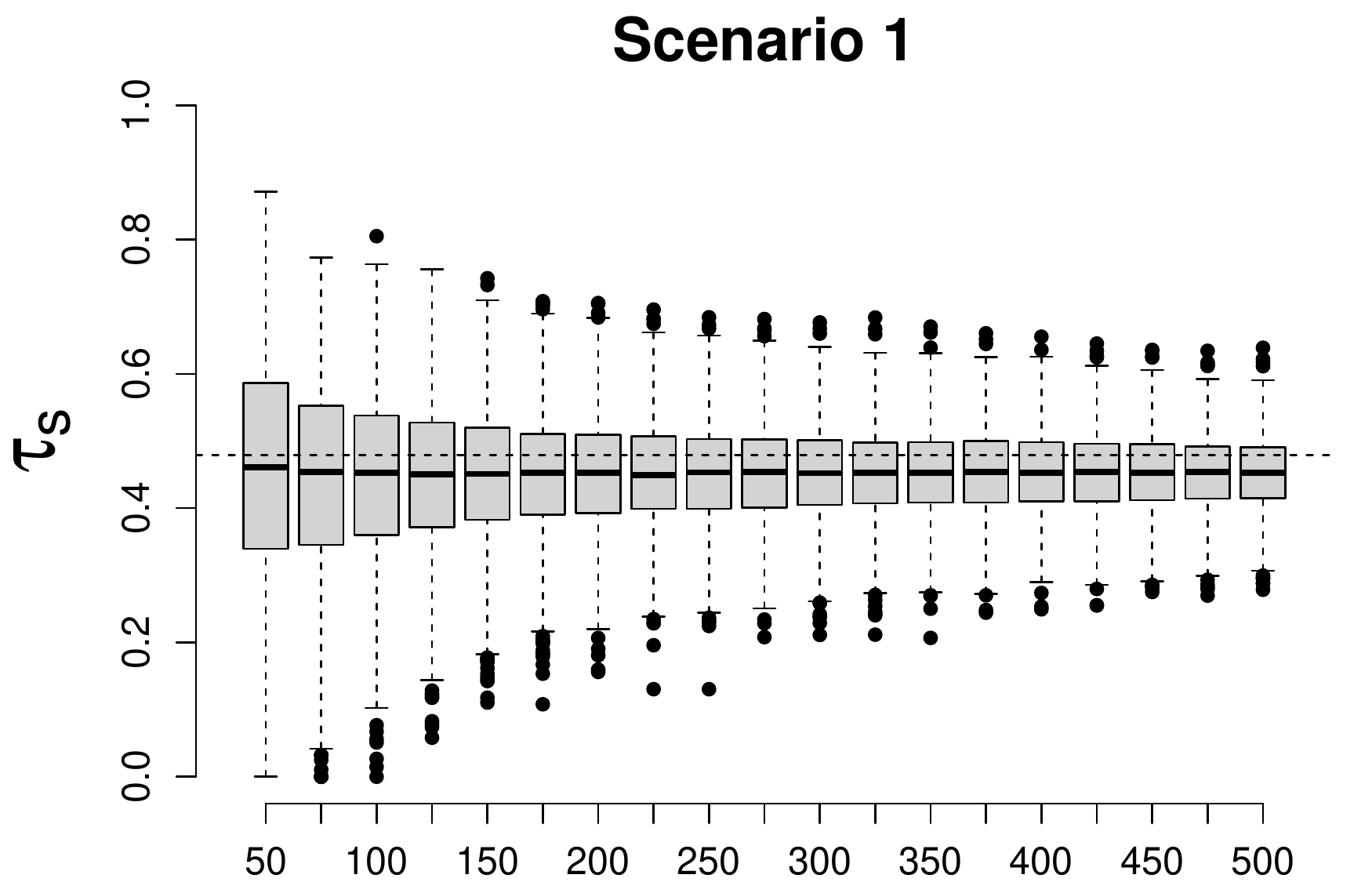}
		\includegraphics[scale=0.3]{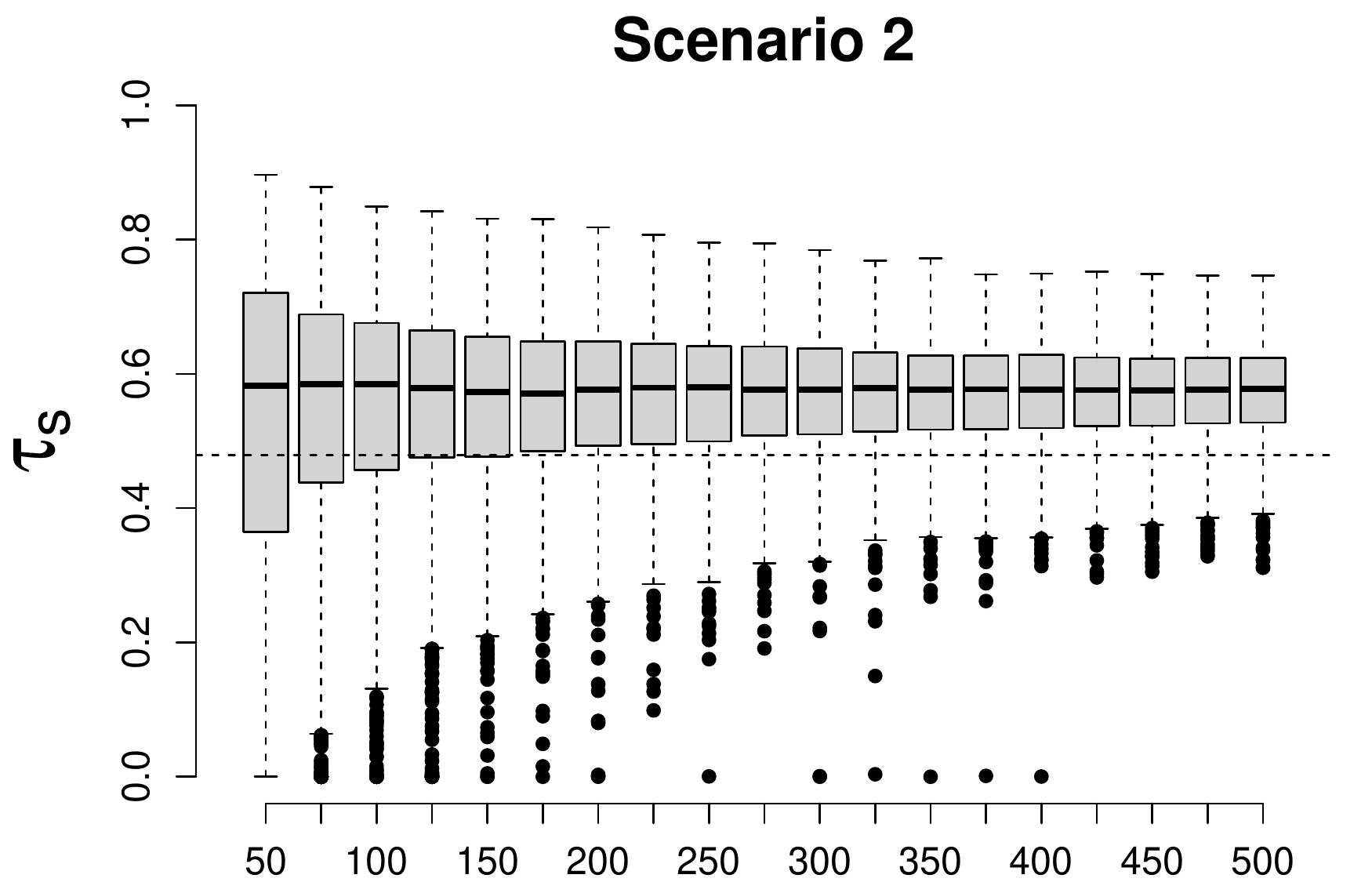}
		\includegraphics[scale=0.3]{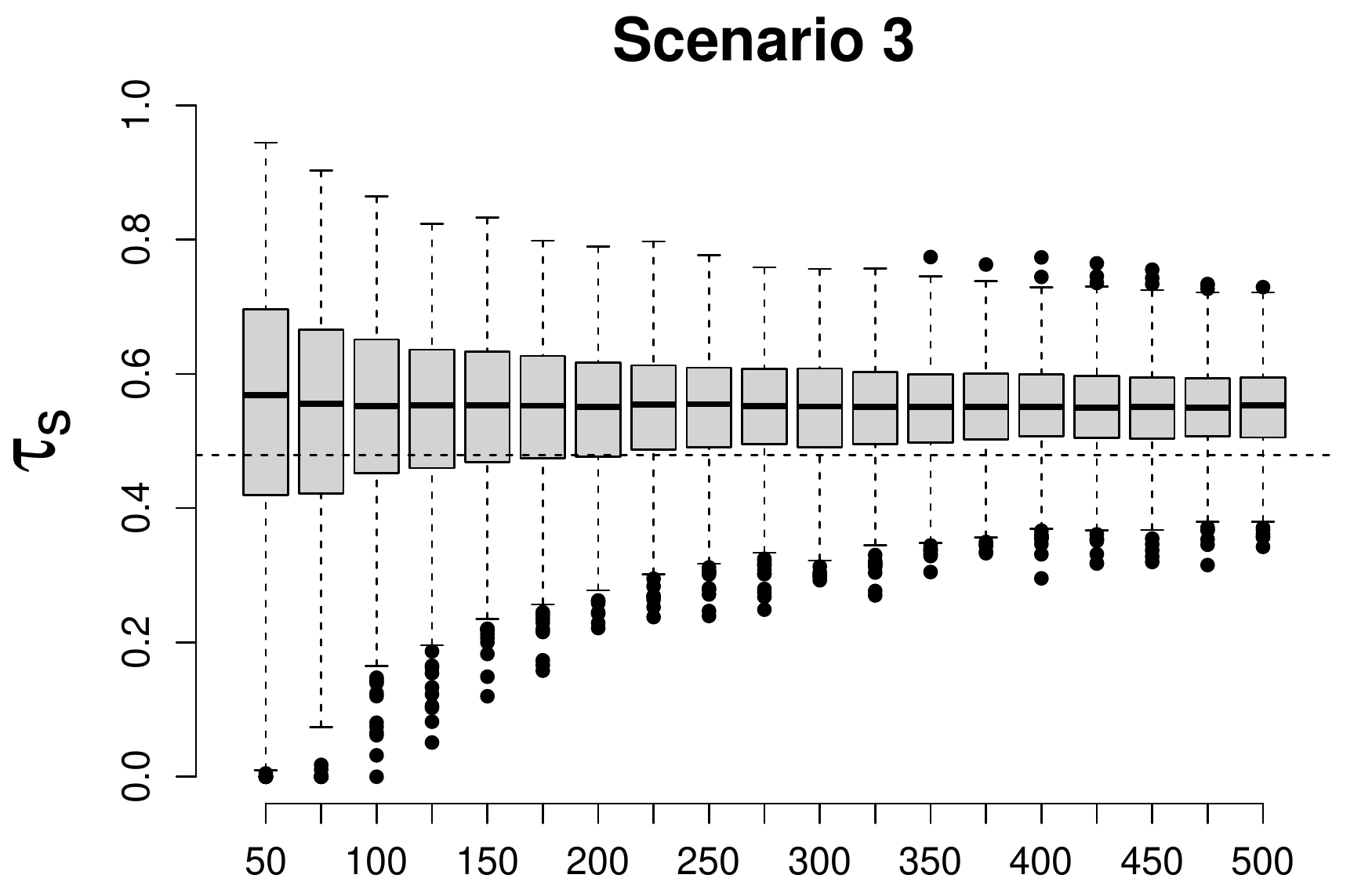}
		\includegraphics[scale=0.3]{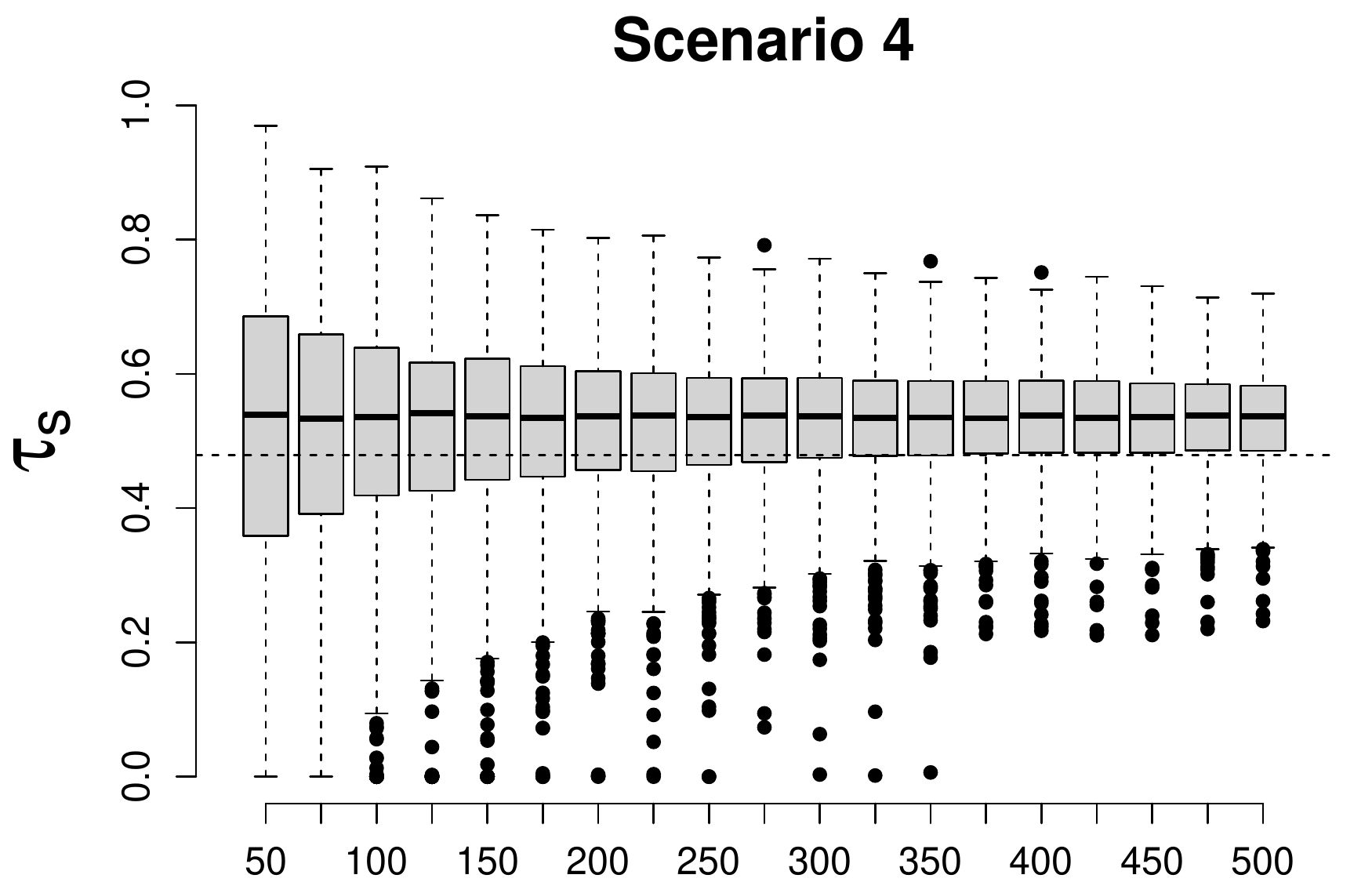}
		\includegraphics[scale=0.3]{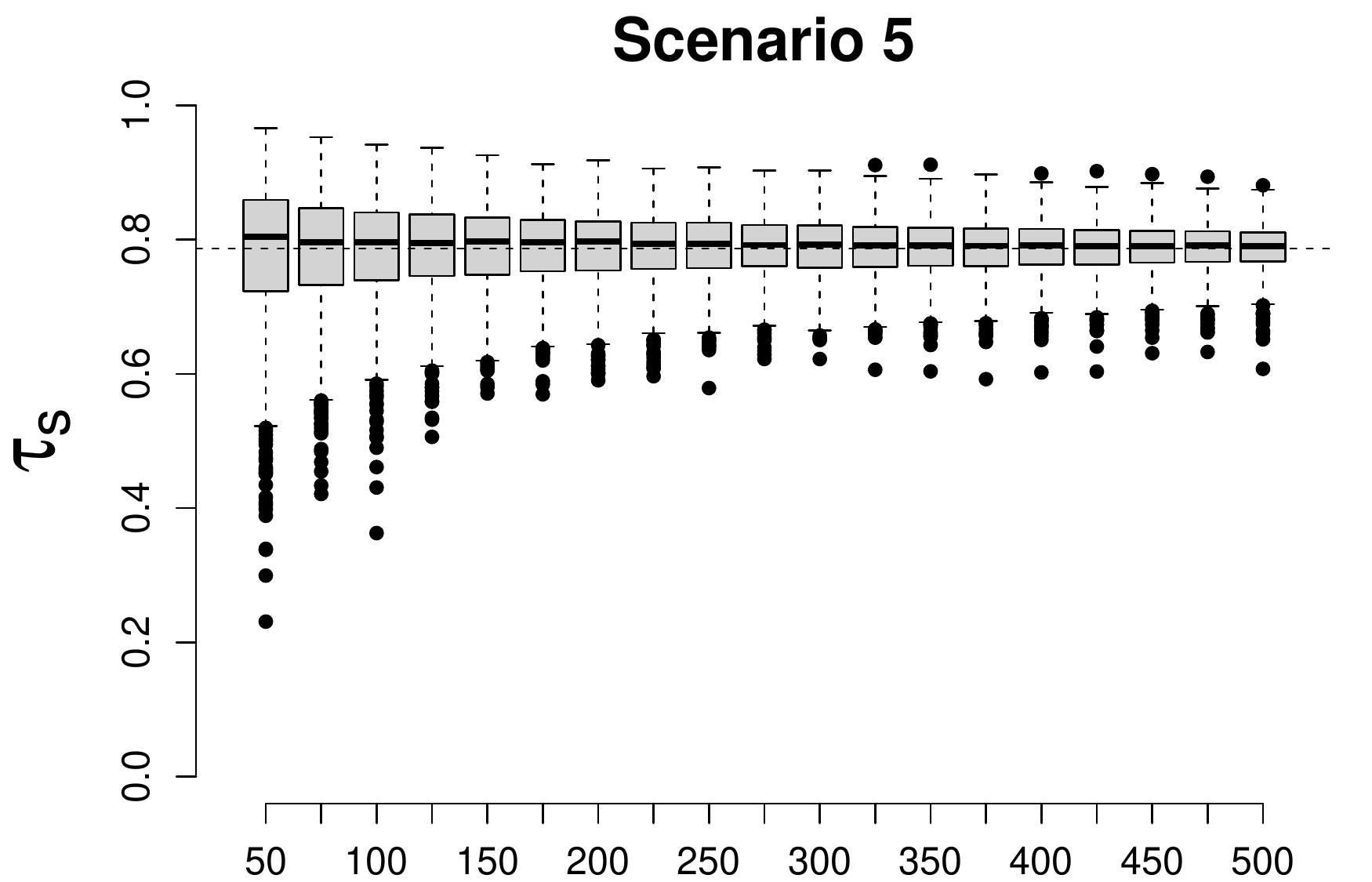}
		\includegraphics[scale=0.3]{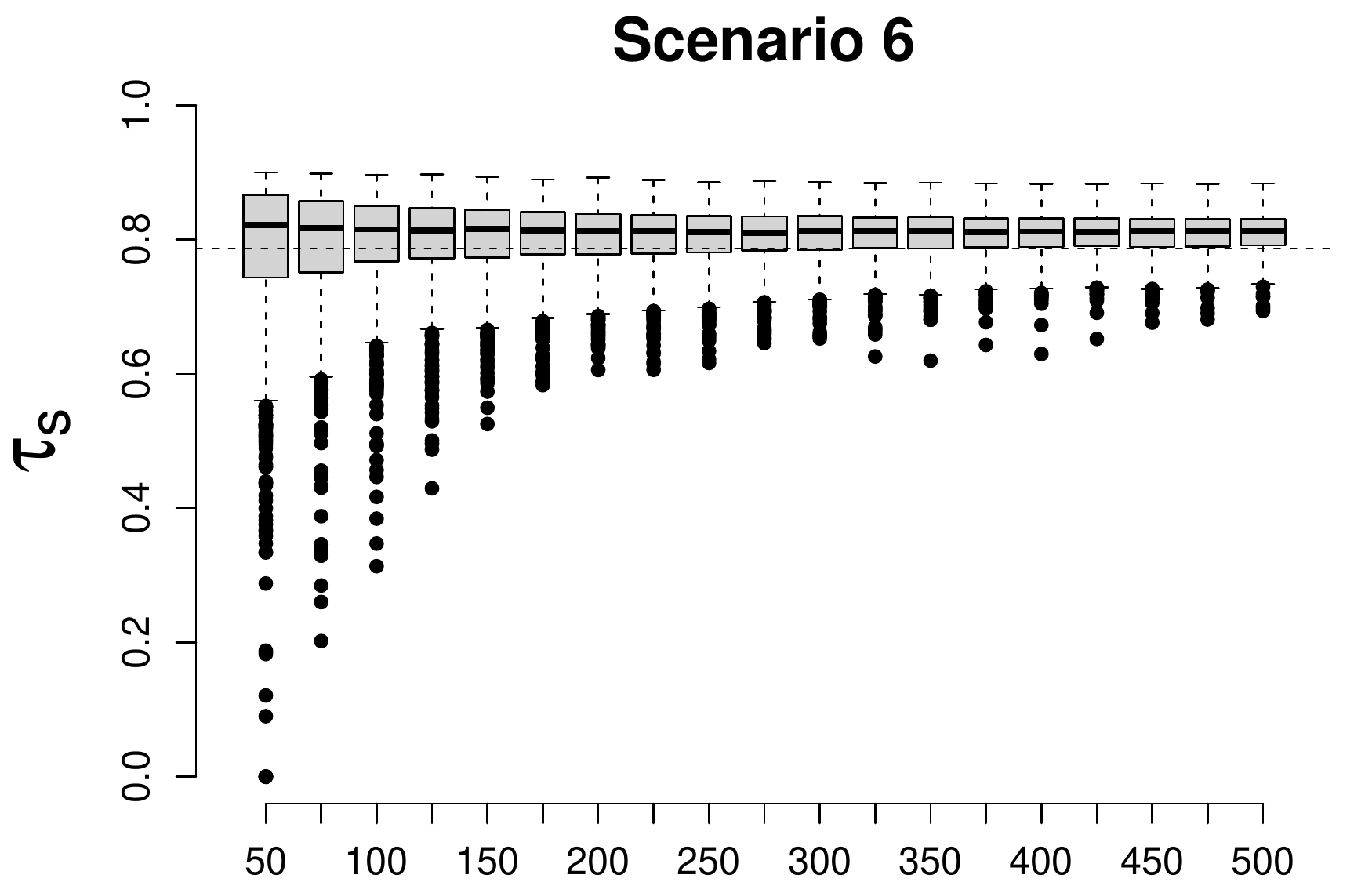}
		\includegraphics[scale=0.3]{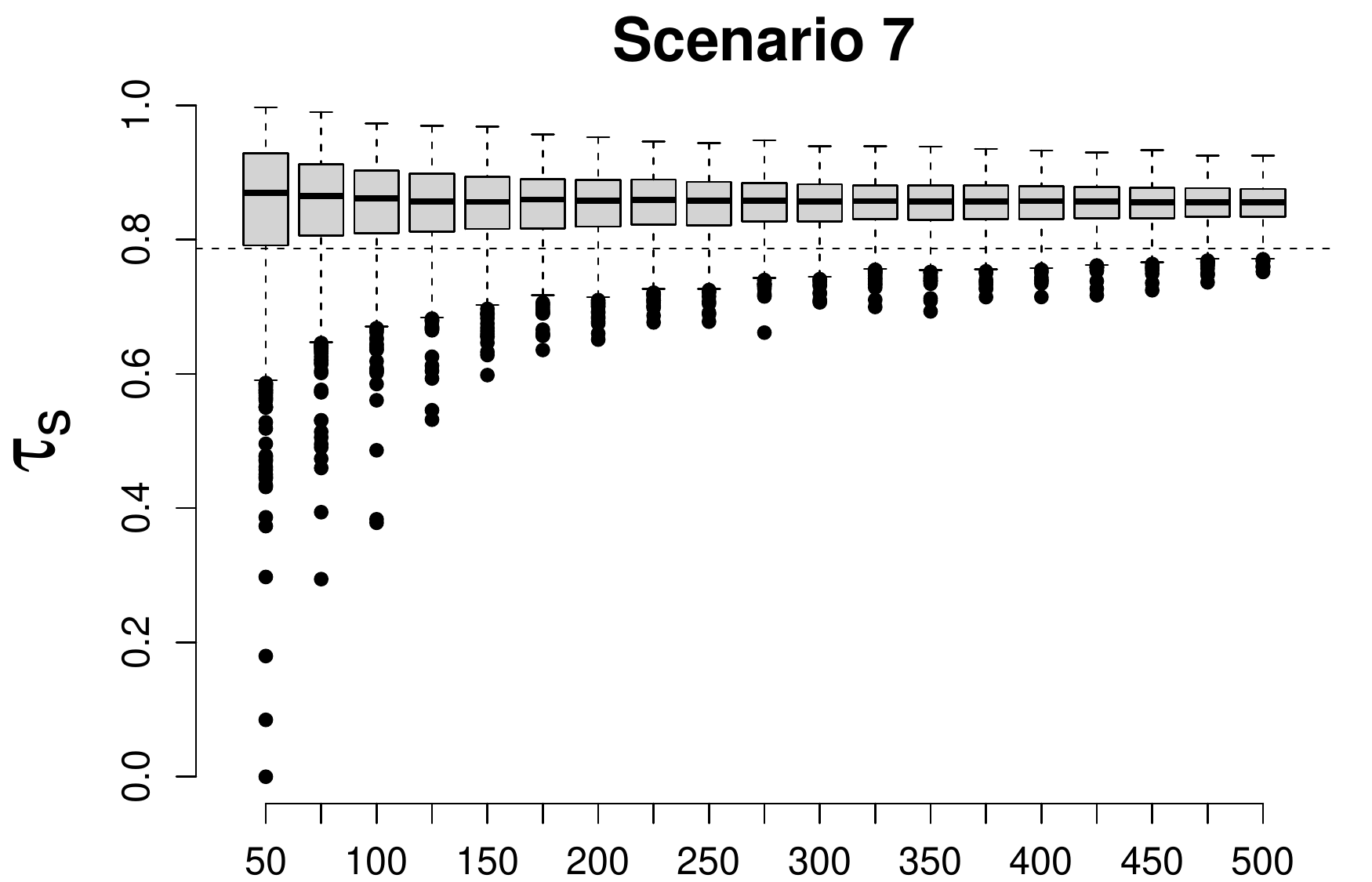}
		\includegraphics[scale=0.3]{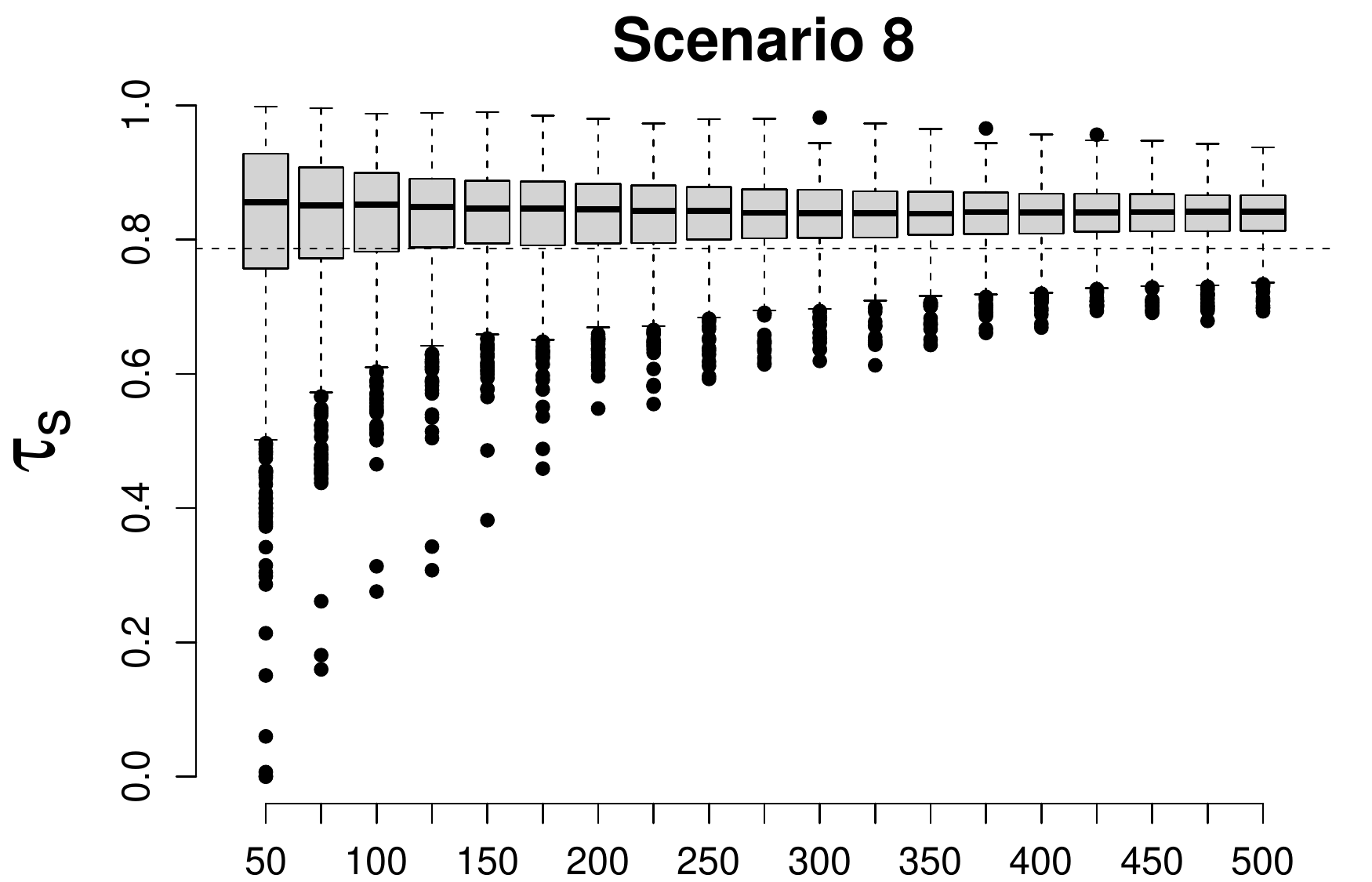}
		\includegraphics[scale=0.3]{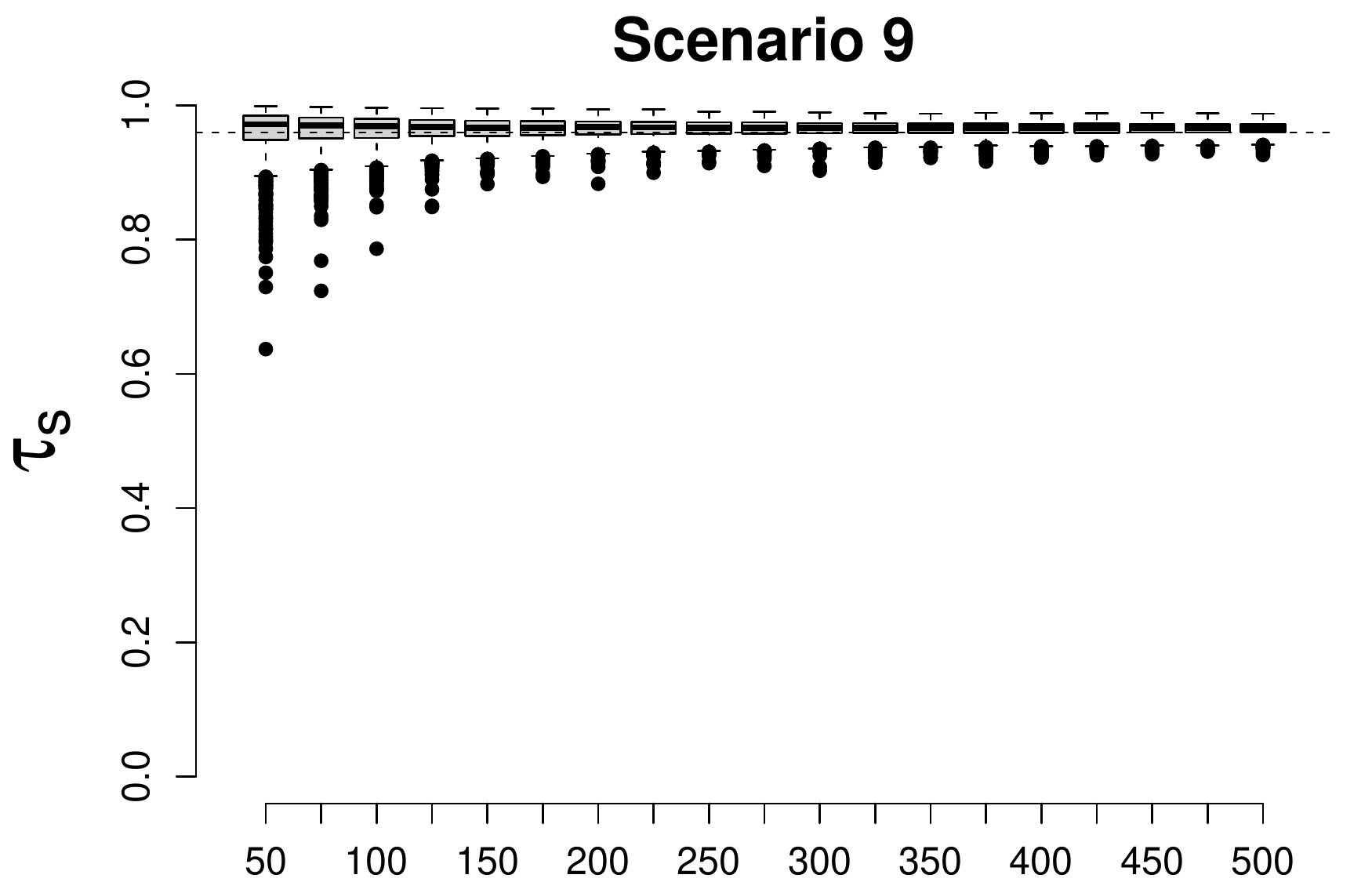}
		\includegraphics[scale=0.3]{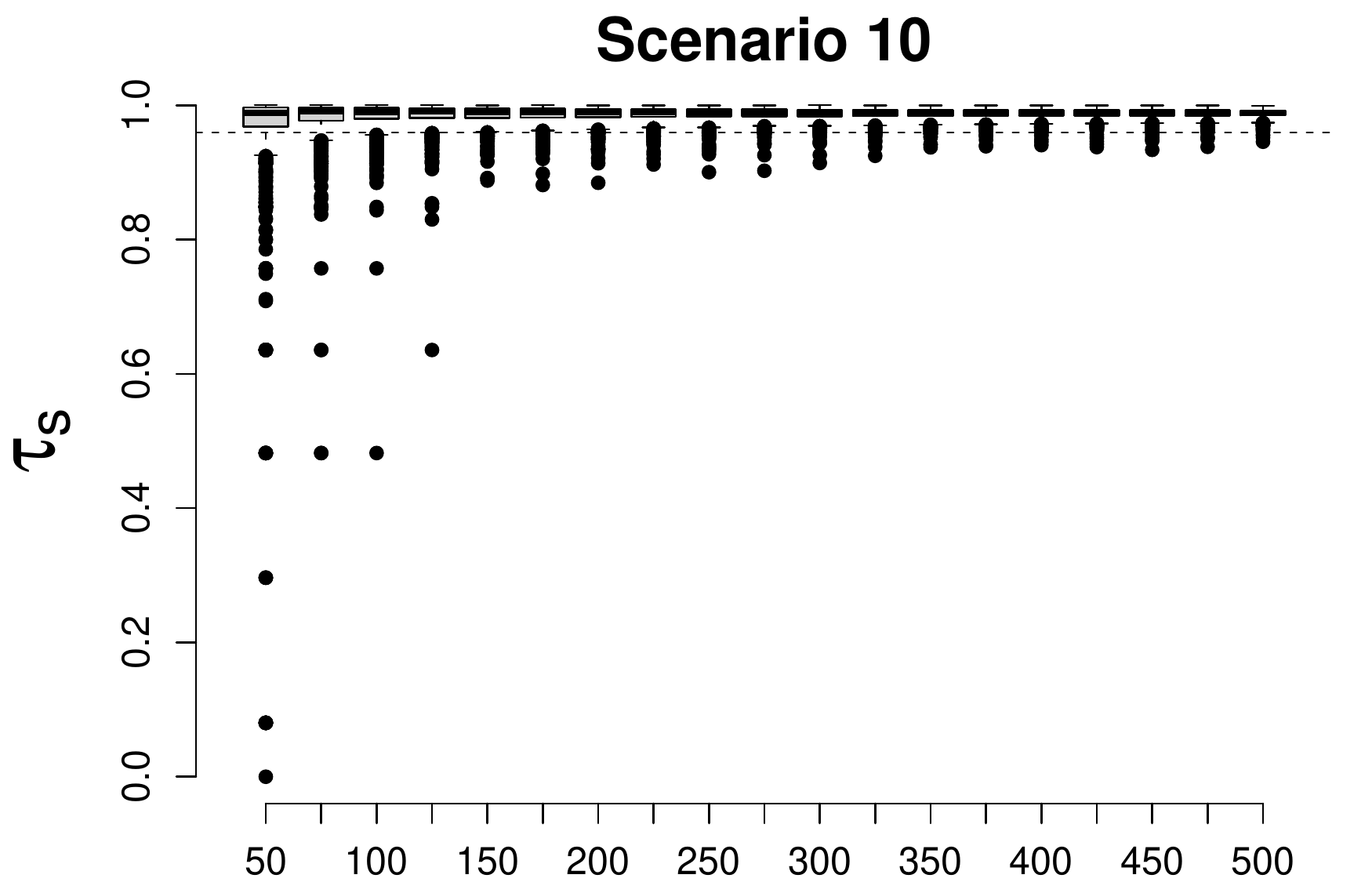}
		\includegraphics[scale=0.3]{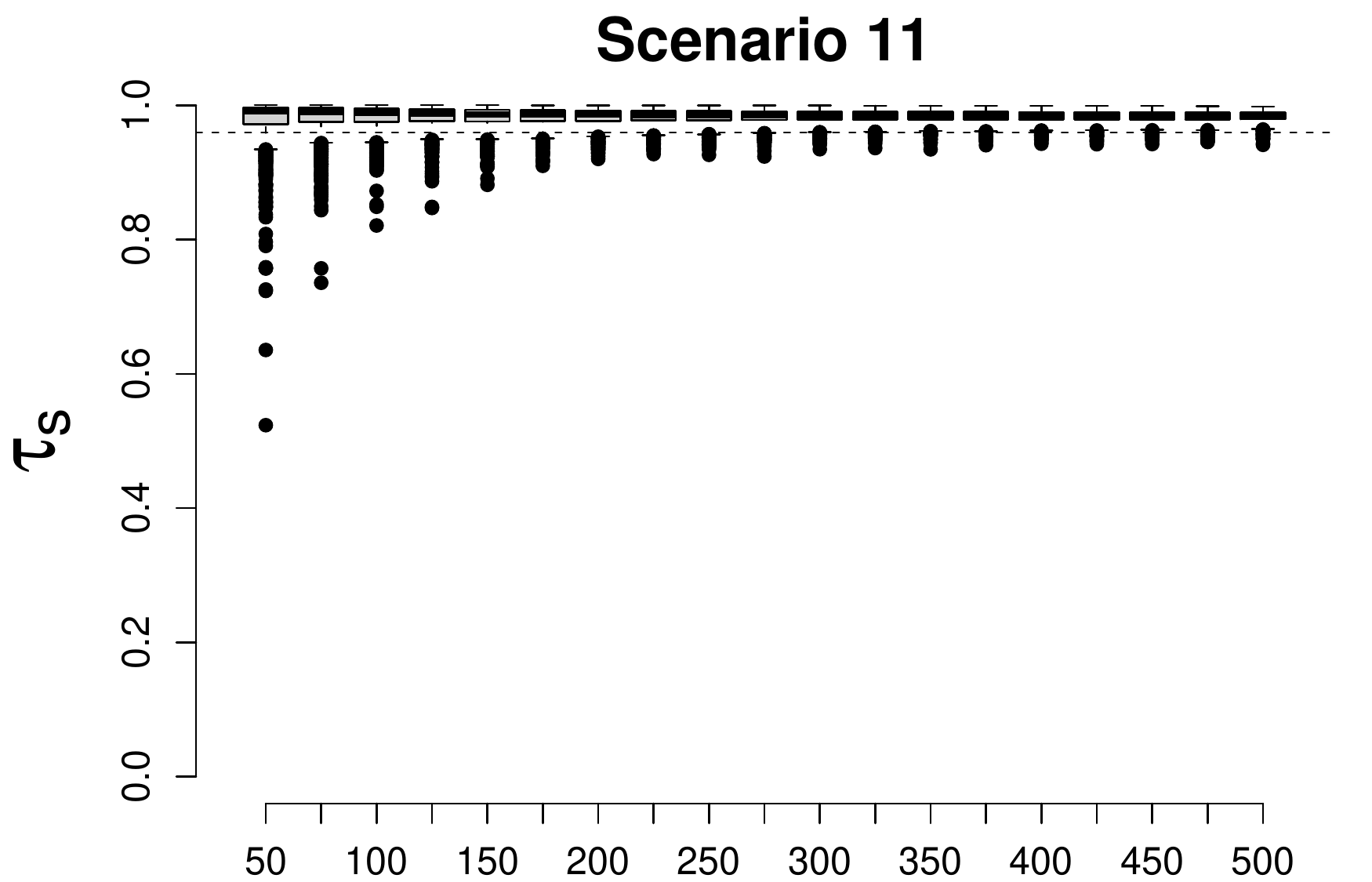}
		\includegraphics[scale=0.3]{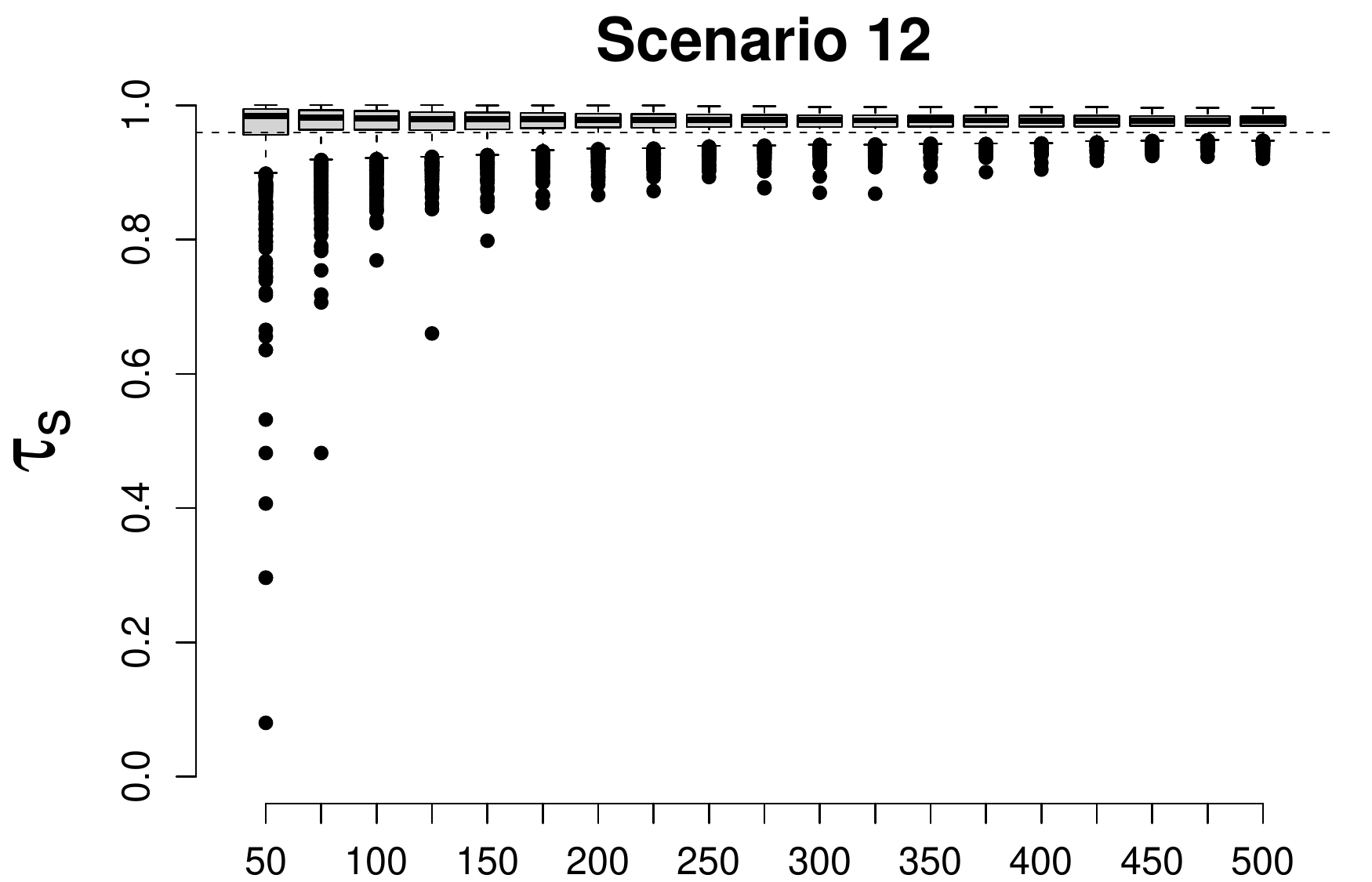}
		\caption{Box-plots of the estimates for $ \tau_s $, considering the maximum likelihood estimation method  in different scenarios and different sample sizes}
		\label{figsp}
	\end{center}
\end{figure}

Figures \ref{figkendal} and \ref{figsp} illustrate the estimates for the Kendall and Spearman correlation coefficients. Despite the difficulty to get an analytical expression for the Spearman correlation, it was obtained using numerical methods (see Section \ref{copulas}). Due to bias of the parameter $ \phi $ (see Figure \ref{figrho}) the Kendall and Spearman correlation measures were quite a bit higher than the expected nominal values for these coefficients. These differences are identified mainly in the scenarios with higher cure rates and with high correlation between $ T_1 $ and $ T_2 $. In general, there is a great variability in the measurements obtained by Kendall and Spearman methods, despite the obtained results being close to the nominal values. However, this does not apply in situations where high Spearman correlation values between $ T_1 $ and $ T_2 $ are observed; in this case, almost all measurements are close to 1.

The confidence intervals may include or not the nominal values of the correspondent parameters. It was defined that the observed coverage probability is the number of times where the nominal value is inside to the corresponding confidence interval. This event can be modeled by a binomial distribution $ Binomial (n,p)  $, where $ n $ is number of simulated samples and $ p $ is the considered nominal coverage probability. In this paper we used $ n=1000 $ and $ p=0.95 $ for each sample size used in the ML estimation, thus rejecting the equality between the nominal expected coverage probability  and the observed coverage probability assuming a significance level of 5\%,  if the  observed coverage probability is outside the range interval $ (0.9365, 0.9635)$.

\begin{figure}[H]
	\begin{center}
		\includegraphics[scale=0.48]{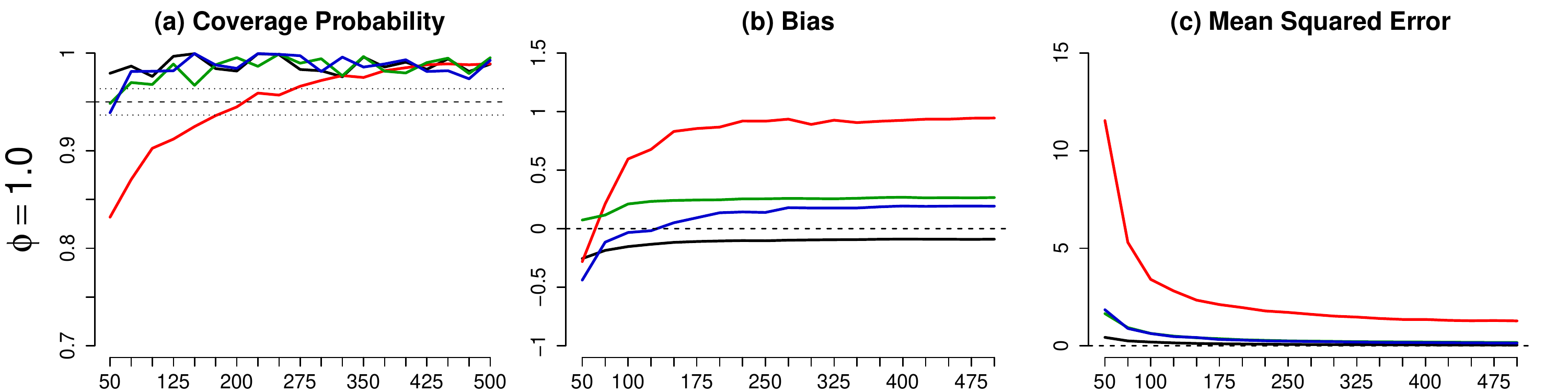}
		\includegraphics[scale=0.48]{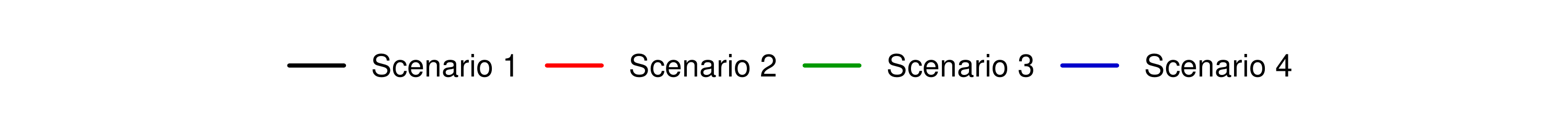}
		\includegraphics[scale=0.48]{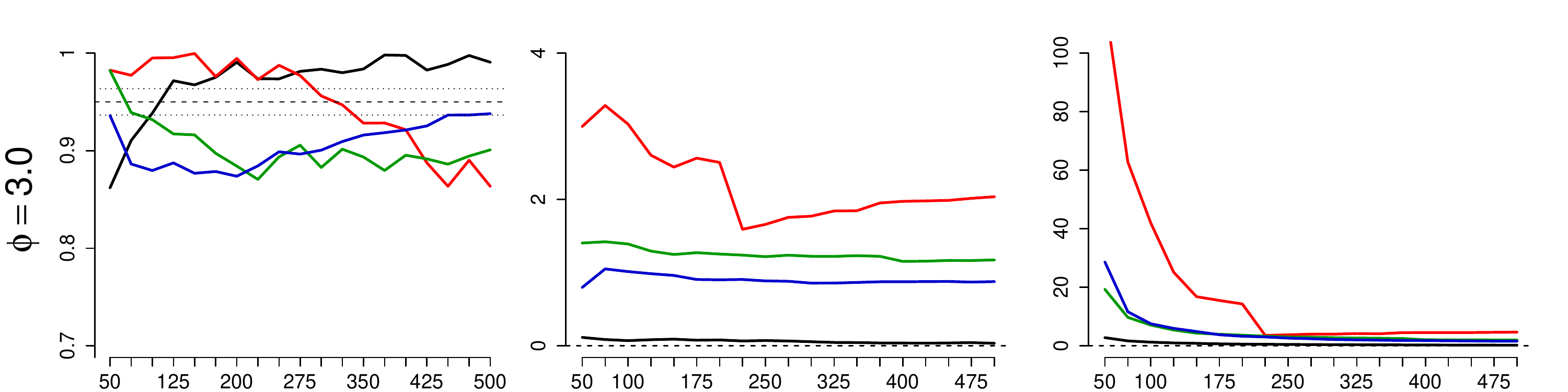}
		\includegraphics[scale=0.48]{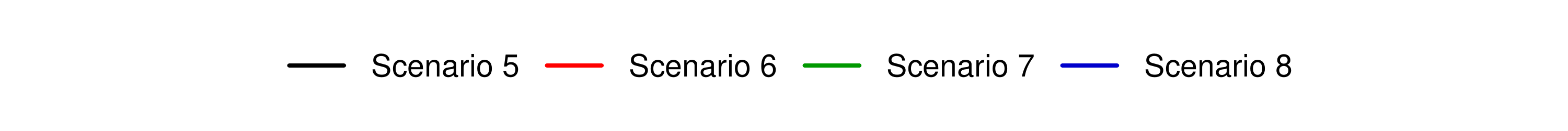}
		\includegraphics[scale=0.48]{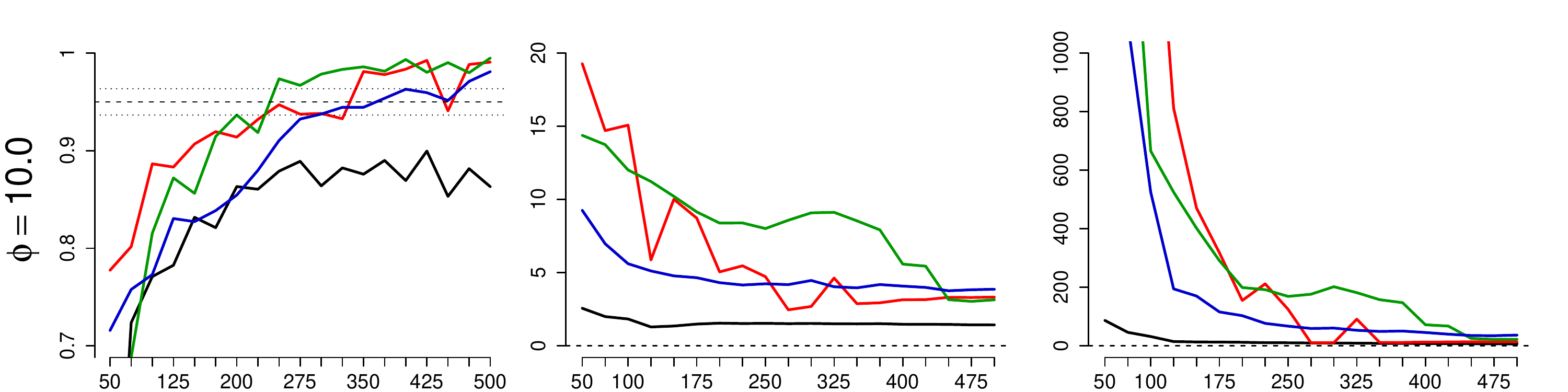}
		\includegraphics[scale=0.48]{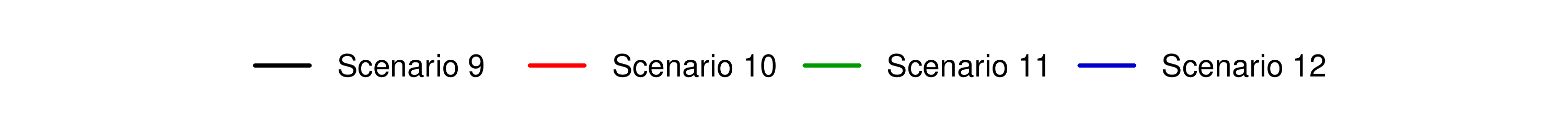}
		\caption{Plots of the coverage probability, biases and MSE for $ \phi $, considering the maximum likelihood estimation method.}
		\label{figPhi}
	\end{center}
\end{figure}

Figure \ref{figPhi} describes the coverage probability, bias and mean squared error for the parameter $\phi$, in each considered scenario. Observing the graphs for $\phi=1.0$ (low correlation between $T_1$ and $T_2$), the coverage probability is close to 95\%. In the other scenarios the coverage probability in general it is greater than 95\%. Besides that, it is possible to observe small biases, except in scenario 2, that considers a higher cure rate and produced a relatively high bias. The same does not apply to the cases $\phi=3.0$ and $\phi=10.0$. The scenarios 2 and 9 do not produce 95\% coverage probability, and there are still large biases. As an important result, we can observe that when $\phi=10.0$ the coverage probability is satisfactory for sample sizes larger than  300. Also it is observed that there are scenarios with high coverage probability when $\phi=10.0$, however, this is due to the high estimated standard error for the parameter $ \phi $. In this case, there is the presence of high bias in the estimated value for the parameter $ \phi $, so the estimated range is not closed to the nominal value. This happens especially in scenarios  with the presence of high cure rate in at least one of the time-to-event variables.

\begin{figure}[H]
	\begin{center}
		\includegraphics[scale=0.48]{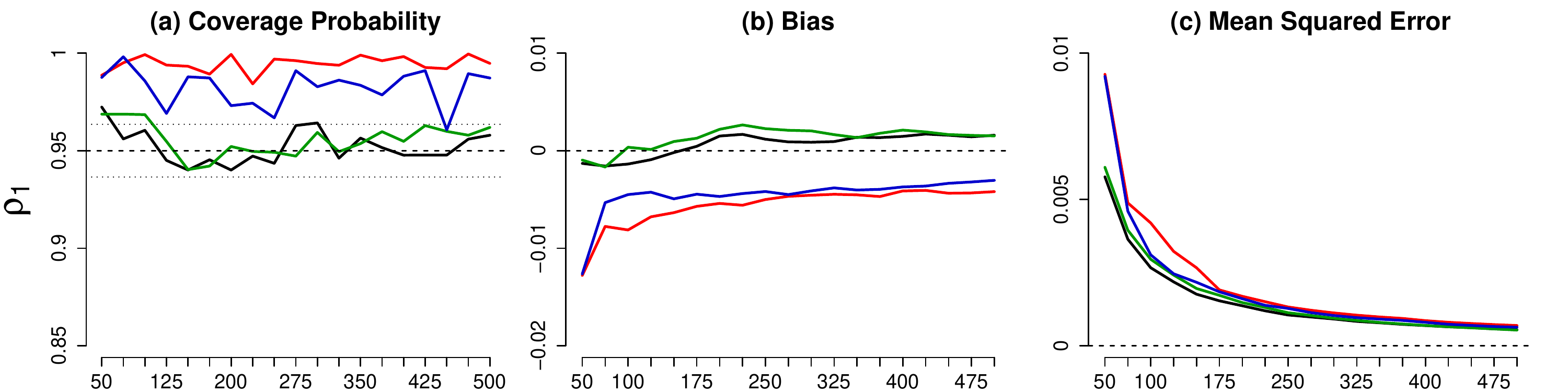}
		\includegraphics[scale=0.48]{legend-1-4.pdf}
		\includegraphics[scale=0.48]{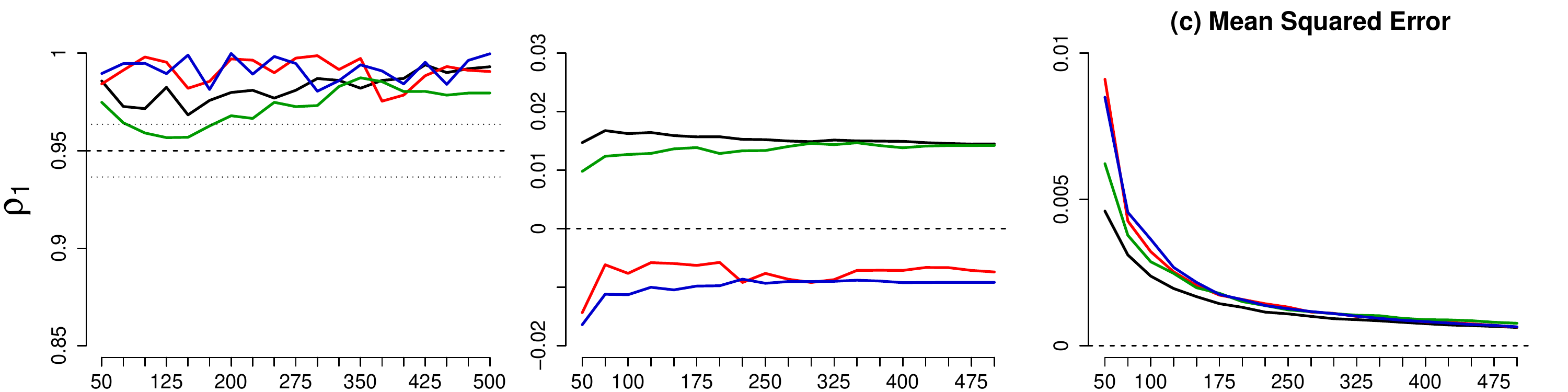}
		\includegraphics[scale=0.48]{legend-5-8.pdf}
		\includegraphics[scale=0.48]{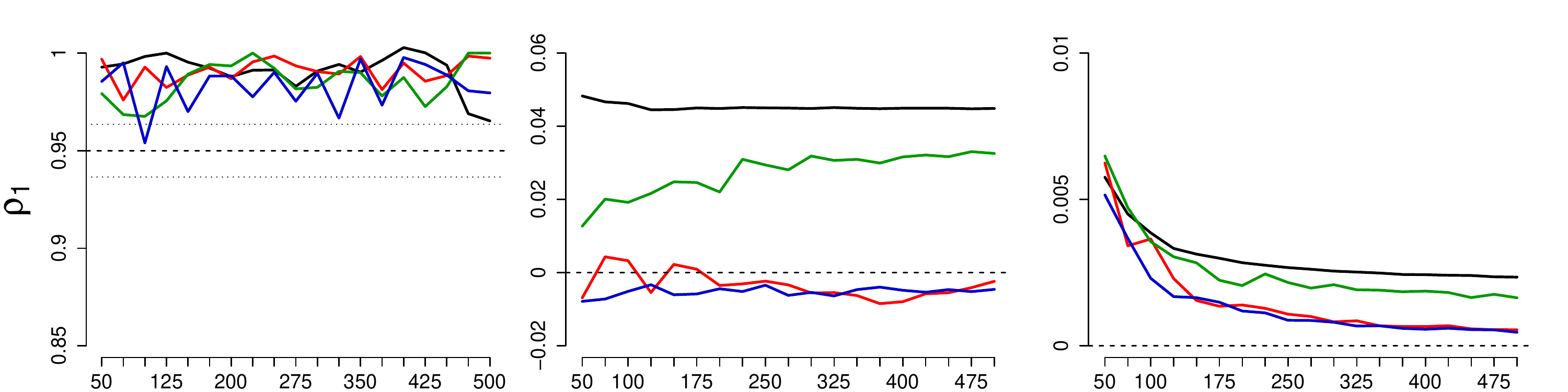}
		\includegraphics[scale=0.48]{legend-9-12.pdf}
		\caption{Plots of the coverage probability, biases and MSE for $ \rho_1 $, considering the maximum likelihood estimation method.}
		\label{figrho1}
	\end{center}
\end{figure}

The standard errors for the estimates of the parameters $ \rho_1 $ and $ \rho_2 $ were calculated using the delta method, since these parameters are obtained as functions of the parameters $ \alpha_1 $, $ \alpha_2 $, $ \beta_1 $ and $ \beta_2 $. Figure \ref{figrho1} shows the coverage probability, bias and mean squared error for the cure rate parameter $\rho_1$, in all considered scenarios. 
The coverage probability is high in all scenarios, given that the estimates for $ \rho_1 $ have low bias and a relatively high standard error. In these situations, probabilities are close to 100\%.In addition, these results reinforce the conclusions previously obtained from Figure \ref{figeta1}, where it is possible to conclude that the ML method adequately estimates the cure rate values.

\begin{figure}[H]
	\begin{center}
		\includegraphics[scale=0.48]{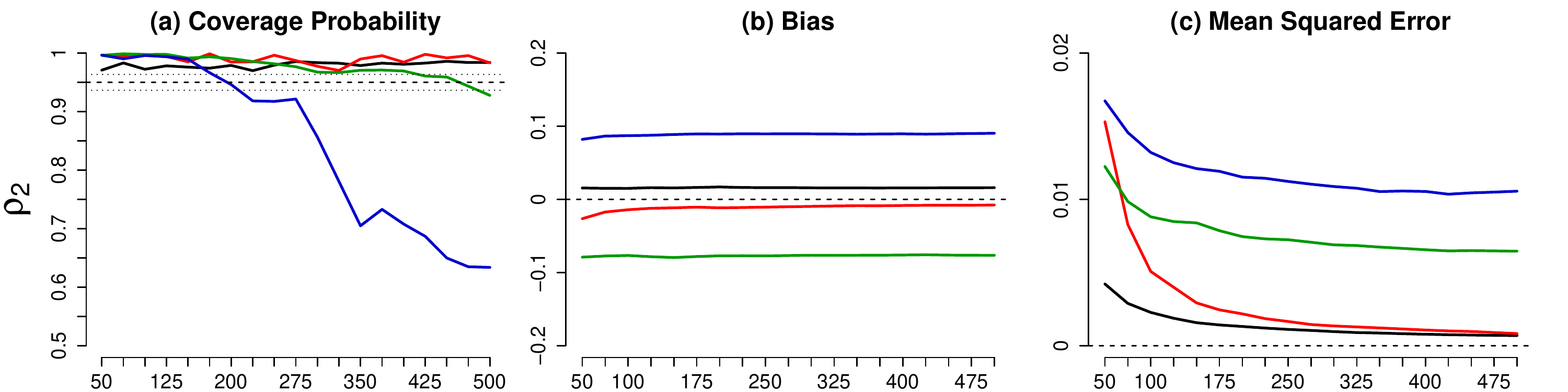}
		\includegraphics[scale=0.48]{legend-1-4.pdf}
		\includegraphics[scale=0.48]{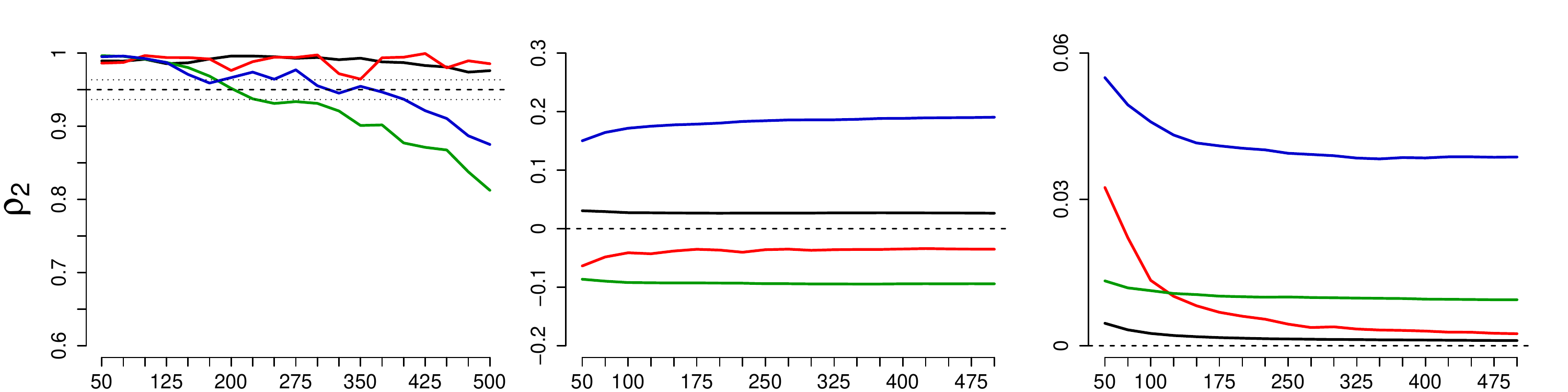}
		\includegraphics[scale=0.48]{legend-5-8.pdf}
		\includegraphics[scale=0.48]{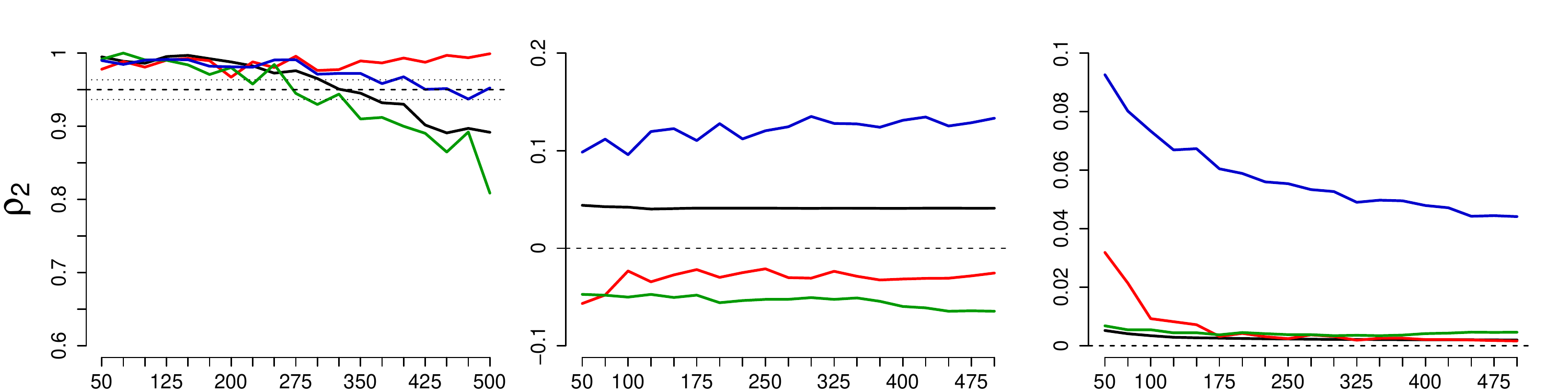}
		\includegraphics[scale=0.48]{legend-9-12.pdf}
		\caption{Plots of the coverage probability, biases and MSE for $ \rho_1 $, considering the maximum likelihood estimation method.}
		\label{figrho2}
	\end{center}
\end{figure}

Figure (\ref{figrho2}) shows the results of the simulation study  considering the parameter $ \rho_2 $. The coverage probability is satisfactorily close to 95\% in almost all scenarios. The same is not observed in the scenarios with different cure rates, in particular when the parameter $\rho_1$ is greater than parameter $\rho_2$, where we observed that the bias of the estimate for the parameter $\rho_2$ does not tend to 0, and the MSE is relatively high. This probably occurred due to the correlation considered in the simulation process of samples are pushing  $\rho_2$  nearest to $\rho_1$. In all other scenarios, the bias of parameter $\rho_2$   is closest  to 0. 

It was noted during the simulation process, the presence of simulated samples that resulted in monotone likelihood functions, mainly when  we considered samples sizes less than 200 and with high value for $ \phi $. 

The coverage probability, biases and MSE for the estimators of the parameters $ \alpha_1 \; \alpha_2, \;  \beta_1$ and $ \ \beta_2 $ were also evaluated. Considering the scenarios with $ \phi = 1.0 $ and $ \phi = 3.0 $, the coverage probability, biases and MSE behave as expected, except for the estimator of the parameter $ \beta_2 $ that exhibit high bias and unexpected coverage probability. The parameters have different behaviors, reacting in different ways for each combination of parameter values, but it is noted that in the scenarios with low cure rate the bias are closer to zero.  In addition, for all parameters, the bias and MSE decrease as the sample size increases, as it is expected.

It is observed in general a high bias related to  the estimated parameters, and this bias is large enough to impair the probabilities of coverage of the correspondent confidence intervals. However, the range of bias were low compared to the parameter estimates. Probably, the previously mentioned problems associated with the parameter estimation are consequences of the method used to generate samples assuming a dependence structure between  $T_1$ and $T_2$. In addition, the fit of BDGD bivariate model was verified for some samples  by comparison of the estimated survival function with the Kaplan-Meier estimator. From these plots it was possible to see that the  estimated survival curves by the BDGD model were satisfactorily closed to Kaplan-Meier curve, for both lifetimes $  T_1 $ and $ T_2 $.

In a brief additional simulation study, it was considered a reparametrization for $\alpha_1,\, \alpha_2, \beta_1 $ and $ \beta_2 $, where: $ \gamma_k=\exp(\alpha_k) $ and $ \lambda_k=\exp(\beta_k) $; also $ \eta_k=\frac{1}{\alpha_k} $ and $ \theta_k=\frac{1}{\beta_k} $, $ k $=1,2. However, no significant changes were observed comparing the obtained inference results with the previously inference results  presented in this section.

\section{Applications to Real Data Sets}
In order to illustrate the proposed model, we present in this section, four applications with real
data sets. In each application,  the Kendall correlation $ \tau_k $ and the Spearman correlation $ \tau_s $, were compared with the empirical correlation between $T_1$ and $T_2$, denoted by $ \tau_e $, obtained by the package \textit{SurvCorr} \citep{survcor}. Also, it was compared the hazard function estimates by the proposed model with the empirical hazard function (obtained using the package ``bshazard" \citep{haz}). 

The Bayesian estimates were based on 2,000 simulated Gibbs samples for the joint posterior distribution of interest recorded by every 50th iteration from 1,000,000 Gibbs samples after a ``burn-in'' period of 50,000 samples deleted to eliminate the effect of the initial values assumed in the simulation procedure. The convergence of the MCMC samples was checked by visual examination of traceplots of the simulated samples and convergence and stationary tests using the  package \textit{coda} \cite{coda}. Approximately  non-informative uniform prior distributions were assumed for the parameters of the BDGD model in almost all applications.

\subsection{Application to a breast cancer data set}
In this first application,  it was considered the data analysis of a data set related to a cohort study, where 97 patients underwent surgical treatment for breast cancer followed up for a period between the year 2000 to 2011. More details about this data set can be found in \cite{shigemizu2017prediction}. For the bivariate lifetime application it was considered  as $ T_1 $ the  disease-free survival time (DFS) and $T_2$ representing the overall survival time (OS). In the dataset, there is 75\% censored data for the disease-free survival time ($ T_1 $) and 80\% censored data for  the overall survival time ($ T_2 $). 

\begin{figure}[H]
	\begin{center}
		\includegraphics[scale=0.6]{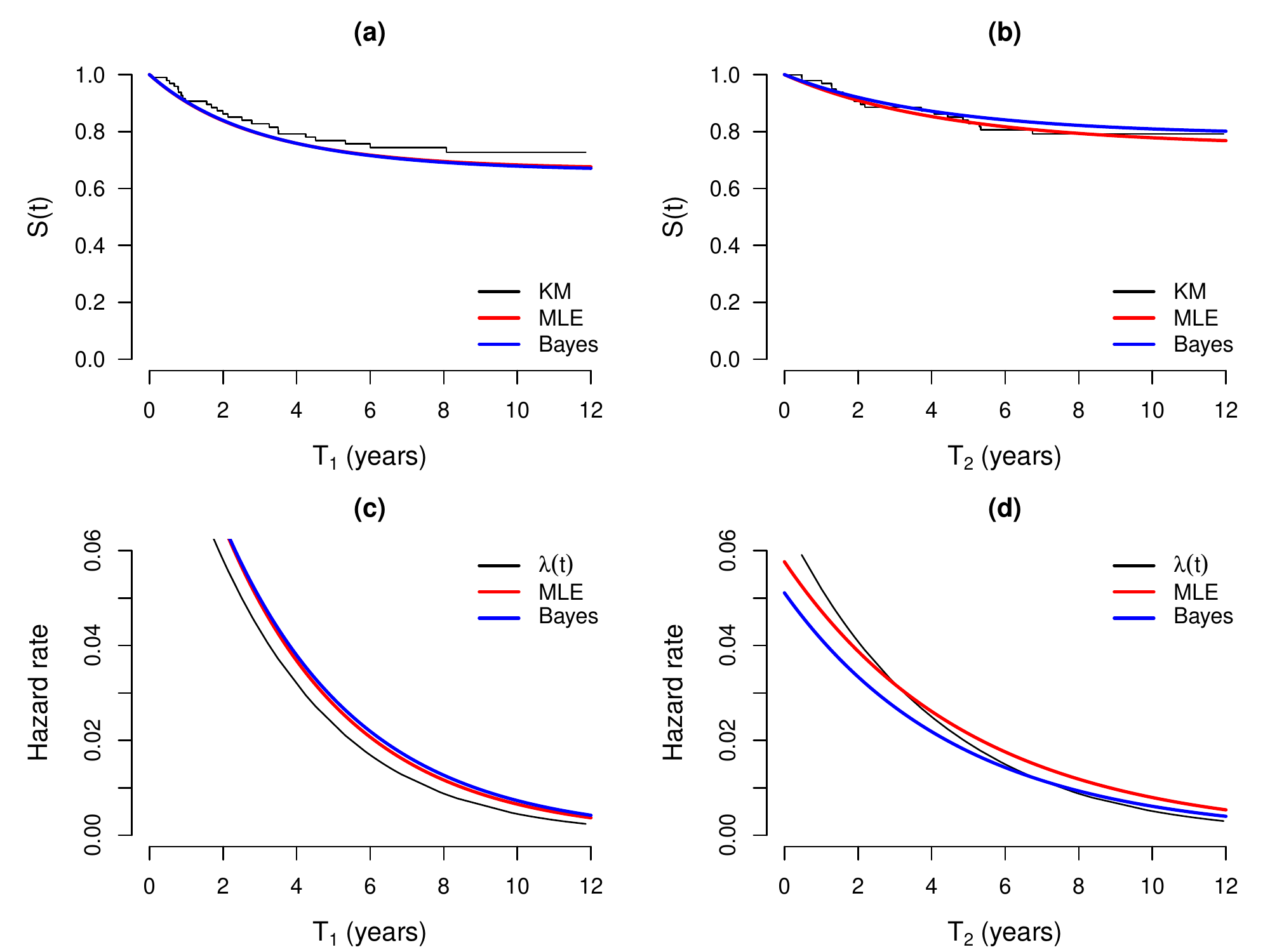}
		\caption{Plots of the survival functions estimated by Kaplan-Meier method and from the BDGD (upper panels) and respective hazard functions (lower panels) for DFS time (panels (a) and (c)) and OS time  (panels (b) e (d)), considering breast cancer data.}
		\label{fig:breast}
	\end{center}
\end{figure}

Table \ref{breast}  shows the ML estimates and the Bayesian estimates for the parameters of the BDGD model considering the breast cancer data. Note that the ML and Bayesian estimates are very close to each other. The estimated values for  $ \tau_s $ obtained by copula functions are greater than the empirical correlation obtained by the R package \textit{Survcorr} ($ \tau_e=0.8702\, (0.5288, 0.9691) $), but we can note that the estimated value of  $ \tau_s $  is contained in the 95\%  confidence interval of $ \tau_e $, as shown in the simulation results. The obtained Bayesian estimates for the parameters $ \alpha_2 $ and $ \beta_2 $ are very close to the ML estimates, although there is a difference for the parameter $ \rho_2 $. This difference can be seen in Figure \ref{fig:breast} showing the estimated survival (upper panels) and hazard (lower panels) curves for BDGD model proposed in this study. On the panel (b) of Figure \ref{fig:breast},  the survival curve estimated by a Bayesian approach diverges slightly from the Kaplan-Meier estimator for the survival function. In additional, on panel (d) of Figure \ref{fig:breast}, the hazard function estimated from ML approach is the closest to the empirical curve estimated by bshazard.
For $ T_1 $ both estimation methods resulted in similar curves. From the Kaplan-Meier plot for the survival function, it can be observed high cure rates in both lifetimes, where in $ T_1 $ there is a plateau close to the value 0.70, and in $ T_2 $ close to the value 0.75. These values are close to those estimated by the BDGD model. For the hazard curve, the model has a satisfactory fit to capture the decreasing shape of the empirical hazard function. 

\begin{table}[H]
	\centering
	\caption{Maximum likelihood estimates for the parameters of the BDGD model for the breast cancer data.}
	\label{breast}
	\begin{tabular}{@{}ccccccc@{}}
		\toprule
		\multirow{2}{*}{Parameters} & \multicolumn{3}{c}{Maximum Likelihood Estimators} &  & \multicolumn{2}{c}{Bayesian Estimators} \\ \cmidrule(l){2-7} 
		& Estimate & \begin{tabular}[c]{@{}c@{}}Standard \\ Error\end{tabular} & 95\%  CI &  & Median & 95\%  CrI \\ \midrule
		$ \alpha_1 $ & 0.1163 & 0.0305 & (0.0858, 0.1470) &  & 0.1139 & (0.0326, 0.1486) \\
		$ \alpha_2 $ & 0.0576 & 0.0190 & (0.0386, 0.0767) &  & 0.0511 & (0.0308, 0.0779) \\
		$ \beta_1$ & 0.2877 & 0.0853 & (0.2025, 0.3731) &  & 0.2747 & (0.2032, 0.3910) \\
		$ \beta_2 $ & 0.1980 & 0.0895 & (0.1085, 0.2877) &  & 0.2125 & (0.1085, 0.2963) \\
		$ \rho_1 $ & 0.6674 & 0.1096 & (0.4525, 0.8823) &  & 0.6752 & (0.5143, 0.8943) \\
		$ \rho_2 $ & 0.7474 & 0.1248 & (0.5028, 0.9921) &  & 0.7865 & (0.5894, 0.8867) \\
		$ \phi$ & 8.2022 & 2.0747 & (4.1358, 12.2686) &  & 7.8601 & (4.1743, 11.7891) \\ \midrule
		$\tau_k $ & 0.8039 & 0.0575 & 0.6912, 0.9167 &  & 0.7986 & (0.6772, 0.8551) \\
		$\tau_s$ & 0.9431 & - & - &  & 0.9401 & (0.8555, 0.9681) \\  \bottomrule
	\end{tabular}
	\caption*{\small 95\%CI: 95\% confidence interval; 95\%CrI: 95\% credible interval.}
\end{table}

\subsection{Application to a diabetic retinopathy data set}
The diabetic retinopathy data used in the second application was introduced by \cite{diabetic1976preliminary}. In this study, 197 diabetic patients patients up to 60 years old were followed-up for a fixed period. Each patient had one eye randomized for laser treatment and the other eye receiving no treatment.  
For a bivariate analysis  $ T_1 $ is the time up to visual loss for the control eye, while $ T_2 $ corresponds to the time up to visual loss for the treatment eye. There was in this study 43\% censored data of not treated eyes and 73\% censored data of treated eyes.

\begin{figure}[H]
	\begin{center}
		\includegraphics[scale=0.6]{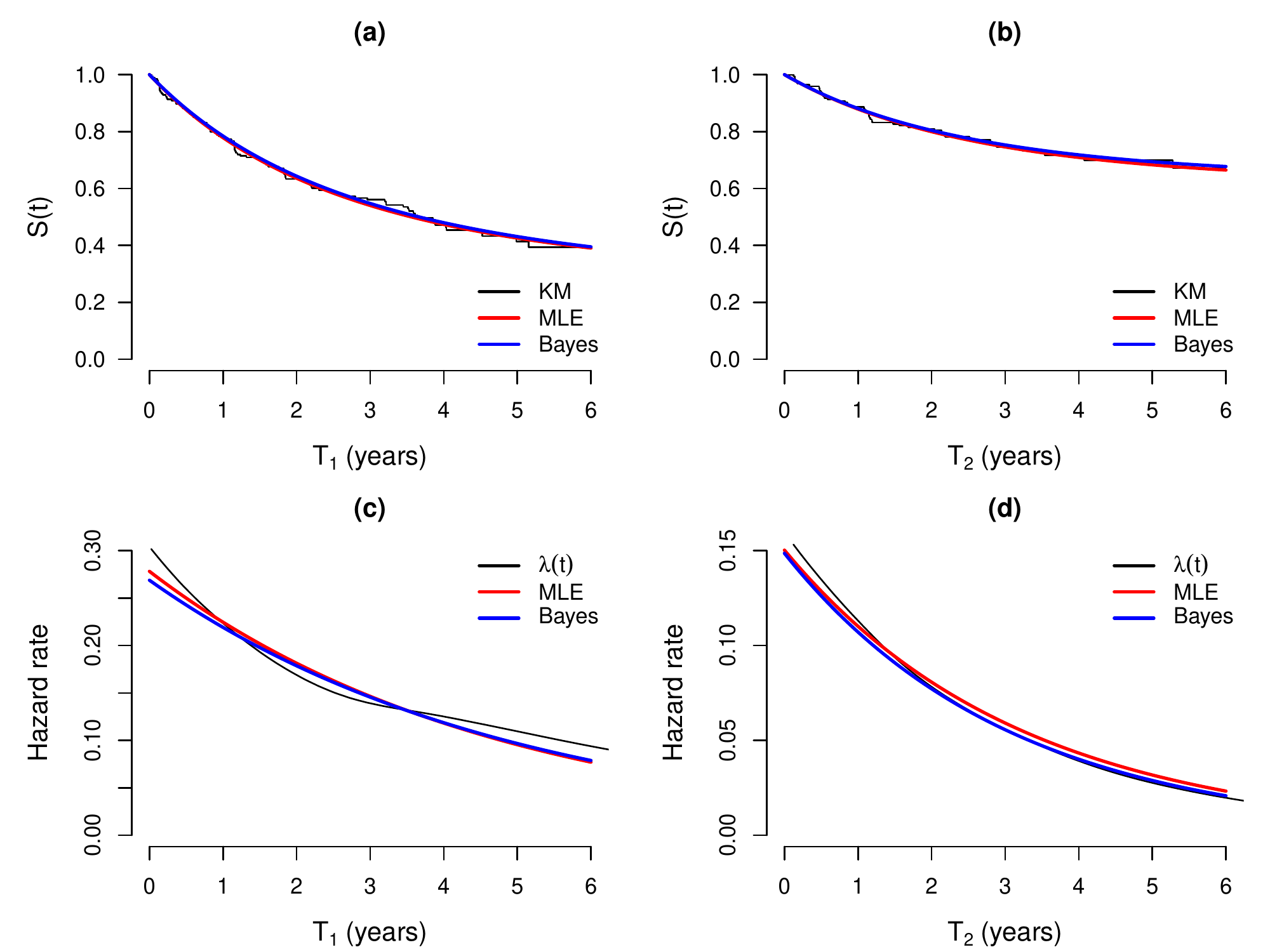}
		\caption{Plots of the survival functions estimated by Kaplan-Meier method and from the BDGD (upper panels) and respective hazard functions (lower panels) for control eye (panels (a) and (c)) and treatment eye  (panels (b) and (d)), considering the diabetic retinopathy data.}
		\label{fig:retino}
	\end{center}
\end{figure}

Table \ref{tab:retino}  shows the ML estimates and Bayesian estimates for the parameters of the BDGD model considering the  diabetic retinopathy data. In this application also it is observed that the ML estimates and Bayesian estimates are very similar. Both approaches estimate the cure rate percentage almost identical.
The Bayesian estimate for $ \tau_s $ is smaller than the estimate obtained in the ML approach. However, the values of  $ \tau_s $ obtained from copula functions are significantly greater than the empirical correlation estimated by the \textit{Survcorr} ($\tau_e=0.3491 \, (0.1071, 0.5522)$), but $ \tau_s $  is contained in the 95\%  confidence interval for $ \tau_e $.  Figure \ref{fig:retino} compares the survival curves $ S(t_i) $  (upper panels) estimated from the Kaplan-Meier method and the empirical hazard function $ \lambda(t_i) $ (lower panels), $i=1,2$,  with the survival and hazard curves fitted by the BDGD model considering the retinopathy data. The ML estimates and Bayesian estimates produced similar plots. From the Kaplan-Meier estimator for the survival function, it was observed that the estimated fitted curves were very satisfactory for both  $ T_1 $ and $ T_2 $. For the hazard curve, the model have a satisfactorily fit to capture the decreasing shape from the empirical hazard function. The estimated hazard curves using copula functions do not follow the total shape of the empirical hazard function for $ T_1 $ (panel (c)); it is possible that a more flexible distribution is needed for a better fitting.
In general, it can note that the proposed BDGD model gives a reasonable fit for the retinopathy  data, with moderate and high cure rate. It is important to say, that for this application it was needed to assume more informative prior distribution for the parameters  $ \beta_i (i=1,2) $ to get better convergence for the MCMC simulation algorithm.

\begin{table}[H]
	\centering
	\caption{Maximum likelihood estimates for the parameters of the BDGD model for the  retinopathy data.}
	\label{tab:retino}
	\begin{tabular}{@{}ccccccc@{}}
		\toprule
		\multirow{2}{*}{Parameters} & \multicolumn{3}{c}{Maximum Likelihood Estimators} &  & \multicolumn{2}{c}{Bayesian Estimators} \\ \cmidrule(l){2-7} 
		& Estimate & \begin{tabular}[c]{@{}c@{}}Standard \\ Error\end{tabular} & 95\%  CI &  & Median & 95\%  CrI \\ \midrule
		$ \alpha_1 $ & 0.2781 & 0.0441 & (0.2339, 0.3223) &  & 0.2688 & (0.2032, 0.3453) \\
		$ \alpha_2 $ & 0.1502 & 0.0325 & (0.1178, 0.1828) &  & 0.1486 & (0.1022, 01973) \\
		$ \beta_1$ & 0.2239 & 0.0789 & (0.1350, 0.2929) &  & 0.2045 & (0.1050, 0.2956) \\
		$ \beta_2 $ & 0.3109 & 0.1104 & (0.2005, 0.4214) &  & 0.3277 & (0.2068, 0.4442) \\
		$ \rho_1 $ & 0.2725 & 0.1324 & (0.0129, 0.5321) &  & 0.2652 & (0.0626, 04649) \\
		$ \rho_2 $ & 0.6267 & 0.1262 & (0.3693, 0.8642) &  & 0.6342 & (0.4372, 0.7708) \\
		$ \phi$ & 0.9500 & 0.3479 & (0.6021, 1.2979) &  & 0.9038 & (0.5228, 1.2805) \\ \midrule
		$\tau_k $ & 0.3220 & 0.0379 & (0.2476 0.3965) &  & 0.3112 & (0.2072, 0.3903) \\
		$\tau_s$ & 0.4634 & - & - &  & 0.4418 & (0.3052, 0.5518) \\ \bottomrule
	\end{tabular}
	\caption*{\small 95\%CI: 95\% confidence interval; 95\%CrI: 95\% credible interval.}
\end{table}

\subsection{Application to a cervical cancer data set}
In this application, it is considered a medical data set from a published study by \cite{brenna2004prognostic} where it was also assumed the BDGD model. In this study 118 women received a standard treatment recommended to invasive cervical cancer. In a bivariate analysis $ T_1 $ is the disease-free survival (DFS), defined	as the time from the date of surgery to the first event of disease recurrence and $T_2 $ is the overall survival (OS), defined as the time from the date of surgery to the death. There is 48\% censored data in $ T_1 $ and 53\% censored data in $ T_2 $.

\begin{figure}[H]
	\begin{center}
		\includegraphics[scale=0.6]{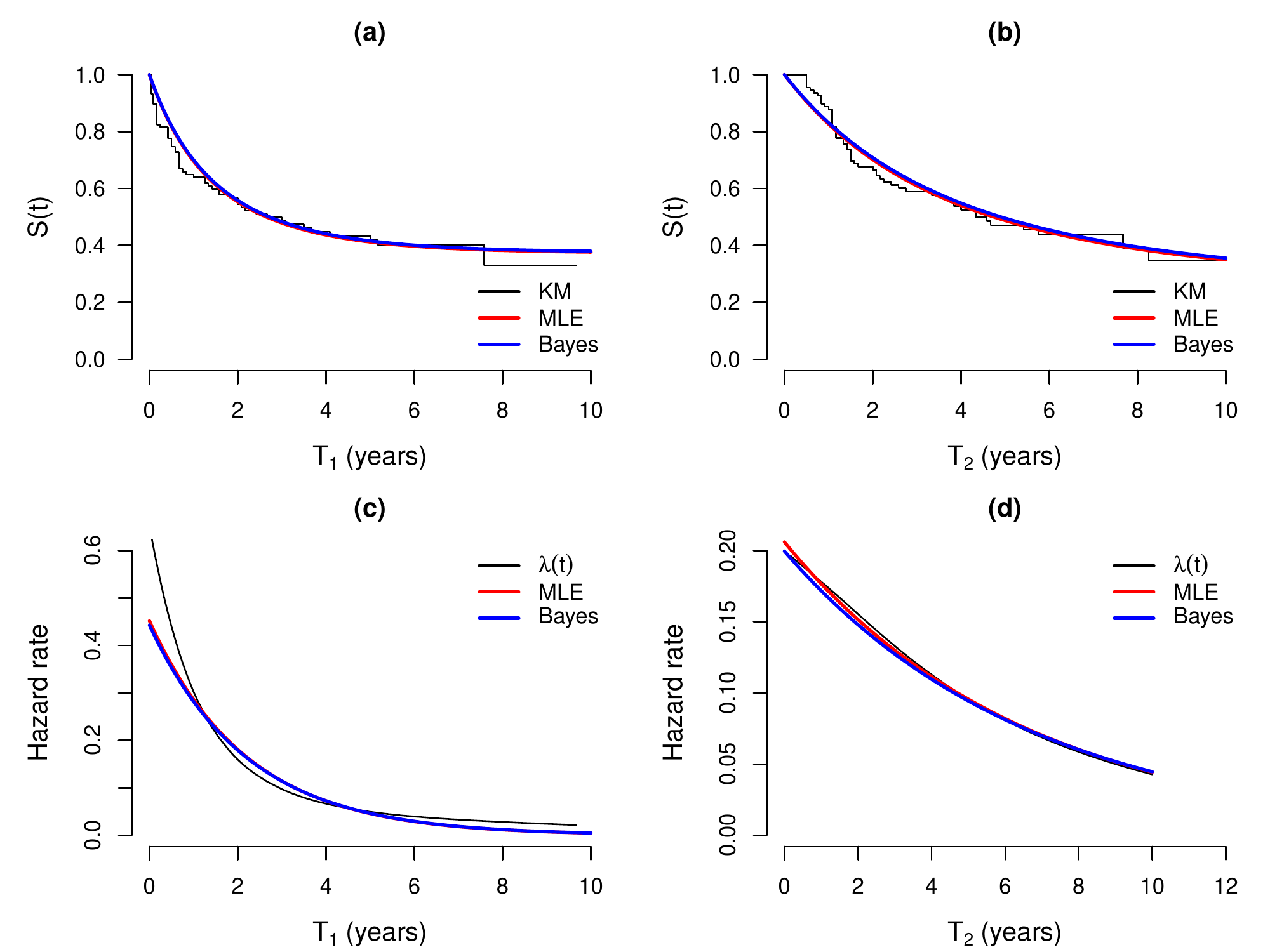}
		\caption{Plots of the survival functions estimated by Kaplan-Meier method and from the BDGD (upper panels) and respective hazard functions (lower panels) for DFS time (panels (a) and (c)) and OS time  (panels (b) and (d)), considering the cervical cancer data.}
		\label{fig:cervical}
	\end{center}
\end{figure}

Table \ref{tab:cervical} presents the ML estimates and Bayesian estimates for the parameters of the BDGD model considering the  cervical cancer data. In this application also  it is observed similar inference results assuming classical and Bayesian approaches. In this application, the values of  $ \tau_s $ obtained by copulas functions are  very close to the empirical correlation obtained by \textit{Survcorr} ($\tau_e=0.9118 \, (0.8477, 0.9498) $), and the estimate ranges for $ \tau_s $ estimated from Bayesian approach and  $ \tau_e $ are very similar. \cite{PERES2020e03961} presented similar results for the correlation between $T_1$ and $T_2$ where the obtained  value for the correlation  between  $T_1$ and $T_2$ was $ 0.8933 $. In general the ML estimates and Bayesian estimates produced close plots for survival function and hazard function, see Figure \ref{fig:cervical}. It was only observed that for the lifetime $ T_1 $ the fitted hazard based on the  BDGD model  there was a slightly change from the empirical hazard curve obtained from  package ``bshazard" (panel (c)). In this real data a cure rate in both $ T_1 $ and $ T_2 $ is moderate. Apparently, in $ T_2 $ the cure rate estimated by the  and Bayesian approaches (26\%) does not follow the plateau close to the value 0.40 indicated from the Kaplan-Meier curve. However, the hazard curves based on the BDGD fitting are very close to the empirical  hazard function (panel (d)). 

\begin{table}[H]
	\centering
	\caption{Maximum likelihood estimates for the parameters of the BDGD model for the  cervical cancer data.}
	\label{tab:cervical}
	\begin{tabular}{@{}ccccccc@{}}
		\toprule
		\multirow{2}{*}{Parameters} & \multicolumn{3}{c}{Maximum Likelihood Estimators} &  & \multicolumn{2}{c}{Bayesian Estimators} \\ \cmidrule(l){2-7} 
		& Estimate & \begin{tabular}[c]{@{}c@{}}Standard \\ Error\end{tabular} & 95\%  CI &  & Median & 95\%  CrI \\ \midrule
		$ \alpha_1 $ & 0.4520 & 0.0775 & (0.3744, 0.5296) &  & 0.4431 & (0.3547, 0.5541) \\
		$ \alpha_2 $ & 0.2060 & 0.0357 & (0.1703, 0.2418) &  & 0.1996 & (0.1525, 0.2473) \\
		$ \beta_1$ & 0.4580 & 0.0859 & (0.3721, 0.5440) &  & 0.4524 & (0.3559, 0.5451) \\
		$ \beta_2 $ & 0.1537 & 0.0501 & (0.1036, 0.2038) &  & 0.1499 & (0.1026, 0.1975) \\
		$ \rho_1 $ & 0.3727 & 0.0935 & (0.1894, 0.5561) &  & 0.3736 & (0.2463, 0.4961) \\
		$ \rho_2 $ & 0.2617 & 0.1296 & (0.0076, 0.5159) &  & 0.2649 & (0.1146, 0,4258) \\
		$ \phi$ & 7.8998 & 1.2166 & (5.5151, 10.2845) &  & 8.0193 & (5.1498, 10.8580) \\ \midrule
		$\tau_k $ & 0.7979 & 0.0158 & (0.7670 0.8290) &  & 0.8003 & (0.7202, 0.8444) \\
		$\tau_s$ & 0.9398 & - & - &  & 0.9411 & (0.8890, 0.9635) \\\bottomrule
	\end{tabular}
	\caption*{\small 95\%CI: 95\% confidence interval; 95\%CrI: 95\% credible interval.}
\end{table}

\subsection{Application to a tobacco-stained fingers data set}
In this application, we performed a retrospective cohort study on a sample of 143 smokers screened between March 2006 and January 2010 in a 180-bed community hospital in La Chaux-de-Fonds, Switzerland. Data on death and hospital admission were collected until June 2014. More details on this data set can found in \cite{tobacco}. In this bivariate study, it is considered as  $ T_1 $ the time before the first hospital readmission in smokers with stains on their fingers which was censored in case of death before the closure date;the lifetime $ T_2 $ is  the survival time of the patient  with tobacco-tar stain on their fingers. There was 26\% censored data in $ T_1 $ and 48\% censored data in $ T_2 $.

\begin{figure}[H]
	\begin{center}
		\includegraphics[scale=0.6]{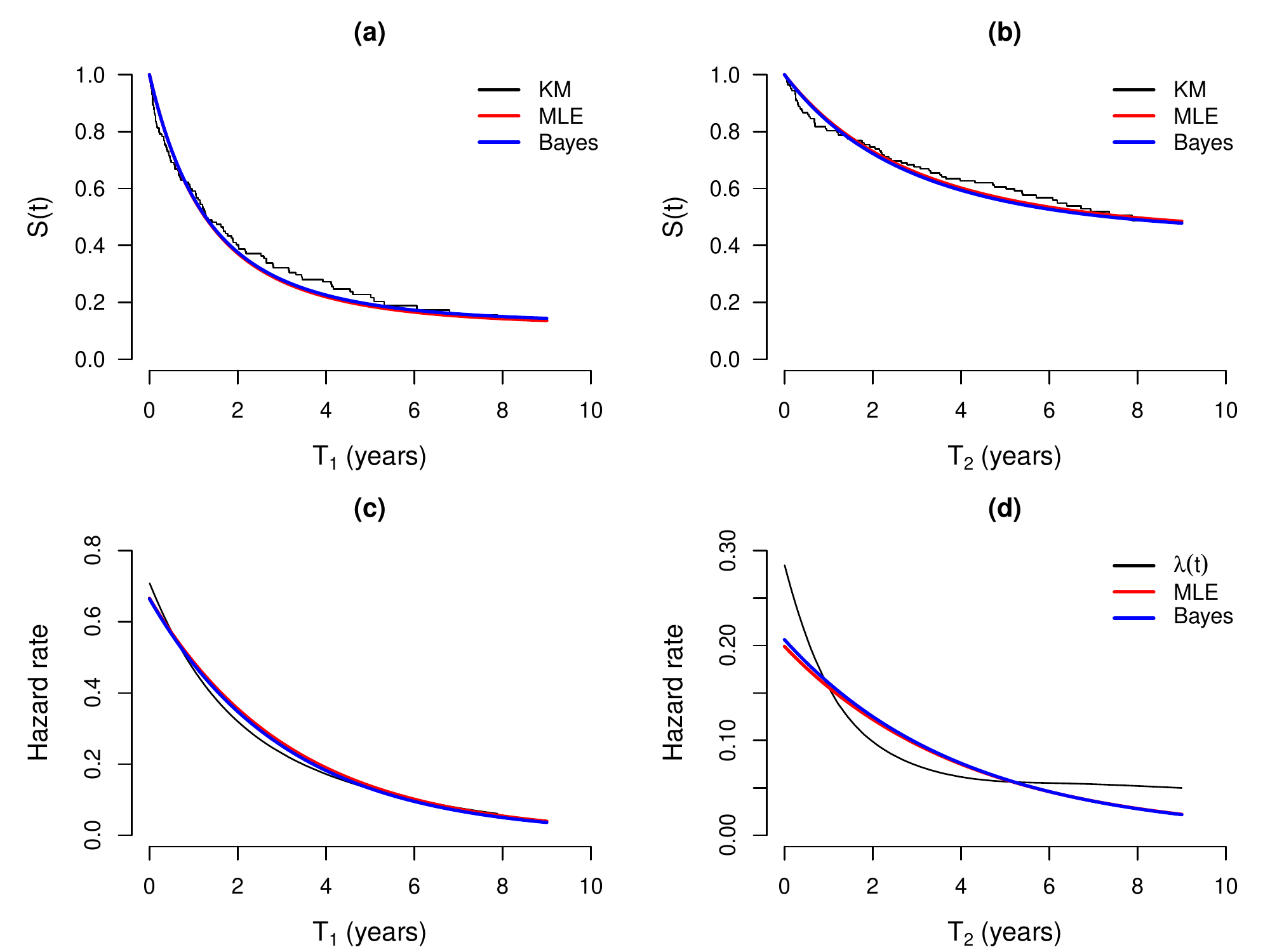}
		\caption{Plots of the survival functions estimated by Kaplan-Meier method and from the BDGD (upper panels) and respective hazard functions (lower panels) for first time hospital readmission (panels (a) and (c)) and survival time  (panels (b) and (d)), considering the tobacco-stained fingers data.}
		\label{fig:tobacco}
	\end{center}
\end{figure}

Table \ref{tab:tobacco}  shows the ML and Bayesian estimates for the parameters of the BDGD model considering the  tobacco-stained fingers.
In this application also was needed to assume more informative uniform prior distributions for the parameters   $ \alpha_1$ and $ \beta_i\; (i=1,2) $. The estimates obtained considering the ML and Bayesian approach produced similar values, except for the parameter $\phi$, where the Bayesian estimates were higher than the ML estimates.
The value of  $ \tau_s $ obtained from ML estimates by copulas functions is equal to the empirical correlation ($\tau_e=0.4998 \, (0.2658, 0.6782)$). Bayesian estimates of  $ \tau_s $ are contained in the 95\%  confidence interval for $ \tau_e $. Similar results were obtained by \cite{de2019discrete}. We can see in Figure \ref{fig:tobacco} that the Kaplan-Meier plot indicates that $ T_1 $ has low cure rate and $ T_2 $ shows moderate cure rate (upper panels). The survival and hazard curves produced by  ML estimates and Bayesian estimates are very similar. Based on the Kaplan-Meier curve for the empirical survival function, it is noted that the estimated curves were satisfatory fitted for $ T_1 $ and $ T_2 $. The BDGD model adequately estimated a cure rate in both ML and Bayesian approaches. Observing the estimated hazard function,  we see that the proposed model captures in a good way , the decreasing shape of the hazard function. However, the hazard function estimated for $ T_2 $ based on the proposed BDGD model does not fully follow the behavior of the hazard function obtained from the package ``bshazard" (panel (d)). It may be needed a more flexible distribution for a perfect fit of the  hazard function in $ T_2 $.

\begin{table}[H]
	\centering
	\caption{Maximum likelihood estimates for the parameters of the BDGD model for the  tobacco data.}
	\label{tab:tobacco}
	\begin{tabular}{@{}ccccccc@{}}
		\toprule
		\multirow{2}{*}{Parameters} & \multicolumn{3}{c}{Maximum Likelihood Estimators} &  & \multicolumn{2}{c}{Bayesian Estimators} \\ \cmidrule(l){2-7} 
		& Estimate & \begin{tabular}[c]{@{}c@{}}Standard \\ Error\end{tabular} & 95\%  CI &  & Median & 95\%  CrI \\ \midrule
		$ \alpha_1 $ & 0.6662 & 0.0923 & (0.5739, 0.7585) &  & 0.6648 & (0.5095, 0.7937) \\
		$ \alpha_2 $ & 0.1990 & 0.0376 & (0.1614, 0.2367) &  & 0.2062 & (0.1532, 02479) \\
		$ \beta_1$ & 0.3141 & 0.0715 & (0.2426, 0.3856) &  & 0.3242 & (0.2538, 0.3962) \\
		$ \beta_2 $ & 0.2444 & 0.0625 & (0.1819, 0.3070 ) &  & 0.2500 & (0.1549, 0.3450) \\
		$ \rho_1 $ & 0.1199 & 0.0590 & (0.0041, 0.2357) &  & 0.1303 & (0.0552, 0.2445) \\
		$ \rho_2 $ & 0.4428 & 0.1148 & (0.2177, 0.6681) &  & 0.4419 & (0.2381, 0.6075) \\
		$ \phi$ & 1.0477 & 0.2795 & (0.4998, 1.5955) &  & 1.3635 & (0.5574, 7.9728) \\ \midrule
		$\tau_k $ & 0.3437 & 0.0315 & (0.2820 0.4056) &  & 0.3889 & (0.2179, 0.4965) \\
		$\tau_s$ & 0.4921 & - & - &  & 0.5463 & (0.3204, 0.6784) \\ \bottomrule
	\end{tabular}
	\caption*{\small 95\%CI: 95\% confidence interval; 95\%CrI: 95\% credible interval.}
\end{table}

\section{Concluding remarks}
A new bivariate lifetime distribution model was proposed in this article based on a defective Gompertz distribution and using the Clayton copula function in presence of cure fraction.  Considering this new model, we performed a comprehensive simulation study to describe the performance  of the inference results under the ML approach. This model is efficient to fit data with weak and strong correlation between lifetimes the $ T_1 $ and $ T_2 $ in several scenarios. However, in the situations where there is a high proportion of cure fraction and small sample sizes ($ n<100 $)a careful use of this model is required. It was  observed that the estimates are more easily obtained if the lifetime variables have values lower than 20, which demands some transformation in the data in some applications. In the application studies it was verified that both the ML and Bayesian methods are suitable approaches to estimate the parameters of the BDGD model. It is important to point out that a suitable choice for the initial values in the ML iterative estimation procedure is required, as well as the Bayesian method depends on adequate hyperparameter values for the prior probability distributions for the parameters of the BDGD model. The applications in simulated and real data evidenced that the BDGD model can be satisfactorily fitted in most cases, considering both the ML and Bayesian approaches. Lastly, we conclude that  the proposed model can be  easily implemented using \textit{R} or {Rjags} softwares.

\bibliographystyle{apalike}
\bibliography{Ref}

\end{document}